\documentclass[lettersize,journal]{IEEEtran}
\usepackage{amsmath,amsfonts}
\usepackage{algorithm2e}
\usepackage{array}
\usepackage[caption=false,font=normalsize,labelfont=sf,textfont=sf]{subfig}
\usepackage{textcomp}
\usepackage{stfloats}
\usepackage{url}
\usepackage{verbatim}
\usepackage{graphicx}
\usepackage{cite}
\usepackage{svg}
\usepackage{graphicx}
\usepackage{longtable}
\usepackage{morefloats}
\usepackage{xurl}
\usepackage{comment}
\usepackage{pdfpages}
\usepackage[dvi]{totpages}
\usepackage[tbtags]{mathtools}
\usepackage[square, sort&compress,comma,numbers]{natbib}
\usepackage{ragged2e}
\usepackage[none]{hyphenat}
\usepackage{xr}
\usepackage{afterpage}
\usepackage{tabularray}
\usepackage{pdflscape} \usepackage{subcaption}
\usepackage{pgfplots}
\usepackage{changepage}
\usepackage{pgf}
\usepackage[utf8]{inputenc}
\usepackage{xstring}
\usepackage{fmtcount}
\usepackage{pgf-umlsd}
\usepackage{amsmath}
\usepackage{float}
\usepackage{caption}
\usepackage{amsmath}
\usepackage{amssymb}
\usepackage{booktabs}
\usepackage{tabularx}
\usepackage{array}
\usepackage{listings}
\usepackage{multirow}

\usetikzlibrary{positioning, patterns, decorations.pathreplacing}
\usetikzlibrary{calc,math}
\usetikzlibrary{arrows,automata,shapes,backgrounds}
\usetikzlibrary{shapes.geometric}
\usetikzlibrary{fit}
\usetikzlibrary{matrix}
\usetikzlibrary[plotmarks]
\usetikzlibrary{patterns}

\definecolor{specialYellow2}{HTML}{cbf542}
\definecolor{specialGreen}{HTML}{34eb74}
\definecolor{specialRed}{HTML}{eb346b}
\definecolor{specialBlue}{HTML}{31d3e8}
\definecolor{specialYellow}{HTML}{acc716}
\definecolor{specialOrange}{HTML}{e06f24}
\definecolor{specialPink}{HTML}{fd6c9e}
\definecolor{specialNavyBlue}{HTML}{0ba6d9}
\definecolor{SpecialBlueConsensus}{HTML}{0080ff}
\definecolor{SpecialRedConsensus}{HTML}{ff0000}
\definecolor{SpecialGreenConsensus}{HTML}{00ff00}
\definecolor{SpecialOrangeConsensus}{HTML}{ffa500}

\tikzset{
    solid/.style={dash pattern=},
    dotted/.style={dash pattern=on \pgflinewidth off 2pt},
    densely dotted/.style={dash pattern=on \pgflinewidth off 1pt},
    loosely dotted/.style={dash pattern=on \pgflinewidth off 4pt},
    dashed/.style={dash pattern=on 3pt off 3pt},
    densely dashed/.style={dash pattern=on 3pt off 2pt},
    loosely dashed/.style={dash pattern=on 3pt off 6pt},
    dashdotted/.style={dash pattern=on 3pt off 2pt on \the\pgflinewidth off 2pt},
    densely dashdotted/.style={dash pattern=on 3pt off 1pt on \the\pgflinewidth off 1pt},
    loosely dashdotted/.style={dash pattern=on 3pt off 4pt on \the\pgflinewidth off 4pt},
}

\usepackage[
    pdfstartview=XYZ,
    bookmarks=true,
    colorlinks=true,
    linkcolor=blue,
    urlcolor=blue,
    citecolor=blue,
    linktocpage=true,
    hyperindex=true
]{hyperref}

\usepackage{orcidlink}

\begin{document}

\title{Securing Edge \& Physical AI: Post-Quantum IoT Blockchain for Mission-Critical Cyber-Physical Agents\\}

\author{K\"{u}r\c{s}at~\c{C}akal  \orcidlink{0000-0002-9116-4910} and~\IEEEmembership{}Suat~\"{O}zdemir  \orcidlink{0000-0002-4588-4538}~\IEEEmembership{}% <-this % stops a space
        % <-this % stops a space
\thanks{K. \c{C}akal is with the Department
of Computer Engineering, Hacettepe University, Ankara, 06800 Turkey e-mail: kursatcakal@hacettepe.edu.tr, kursat.cakal@roketsan.com.tr}% <-this % stops a space
\thanks{S. \"{O}zdemir is with the Department
of Computer Engineering, Hacettepe University, Ankara, 06800 Turkey e-mail: ozdemir@cs.hacettepe.edu.tr}% <-this % stops a space
\thanks{Manuscript received August 3, 2023; revised October 24, 2026.}} 

%\markboth{IEEE COMMUNICATIONS SURVEYS \& TUTORIALS, ~VOL.~xx, NO.~xx, SEPTEMBER~2026}%
%\markboth{IEEE INTERNET OF THINGS JOURNAL, ~VOL.~xx, NO.~xx, SEPTEMBER~2026}%
\markboth{}%
{K\"{u}r\c{s}at and Suat \MakeLowercase{\textit{}}: Article with IEEEtran.cls for IEEE Journals}

\maketitle
\begin{abstract}
Distributed Cyber-Physical Systems (CPS) and Edge AI networks increasingly depend on public-key authentication and key-establishment mechanisms that are vulnerable to cryptographically relevant Quantum Computers (QC). Shor's algorithm threatens widely deployed asymmetric primitives, while Grover's algorithm reduces the effective security margins of symmetric and hash-based mechanisms. Although Post-Quantum Cryptography (PQC) provides a migration path, its larger public keys, ciphertexts, and signatures introduce additional computation, memory, communication, storage, and packet-fragmentation costs that can constrain dynamic multi-agent systems. In this article, we introduce rotational secrecy, an architectural objective that limits temporal key exposure through authenticated transitions between cryptographic epochs. We present Private Authentication for Rotational Secrecy (PARS), a reference mechanism that coordinates policy-triggered key establishment, rotation, and reconciliation among authorized Physical AI agents. For deployments that need distributed agreement on security state, we further develop the PARS-Enabled Edge Security Ledger (PARS-ESL), an optional permissioned architecture maintained by authorized validators or capable edge gateways. PARS-ESL does not store high-volume operational transactions. Instead, it records authenticated state for key epochs, group membership, revocation, algorithm-suite policies, challenge validation, and recovery checkpoints. As a result, resource-constrained agents can take part as lightweight clients, and low-latency operational traffic stays outside the ledger consensus path, protected by the currently active symmetric epoch keys. Our results show that ledger-assisted post-quantum key-epoch management is feasible for selected Edge AI and CPS security-control workloads under the stated trust, compromise, connectivity, and resource assumptions.

\end{abstract}

\begin{IEEEkeywords}
Physical AI, post-quantum cryptography, cyber-physical systems, blockchain, cybersecurity.
\end{IEEEkeywords}

\section{Introduction} \label{Introduction}

\begin{figure*}[!htb]
\centering
\includegraphics[width=1.0\linewidth]{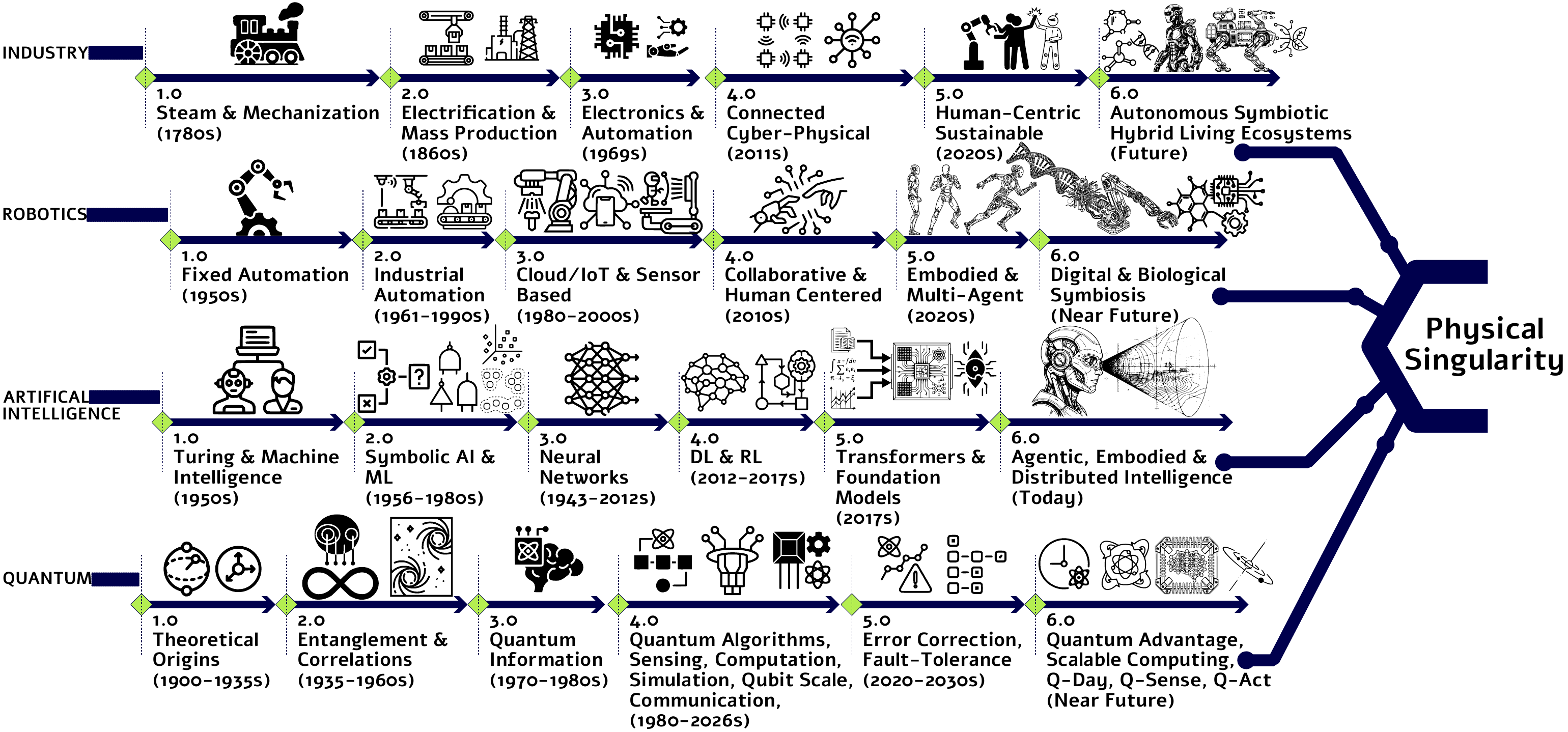}
\caption{Evolution of Physical AI Systems \cite{b11-robinmurphy,b11-zamalloa2017,b11-huyxiejain2023}.}
\label{fig_HistoricalEvolution}
\end{figure*}

\IEEEPARstart{C}{yber}-Physical Systems (CPS) increasingly integrate computation, communication, sensation, perception, actuation and learning at Edge with Physical AI by enabling robots, vehicles, industrial controllers, unmanned platforms, and other autonomous agents to perceive their environments, make local decisions, and coordinate at the network edge over an interior Act, Sense, Compute, Analysis, Learn and Decide (ASCAL-D) loop. These agents must authenticate peers, manage dynamic membership, and update cryptographic \IEEEpubidadjcol keys for secure data transfer. Unlike cloud-centric services, these systems often operate under stringent processing, memory, energy, timing, and communication constraints while maintaining mission/safety-critical functions during intermittent or degraded connectivity. Their security therefore depends not only on protecting individual devices, but also on preserving authenticated communication, consistent membership information, and synchronized cryptographic state across the broader multi-agent network. Cryptographically relevant Quantum Computers (QCMPs) threaten widely deployed asymmetric mechanisms and reduce the security margins of selected symmetric and hash-based constructions on which these trust relationships depend. In the physical edge where AI meets robots, space, defense and more, these threats will not only affect our digital existence but also physical existence. The migration to Post-Quantum Cryptography (PQC) introduces larger public keys, ciphertexts, and signatures that can increase packet fragmentation, transmission latency, memory demand, and energy consumption. The combination of technological bounce at Quantum, Robotics, Industry and AI evolution (see Fig. \ref{fig_HistoricalEvolution}) effects outcome at the edge, where security-control operations must coexist with latency-sensitive sensing, inference, coordination, and actuation.

Throughout the annals of human history, warfare has stood as a persistent force, leaving an indelible mark on nations, cultures, and societies. The evolution of conflict, from ancient battles characterized by swords and shields to contemporary confrontations propelled by advanced technology, reflects a dynamic transformation in the nature of warfare. The intersection of war and technology has ushered in a new epoch, fundamentally reshaping the very essence of the battlefield. Modern warfare is now on the verge of a new transformation. The CPS-Oriented Advanced Autonomous Agents and Swarms play a significant and evolving role in the new phase of modern warfare, leveraging advancements in technology to enhance strategic, tactical, and operational capabilities over C4ISR (Command, Control, Communications, Computers, Intelligence, Surveillance and Reconnaissance) architectures. In the rapidly evolving landscape of advanced technologies, authorities such as DARPA, NASA, NATO, U.S. DoD, NSA, NCI, and other key organizations are recognizing the transformative potential of quantum computers, particularly in enhancing sensing paradigm and security architectures. Blockchain, a decentralized and immutable ledger, is being explored as a solution to ensure data integrity and prevent unauthorized actions on mission critical systems.

DARPA, known for pioneering technological advancements, is investing in blockchain technology to bolster data integrity in critical military systems. Blockchain could serve as a crucial tool to detect unauthorized modifications in data, particularly in areas where information integrity is paramount like defense and space, such as Nuclear Command and Control(C2), Satellite C2, and Tactical C2 operations. The agency has awarded contracts, including a \$1.8 million \cite{DARPA2016Quartz} to Galois for formal verification of blockchain technology supplied by Guardtime, emphasizing the importance of nearly unhackable code in its security approach. Moreover, the agency blockchain program manager states \cite{IndraStraOpenAccess} the employed weapons raise concerns about the integrity and it is incredibly important for nuclear, satellite and tactical C2 systems.

Similarly, NATO has taken a proactive stance with its first-ever quantum strategy \cite{NATO2024QuantumStrategySummary}, approved by NATO Foreign Ministers. The strategy outlines \cite{NATO2024QuantumStrategy} how quantum technologies can be applied to defense and security, spanning areas such as sensing, imaging, navigation, and data communications. Recognizing the implications of quantum technologies for defense and security, NATO is fostering cooperation with industry to develop a transatlantic quantum technologies ecosystem. This strategic move aims to prepare NATO to defend against potential malicious use of quantum technologies, aligning with the organization's broader focus on Emerging and Disruptive Technologies (EDTs).

Quantum technologies are already integral to NATO's innovation efforts \cite{NCIAgencyInnovationChallange}, with selected companies contributing innovations in next-generation cyber defence, CPS-infrastructures and cutting edge sensor applications for defense and space context such as satellite connectivity, military applications of blockchain for challenging environments. According to building milestones on this momentum, NATO plans to establish a Transatlantic Quantum Community by engaging government, industry, and academia to advance quantum initiatives. NATO plans to enhance operational resilience by Secure C4ISR Architectures \cite{NATO_C4ISR_AtlanticCouncil, NCIAgencyCommunicator} for Space, Cyberspace, Air, Land, and Maritime Systems with Collaborative NATO-Industry Partnerships.

As part of the strategic concept, NATO recognizes the critical role of technology, including EDT \cite{NATO2024EDT} under the DIANA \cite{NATO2024DIANA} which is responsible for shaping the future of the NATO alliance with leading researchers and entrepreneurs in the Alliance. The Defence Innovation Accelerator for the North Atlantic (DIANA) is designed to leverage the opportunities arising from EDTs, enhancing NATO's competitive advantage in collective defense and security. The development of new capabilities aims to strengthen the Alliance's responsiveness not only to conventional threats but also to challenges posed by these evolving technologies. Likewise, the Science for Peace and Security (SPS) Programme \cite{NATO2024SPS} supports research activities addressing technological trends, with a particular focus on quantum technologies. The integration of Quantum Key Distribution (QKD) and Post-Quantum Cryptography (PQC) is being explored to holistically improve infrastructure security.

The U.S. Space Operations Command states \cite{NASAspacetechUSSpaceOperations} highest concern is protecting weapon systems. Genuine concern about our weapons systems arises due to the interconnected nature of each system communicating with another. Ensuring the security of this network is imperative, as highlighted by Maj. Gen. Douglas Schiess, the commander for U.S. Space Command’s Combined Force Space Component Command, during discussions with reporters at annual conference \cite{US_AFA_Space_Order_of_Battle} of the Air \& Space Forces Association. In addition, according to the open sources \cite{NATO_C4ISR_AtlanticCouncil} destructive cyber malware targeting government and military increasing C2 systems and rendering multi-domain operations inoperable on the battlefield. 

\begin{figure} [!htb]
\includegraphics[width=\columnwidth]{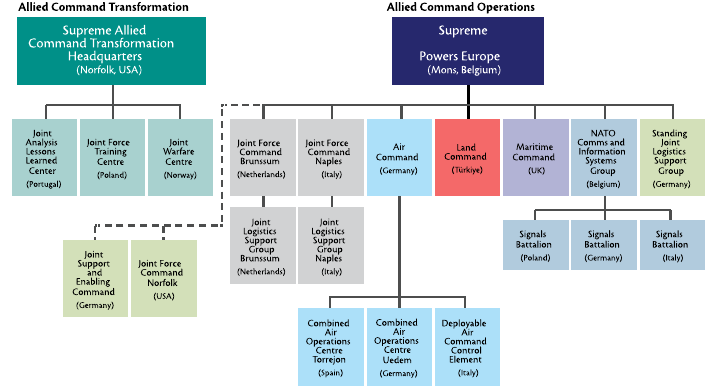}
\caption{North Atlantic Treaty Organization Command Structure}
\label{fig_NATO_Command_Structure}
\end{figure}

The evolution of these technologies highlights the critical requirement for secure communications during the physical agent age, especially given the escalating volume of information transmitted via wireless and optical fiber networks. From everyday online transactions to critical scenarios involving sensitive data transmission between countries under emergency conditions, the imperative of absolute communication security is underscored, considering the ever-present risks of cyber attacks. In a nutshell, military agents and swarms contribute to modern warfare by providing enhanced capabilities in information gathering, surveillance, reconnaissance, coordination, and adaptability. Their deployment helps military forces operate more effectively, efficiently, and safely in complex and dynamic environments. The main motivation of the proposed thesis is ensuring a pioneering robust security backbone for cumulative operability of AI, Data, Autonomy, Space, and Quantum-Enabled Technologies which are 5 of 9 prioritized areas by NATO between Allies (see Fig. \ref{fig_NATO_Command_Structure} \cite{NATO_C4ISR_AtlanticCouncil}).

This study treats post-quantum migration as a cross-layer communications and systems-engineering problem rather than a direct substitution of one cryptographic primitive for another. It introduces rotational secrecy as a general design objective for limiting temporal key exposure through authenticated, policy-controlled transitions between cryptographic epochs. The PARS framework is used as a reference mechanism for coordinating time, action, and group triggered key evolution in distributed Edge AI and CPS environments. An optional permissioned Edge Security Ledger (ESL) complements PARS by maintaining shared security state, including key epochs, membership changes, revocation information, algorithm identifiers, and validation outcomes, without recording high-rate application traffic. The ESL is therefore positioned as a local security-state coordination mechanism rather than a conventional transaction-oriented blockchain. It is not required for every deployment and suitability depends on the trust model, audit requirements, connectivity conditions, validator assumptions, and available resource budget. In the bounded environments considered here, blockchain-based mechanisms are attractive because they provide replicated ordering, tamper evidence, and a shared view of security state across authorized participants, properties previously explored in high-assurance CPS settings ranging from defense, space and other industries that operate safety-critical C4ISR and ASCAL-D environments \cite{DARPA2016Quartz}.

QC has advanced from theoretical possibility toward a disruptive, security-relevant technology \cite{b2-2019QuantumTransformation,b3-2018Huang}, so migration need is no longer speculative but procedural. NIST finalized its first post-quantum standards FIPS 203 (ML-KEM), FIPS 204 (ML-DSA), and FIPS 205 (SLH-DSA) in 2024 \cite{NIST2024FIPS203,NIST2024FIPS204,NIST2024FIPS205} and added the code-based HQC as a backup KEM in 2025 \cite{nistIR8545}, fixing a concrete migration horizon that cascades from federal systems to the broader CPS supply chain. In the most safety-critical sectors, the timeline is mandated rather than advisory with NSA's Commercial National Security Algorithm Suite (CNSA) 2.0 \cite{nsaCNSA2} and the Committee on National Security Systems (CNSS) advisory memorandum on post-quantum cybersecurity bind national-security systems to quantum-resistant algorithms, and comparable readiness pressure is propagating across critical CPS. The underlying threat is hardening fast, resource estimates for factoring a 2048-bit RSA key have fallen from roughly 20 million noisy qubits over eight hours \cite{gidneyEkera2021} to fewer than one million in under a week, while experimental milestones (from programmable quantum advantage \cite{arute2019} to below-threshold error correction on the 105-qubit Willow processor \cite{acharya2024Willow}, and even demonstrations of quantum-secure transactions \cite{cao2024QEcommerce}) mark measurable progress toward fault-tolerant attackers. Market forecasts reinforce the scale of disruption, McKinsey projects a \$1.3 to \$2.7~trillion quantum-computing economic value by 2035 \cite{2026mckinseyquantumcomputers}, but it is the standardization deadlines, not market size, that compel action now. This urgency is organizational as well as procedural for government, standards bodies, innovation initiatives, space programs, and defense alliances have launched dedicated strategies \cite{NATO2024QuantumStrategy,NATO2024QuantumStrategySummary,NATO2024EDT,NATO2024DIANA,NATO2024SPS,NATO_C4ISR_AtlanticCouncil,NCIAgencyInnovationChallange,NCIAgencyCommunicator,US_AFA_Space_Order_of_Battle,NASAspacetechUSSpaceOperations,IndraStraOpenAccess} to prepare critical systems for the post-quantum transition.

\begin{figure}[!htb]
\includegraphics[width=1\columnwidth]{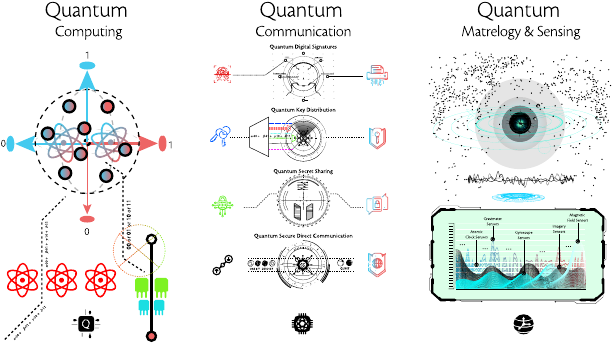}
\caption{Quantum Information Science Overview}
\label{fig_Introduction_Quantum}
\end{figure}

\begin{figure}[htb]
\includegraphics[width=1\columnwidth]{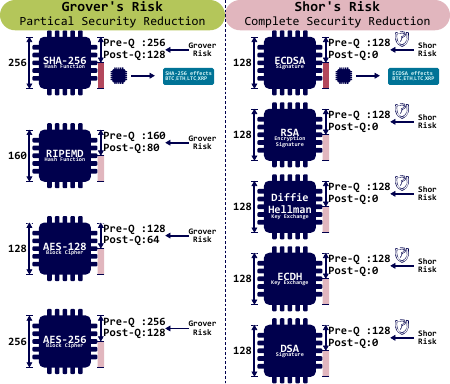}
\caption{Security Levels of Generic Algorithms for Blockchain}
\label{tab_BlockchainSecurityLevels}
\end{figure}

The post-quantum threats and opportunities rests on (see Fig. \ref{fig_Introduction_Quantum}) quantum information science. Two algorithms defines the threat models \cite{b4-2018Suomalainen}. Shor's algorithm \cite{shor1994} breaks the asymmetric primitives behind RSA, DSA, Diffie--Hellman, and elliptic-curve cryptography \cite{b5-2021Asif, b6-2023Chen} the exposure of ECC-secured wallets are the canonical illustration \cite{b7-2020Ilie,b19-2023Hayward} and Grover's algorithm \cite{grover1996} halves the effective strength of the SHA, RIPEMD, and AES primitives used by blockchain (see Fig. \ref{tab_BlockchainSecurityLevels} \cite{b15-2024kcso}), degrading the keys and hashes that protect Bitcoin, Ethereum, and comparable ledgers. Countermeasures fall into quantum security schemes \cite{b8-2020Fernandez, b9-2021Balamurugan, b10-2019Fernandez} (see Fig. ~\ref{fig_Post_Quantum_Cryptosystems}), spanning quantum-mechanical primitives such as quantum key distribution (QKD) and classical PQC. Because CPS endpoints must run on conventional, often deeply embedded hardware, this work address PQC, specifically the Kyber, Dilithium, and SPHINCS+ families benchmarks and novel approach without PQC. Crypto-agility is essential rather than optional, the SIKE isogeny scheme was broken in under an hour and the Rainbow multivariate scheme \cite{b16-2021Chen} was dismantled shortly after deployment, so no single algorithm can be trusted for a system's service life \cite{josephTransition2022}. In addition, there is no evidence for backdoors of NIST algorithms. A final obstacle is spatial and especially punishing at the edge, lattice signatures are large (an ML-DSA-65 signature is 3{,}293 bytes against ECDSA's 64), fragmenting packets over contested, low-bandwidth links and straining storage on
embedded nodes \cite{gharavi2024PQBlockchainIoT}, which motivates lightweight, rotational-key ledgers. 

%\vspace{-2mm}
%\vspace{-2mm}

Recent surveys \cite{yang2024PQBlockchainSurvey, naik2023portfoliooptimizationquantumblockchain, chhetri2025PQCSurvey} consolidate the resulting post-quantum blockchain work for CPS security. Nevertheless, four limitations remain central to ensure security of CPS, Embodied/Physical AI or Edge AI. First, transaction-oriented blockchains commonly retain application histories whose growth is unnecessary for cryptographic control-plane coordination. Second, periodic key replacement is often discussed without distinguishing it from formally defined forward secrecy, backward secrecy, or post-compromise security. Third, many proposals assume that every endpoint can maintain a full ledger, execute expensive validation procedures, or remain continuously connected. Fourth, quantum level hardware requirements at the end points which cause excessive cost and migration period.

After thoroughly reviewing the research literature, this study presents, under the PARS mechanism, a case study outlining the most feasible, cost-effective, mission-critical performance-enhancing, and scalable approach for swarms of edge-physical-embodied AI solutions. The rotational secrecy denotes the broader objective of limiting temporal key exposure through authenticated changes in cryptographic state or transactions between Edge AI nodes. PARS provides the corresponding reference framework by coordinating key updates among authorized peers according to time-based, action-based, and group-based triggers. The optional ESL records the resulting key epochs, membership changes, algorithm identifiers, revocation information, and security-policy state while avoiding the accumulation of complete sensor, telemetry, or control-message histories. The intended benefit is not that key rotation alone guarantees strict forward secrecy or post-compromise recovery. Rather, explicit key epochs and periodic reconciliation can reduce compromise exposure when they are combined with fresh secret contributions, secure state erasure, replay protection, authenticated transitions, and consistent membership management. These assumptions and properties are formalized later through the system, adversary, and protocol models.

To the best of our knowledge, and under the bounded criteria adopted in this article, PARS is an early framework that combines post-quantum-capable rotational key management with an optional permissioned security-state ledger for autonomous multi-agent CPS environments. This statement does not claim priority over post-quantum blockchains, group key-management protocols, crypto-agile security architectures, or PQC algorithms in general. Instead, it substantially researches, investigates and presents the required action stack for securing physical artificial intelligence, with earlier study \cite{b15-2024kcso}, which introduced the initial concept, by providing a broader methodology, a more complete PARS specification, the ESL architecture, a structured quantum-relevant CPS threat taxonomy, and an expanded empirical evaluation. The principal contributions are summarized as follows:

\begin{itemize}

    \item \textbf{A cross-layer engineering methodology for quantum-resilient Edge AI and CPS security:}
    The article connects quantum-threat assessment, PQC primitive selection, hybrid KEM--KDF--AEAD construction, authentication, key evolution, distributed trust, and migration planning within a unified workflow.

    \item \textbf{A structured taxonomy of quantum-relevant threats for physical ai world and architectural alternatives:}
    The study maps adversarial capabilities to affected cryptographic primitives, communication functions, distributed-system layers, and possible CPS consequences. It also distinguishes attacks addressed by cryptography from jamming, malicious firmware, false sensor data, physical capture, and other threats requiring complementary controls and  operational layers over evolving asymmetic space and defense cases.

    \item \textbf{Rotational secrecy as a general Edge AI security objective:}
    The article formalizes authenticated key-epoch evolution as a reusable design principle for limiting temporal compromise exposure in dynamic multi-agent systems, while distinguishing that objective from unconditional guarantees of forward secrecy or post-compromise security. It present PARS as a reference framework for dynamic key evolution, PARS coordinates time/action/group-triggered key updates using classical, hybrid, or post-quantum algorithm suites. Its security properties are stated relative to explicit assumptions concerning identity provisioning, secure erasure, fresh entropy, membership consistency, and network synchronization. 

    \item \textbf{An optional Edge Security Ledger for distributed security-state coordination:}
    The ESL provides a reference architecture for maintaining key epochs, membership records, revocation information, algorithm policy, validation outcomes, and audit-relevant control state without storing high-rate application transactions. The study also identifies when such a ledger is justified and when centralized or replicated non-blockchain alternatives are preferable.

    \item \textbf{An empirical methodology for resource-constrained evaluation:}
    The article evaluates security and communication protocol costs and benchmarking on resource-constrained platforms, representative of Edge AI/CPS endpoints across latency, memory, storage, bandwidth, scalability, and operational reliability, and uses the resulting measurements to guide algorithm-suite and deployment decisions for heterogeneous Edge AI and CPS platforms.  

    \item \textbf{A critical analysis of adaptive challenge mechanisms and consensus:}
    The study examines the proposed adaptive validation mechanism, including its challenge complexity, verification cost, and relationship to Zero-Knowledge Proof (ZKP) terminology. Rather than assuming that increased computational difficulty directly yields stronger zero knowledge, the article separates authentication cost, proof soundness, consensus behavior, and quantum-security assumptions.

\end{itemize}

Collectively, these contributions position PARS and the ESL not as universally required solutions, but as a reusable reference for engineering post-quantum security-control planes under constrained computation, communication, and trust conditions. The remainder of this article first establishes the Edge/Embodied/Physical AI, distributed-ledger, quantum-threat, and PQC background required to analyze rotational secrecy. It then reviews related quantum-safe edge architectures, presents the PARS and PARS-ESL designs, develops the attack and security analysis, evaluates the reference implementation, and concludes with deployment guidance, limitations, and open research directions.

The trust assumptions underpinning distributed CPS and Edge-AI nodes can be invalidated quickly and at minimal cost, as adversaries exploit intelligence gathered clandestinely about how critical systems are designed and operated, an exposure that cryptographically relevant QCMPs will sharpen through expanded SIGINT, as illustrated by the quantum attack scenarios. Edge deployments are particularly exposed: fleets of drones, satellites, robots, vehicles, and industrial controllers must authenticate each other continuously under strict compute, memory, and bandwidth budgets, yet remain reachable by remote adversaries. Consequently, such systems should adopt modular and dynamic security measures that do not depend solely on communication-flow control imposed by designers and developers at deployment time. A dynamic consensus and security-renewal mechanism, together with cryptographic agility across analytically grounded quantum security levels, is therefore essential rather than optional. ESL, powered by the PARS protocol, addresses both quantum attacks and environmental vulnerabilities through a granular layered architecture that safeguards operational flows and raw data against compromise. While the design generalizes across Edge-AI, Robotics, CPS, and IoT domains, most demanding and best-fit setting remains defense and space operations.
\section{BACKGROUND OVERVIEW} \label{Background}

This section establishes the minimum technical bridge required to understand post-quantum security engineering for distributed Edge AI and Cyber-Physical Systems (CPS). It first characterizes the communication, computation, and trust requirements of embodied edge agents. Then it reviews distributed-ledger mechanisms, the quantum-computing concepts relevant to cryptographic migration, and the principal families of Post-Quantum Cryptography (PQC). The objective is not to provide an exhaustive treatment of Edge AI, blockchain, quantum computing, or cryptography in isolation. Rather, the section identifies the concepts that directly influence authenticated key establishment, key-epoch evolution, packet size, resource consumption, distributed security-state coordination, and the design of a generic security mechanism for edge systems which generalized as PARS and ESL.

\subsection{Edge AI Agents and Intelligent Cyber-Physical Systems} \label{ref_label_edge_agents_and_swarms}

Embodied Edge AI agents combine sensing, local inference, communication, decision-making, and physical actuation within resource-constrained CPS platforms. By processing high-throughput sensor streams directly at the operational edge and close to their physical source, these agents execute real-time decentralized inference and autonomous decision-making without high-latency cloud dependency. This edge paradigm maximizes situational awareness within volatile, hazardous, or infrastructure-degraded environments where human deployment is high-risk. Equipped with multi-modal sensor arrays and on-device hardware accelerators, embodied agents achieve high-fidelity object detection and spatial tracking while minimizing classification uncertainty. These agents can reduce cloud dependence, communication latency, and external bandwidth demand while maintaining local operation during intermittent or degraded connectivity. Their hardware commonly integrates heterogeneous sensors, real-time processors, embedded accelerators, communication interfaces, and actuator-control subsystems. Consequently, the security of an Edge AI agent depends not only on the confidentiality of its data, but also on the authenticity, freshness, and consistency of the information used for inference, coordination, and actuation. Furthermore, they facilitate self-healing, ad-hoc peer-to-peer (P2P) communication networks and streamline automated logistics, significantly enhancing the operational throughput and structural safety of distributed CPS.

Decentralized swarms extend this paradigm by enabling multi-agent coordination and collaborative intelligence among autonomous nodes. Through localized collective intelligence, swarms execute complex multi-modal tasks and dynamically exchange state and cryptographic data, achieving a functional multiplier effect that overcomes individual monolithic resource constraints. This fully decentralized topology eliminates single-point-of-failure (SPoF) vulnerabilities inherent to traditional centralized command-and-control frameworks, bolstering system-level resilience against network partitions, physical tampering, or adversarial cyber attacks. Local collaboration, in practise, can support cooperative mapping, environmental sensing, distributed spectrum observation, coordinated navigation, and fault-tolerant task execution. Their architectural redundancy provides high fault tolerance, ensuring that individual node dropouts or adversarial corruption do not compromise the global mission consensus. Consequently, synchronized swarms redefine distributed systems engineering by seamlessly blending on-device edge intelligence with robust structural adaptability.

%\vspace{-2mm}
\begin{figure} [!ht]
\includegraphics[width=1\linewidth]{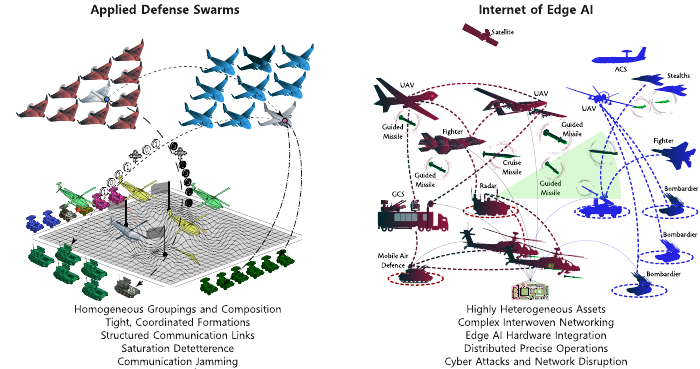}
\caption{Conceptual Overview of Military Edge AI}
\label{fig_Introduction_IoMT}
\end{figure}
%\vspace{-2mm}

The convergence of Edge AI and CPS defines an interconnected network of autonomous, embodied things embedded with heterogeneous sensors and processing units capable of real-time environmental interaction, as illustrated in Fig. \ref{fig_Introduction_IoMT}. This structural framework shifts the operational paradigm away from centralized, cloud-reliant infrastructures toward decentralized machine intelligence capable of distilling high-bandwidth sensory streams into localized insights. Driven by requirements for robust autonomy in infrastructure-deprived or highly adversarial environments—such as autonomous robotic swarms and UAVs—these edge systems mitigate cognitive and physical overloads by deploying localized consensus, adaptive tasks, and automated path planning which is conceptualized within defense sectors as Internet of Military/Battlefield Things (IoM/BT), contemporary frameworks unify these architectures under a comprehensive, secure Embodied Edge AI paradigm.

\begin{figure*} [!htb]
\includegraphics[width=1\linewidth]{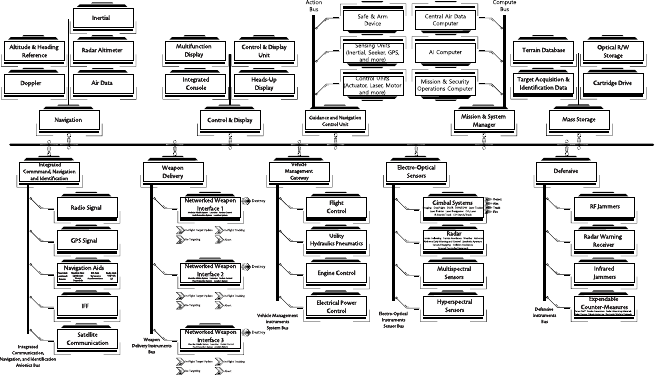}
\caption{Cyberspace Attributes for Mission-Critical Edge AI Avionics}
\label{fig_Avionic_Bus_Architecture}
\end{figure*}

Autonomous edge nodes handle hundreds of physical and operational routing parameters across heterogeneous communication buses, sensor interfaces, and control networks to maintain deterministic functional capabilities for various internal and external data attributes (see Fig. \ref{fig_Avionic_Bus_Architecture}). This deep integration scales the systemic cyberspace attack surface, introducing severe risks regarding signal intrusion and malicious control-loop alteration. Within collaborative swarms executing distributed tasks, safeguarding inter-node telemetry streams against cascading exploits is vital. Potential threat tensors must be analyzed extensively including IFF spoofing, multi-node resource exhaustion, radar warning disruption, Guidance, Navigation and Control (GNC) manipulation, collision avoidance circumvention, flight/engine control subversion and weapon/payload related actions/attributes (acceleration, roll, pitch, yaw, surveillance, target engagement/guidance, operational termination, firing decision, optical mode, optical laser designation/pointing/illumunation, optical pulse rate, target range, ground/air speed, latitude, longitude, altitude, boresight, moment arm, nozzle thrust and more) in terms of aerodynamic, hydrodynamic, thermodynamic, electromagnetic, geodezic, pyrotechnic, fuel kinetic and among many others. These attributes can be grouped into a smaller set of security-relevant categories: platform identity and membership; navigation and motion state; environmental observations; mission or task state; health and diagnostic information; control commands; payload or actuator state; and cryptographic metadata. An adversary need not compromise every attribute to create a physical effect. Manipulation of a small number of authenticated but safety-relevant parameters may be sufficient to disturb coordination, navigation, resource allocation, or actuator behavior.

%\vspace{-2mm}
%\vspace{-2mm}

Edge AI platforms can consequently be characterized along several interacting technical axes. First, pervasive sensing and local analytics support condition monitoring, predictive maintenance, environmental perception, and platform-health estimation. Second, low-latency communication and localized preprocessing shorten decision loops and reduce dependence on remote infrastructure. Third, collaborative intelligence allows groups of agents to distribute sensing, inference, and task execution. Fourth, cryptographic mechanisms must provide authentication, integrity, confidentiality, freshness, and key management within the processor, memory, energy, and link budgets of the physical platform \cite{navigation1982,lockhead1998,2018-ieeesocietyiobt}. PQC introduces an additional systems constraint because post-quantum public keys, ciphertexts, and signatures can be substantially larger than their classical counterparts.

\begin{figure*} [!ht]
\includegraphics[width=1\linewidth]{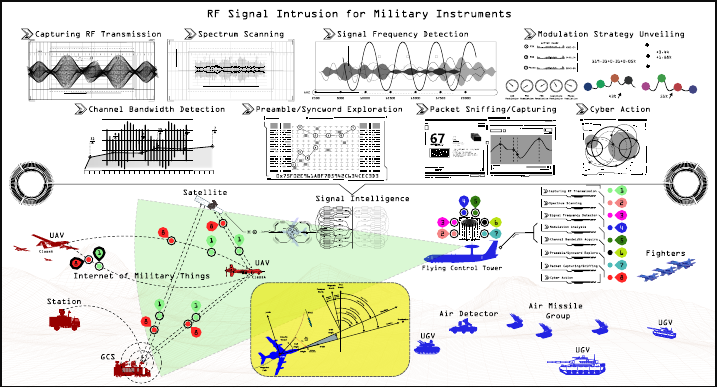}
\caption{Radio Frequency SIGINT in Space and Defense Environments}
\label{fig_RF_Operational}
\end{figure*}

High-assurance data exchange across embodied edge networks encompasses three primary operational dimensions: decentralized Command and Control (C2) directives defining mission constraints; tactical data link protocols transmitting synchronized radar, sonar, and tracking streams; and ultra-low-latency fire control parameters for precise physical actuation. In autonomous robotic and drone systems, these exchanges rely on geospatial state vectors processed natively via onboard avionics such as INS and GPS. Tracking these dynamic states involves the continuous synchronization of coordinate-aligned position vectors, velocity vectors for trajectory forecasting, acceleration vectors for maneuver response estimation, wind vectors for environmental drift compensation, and true/magnetic north vectors for absolute directional alignment. These parameters are further augmented by ground speed vectors for path optimization, heading vectors for orientation monitoring, and slant-range variables that mathematically establish the exact spatial geometry between electro-optical imaging payloads and target objects (see Fig. \ref{fig_RF_Operational}). Securing this comprehensive attribute cluster via P2P cryptographic validation is paramount to preserving collective swarm resilience, because the mentioned attributes has only the small set of effect as well as rest of the attribute space. This reveals the need for a wider principle: the cryptographic architecture should protect the origin, integrity, freshness, and authorization of data without placing expensive public-key operations in every high-frequency control loop.

Deploying Edge AI in physical environments also expands the attack surface. Relevant threats include passive interception, traffic analysis, message replay, identity impersonation, sybil behavior, packet modification, delayed delivery, network partitioning, denial of service, sensor-data manipulation, malicious firmware, control-loop subversion, and physical capture. Standardized or predictable packet structures protocol bases like MIL-STD-XXXX, UART, CAN and more \cite{chirupphapa2022unsupervised, Donmez2021, Zhang2023, MIL-STD-1840C,MIL-HDBK-1553A} can assist an adversary in identifying message types, timing patterns, and system roles within the telemetric packet structures mapped over data links. Cryptographically relevant quantum computers would add a further threat by weakening classical public-key mechanisms based on factorization or discrete logarithms. This vulnerability allows adversaries to manipulate system command structures, bypass authentication blocks, or completely neutralize autonomous inventories via quantum-accelerated attacks without physical kinetic contact. However, quantum capability alone does not create radio access, compromise an endpoint, forge sensor truth, or overcome properly configured symmetric cryptography and PQC. The attack model must therefore distinguish quantum-enabled cryptanalysis from network, software, physical, and operational attacks.

Furthermore, contemporary Radio Frequency (RF), Satellite Communication (SATCOM), and tactical wireless infrastructures remain highly exposed to quantum-era spectral monitoring, a threat compounded by the processing space spectrum as outlined in Fig. \ref{fig_RF_Operational}. Adversaries leverage automated Signal Intelligence (SIGINT) \cite{SIGNINT_NSA_Overview} paradigms to execute passive signal interception across uplinks and downlinks, track transmission patterns to reverse-engineer proprietary protocols, and extract command data to infer operational intent. Over extended exposure windows, intruders decode FHSS sequences to execute eavesdropping, store-analyze-and-forward, adaptive and selective real-time signal injection attacks or subject to harvest-now-decrypt-later collection if vulnerable public-key mechanisms are used to establish session keys.

Mitigating these multi-spectral exploits requires moving away from legacy protocols toward quantum-resistant layered architectures and decentralized cryptographic trust models like the PARS and ESL ensuring absolute operational security under asymmetric threats \cite{US_AFA_Collaborative_Combat_Aircraft_Vectors, US_AFA_Quantum_Technology_and_Defense_Understanding_the_Imperative}. High-rate application traffic should ordinarily be protected using symmetric Authenticated Encryption with Associated Data (AEAD), while authenticated classical, hybrid, or post-quantum key-establishment mechanisms operate in the lower-frequency security-control plane. PARS addresses authenticated key-epoch evolution, while an optional ESL can coordinate membership, revocation, algorithm policy, challenge outcomes, and other shared security state. Neither PARS nor the ESL can guarantee absolute operational security. Their effectiveness depends on trusted provisioning, correct implementation, secure randomness, key erasure, bounded validator compromise, endpoint integrity, and complementary protections against jamming, malicious software, sensor deception, and physical capture.

\subsection{Blockchain} \label{ref_label_blockchain}

\begin{figure} [!h]
\centering
\includegraphics[width=1\linewidth]{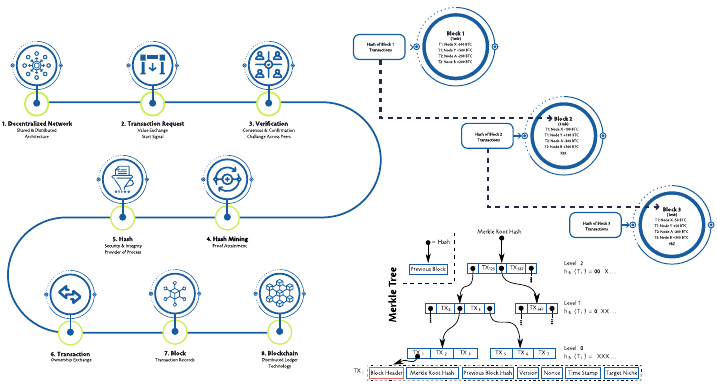}
\caption{Blockchain Pipeline Basic Overview}
\label{fig_Blockchain_Overview_1}
\end{figure}

\begin{figure} [!htb]
\centering
\includegraphics[width=1\linewidth]{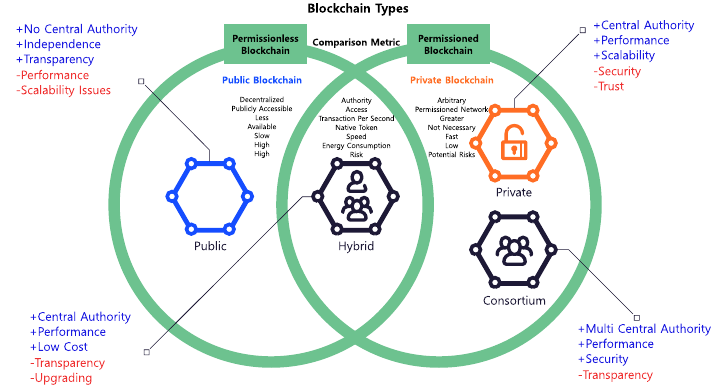}
\caption{Blockchain Types \cite{cansoWhitePapers}}
\label{fig_Blockchain_Types}
\end{figure}

\begin{figure} [!htb]
\includegraphics[width=1\linewidth] {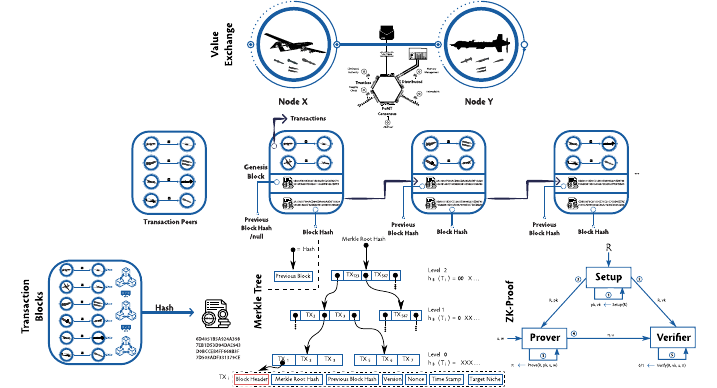}
\caption{PARS-Enabled ESL Overview}
\label{fig_Blockchain_Overview_2}
\end{figure}

A blockchain is an append-oriented replicated ledger in which participating nodes decentralized, distributed P2P ledger that records transactions across an immutable network of nodes, agree on an ordered sequence of state transitions, eliminating single-point-of-failure liabilities inherent to centralized trust anchors \cite{distributedComputingBlockchain}. Each block sequence encapsulates a cryptographic hash of its predecessor, a chronological timestamp, and a nonce, establishing a verifiable and tamper-evident data chain. As illustrated in Figs.~\ref{fig_Blockchain_Overview_1} and \ref{fig_Blockchain_Overview_2}, cryptographic hashes and Merkle commitments provide tamper evidence and efficient membership verification \cite{MerkleTree_Auvolat}, digital signatures authenticate submitted transactions or validator messages, and consensus determines how updates are validated and finalized \cite{Xiao2020Consensus}. Ledger history is therefore not unconditionally immutable: its integrity depends on the cryptographic assumptions, consensus rules, validator threshold, network model, and implementation, including their resistance to emerging quantum threats \cite{gharavi2024PQBlockchainIoT}. Moreover, the permissionless, consortium, and permissioned architectures (see Fig.~\ref{fig_Blockchain_Types}) differ in membership, governance, privacy, finality, scalability, and resource consumption \cite{cansoWhitePapers,Ali2019BlockchainIoT}. Permissioned edge deployments can employ authenticated Byzantine Fault-Tolerant (BFT) consensus without proof-of-work mining; however, ledger integrity and auditability do not inherently provide confidentiality or access control \cite{Salman2019SecurityServices}. Similarly, although Zero-Knowledge Proofs (ZKPs) can verify a statement without exposing its witness \cite{zkProofEmbedded}, a resource puzzle, proof of possession, or generic challenge--response procedure should not be classified as a ZKP unless its completeness, soundness, and zero-knowledge properties are formally established.

The operational efficacy of distributed ledgers stems from five core architectural attributes: decentralization, which distributes ledger validation across heterogeneous nodes via consensus protocols; transparency, which grants verifiable history to all network participants to maximize systemic accountability; immutability, which safeguards data integrity against post-hoc manipulation; smart contracts, which enforce automated, deterministic programmatic execution without intermediaries; and multi-layered cryptographic security, which hardens the underlying network fabric against adversarial exploitation.

At the primitive layer, blockchain utilizes Merkle Trees to enable high-efficiency, sub-linear integrity verification of massive datasets without requiring full data exposure \cite{MerkleTree_Auvolat}, as conceptualized across Fig. \ref{fig_Blockchain_Overview_1} and Fig. \ref{fig_Blockchain_Overview_2}. This data layer relies heavily on deterministic cryptographic hash functions characterized by high avalanche metrics, alongside asymmetric and symmetric cryptographic primitives for robust digital signature verification and data access management. Network coordination is sustained through hardcoded genesis blocks, cryptographic nonces embedded within computationally intensive consensus mechanisms, monotonic timestamps, and structured P2P synchronization routing. To resolve the tension between validation and confidentiality, ZKPs are integrated into the architecture, enabling a proving entity to cryptographically substantiate statement validity or state correctness to a verifier without disclosure of data \cite{zkProofEmbedded}.

\begin{figure} [!htb]
\centering
\includegraphics[width=1\linewidth]{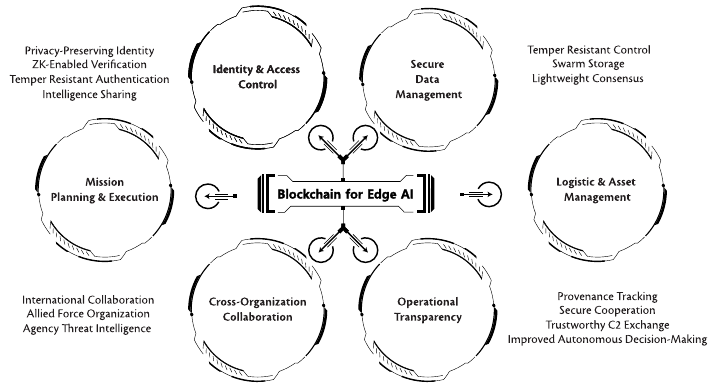}
\caption{Blockchain Use Cases for Edge AI}
\label{fig_Blockchain_Military_Use_Cases}
\end{figure}

While commercial applications cross-cut banking, supply chain logistics, healthcare informatics, and identity federation, the decentralized ledger framework introduces critical architectural primitives to specialized defense systems. Within the scope of this study, the performance and security capabilities of these decentralized features are explicitly mapped onto resource-constrained CPS architectures and tactical Edge AI applications as illustrated in Fig. \ref{fig_Blockchain_Military_Use_Cases}.

As contextualized by Fig.~\ref{fig_Blockchain_Military_Use_Cases}, distributed ledgers can support shared identity, access-control, provenance, and audit functions across Edge-AI domains \cite{Salman2019SecurityServices,Yang2019BlockchainEdge}. In the present architecture, however, the Edge Security Ledger (ESL) has the narrower role of maintaining low-rate security-control state---including key epochs, membership and revocation records, cryptographic-suite identifiers, policy transitions, challenge/proof state, and validation results---rather than continuous sensor streams, inference outputs, telemetry, or real-time control traffic. Such replication is justified when multiple authorized participants require a consistent and auditable security state but cannot continuously depend on a single online authority. Otherwise, centralized key management, hierarchical Public-Key Infrastructure (PKI), or a replicated directory may provide lower latency and complexity. The ESL is therefore optional, and its suitability must be evaluated against validator compromise, consensus and communication overhead, storage growth, finality, partition behavior, privacy, governance, and recovery requirements \cite{Yang2019BlockchainEdge,Xiao2020Consensus}.

\subsection{Quantum Computers} \label{ref_label_quantum_computers}

\begin{figure} [!ht]
\includegraphics[width=1\linewidth]{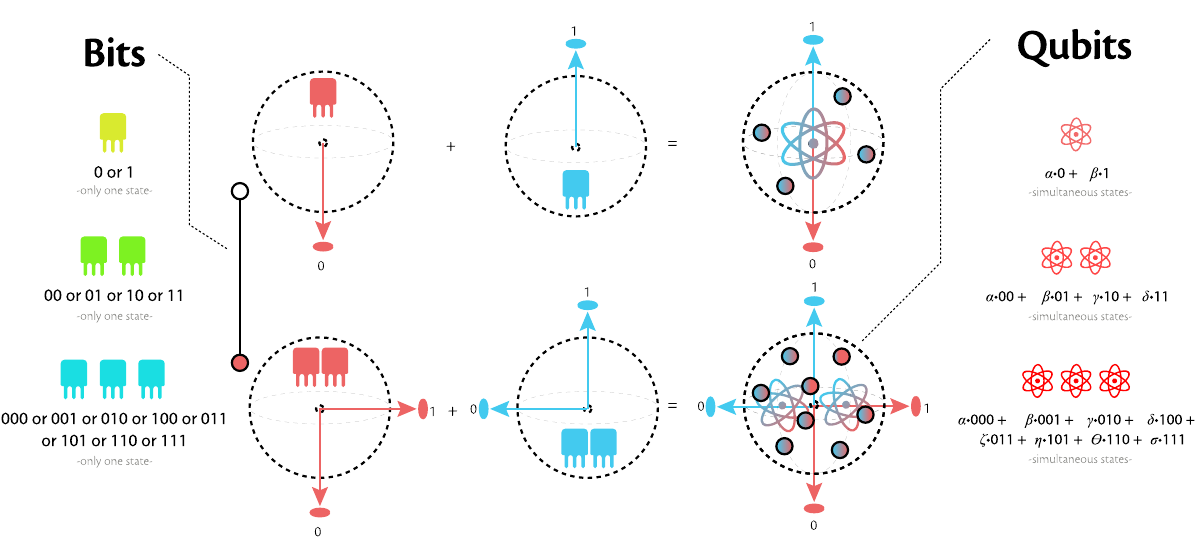}
\caption{Quantum vs Classical Computing Base}
\label{fig_Bits_Qubits}
\end{figure}

Quantum computing transitions from classical binary logic to quantum mechanical principles by replacing classical bits with quantum bits (qubits). Leveraging superposition, qubits exist in multiple linear combinations of states simultaneously, allowing quantum systems to execute complex calculations far more efficiently than classical architectures. While large-scale, fault-tolerant QCMPs remain under active development, their emerging capability to solve classically intractable problems introduces profound implications for modern cybersecurity. This paradigm shifts defensive focus toward critical areas such as Quantum Key Distribution (QKD) for secure, eavesdropping-resistant key exchange, enhanced security protocols integrating post-quantum primitives, and quantum-secure networks establishing robust communication foundations. Fig. \ref{fig_Bits_Qubits} illustrates the structural divergence between classical and quantum processing layers.

\begin{figure} [!htb]
\includegraphics[width=1\linewidth]{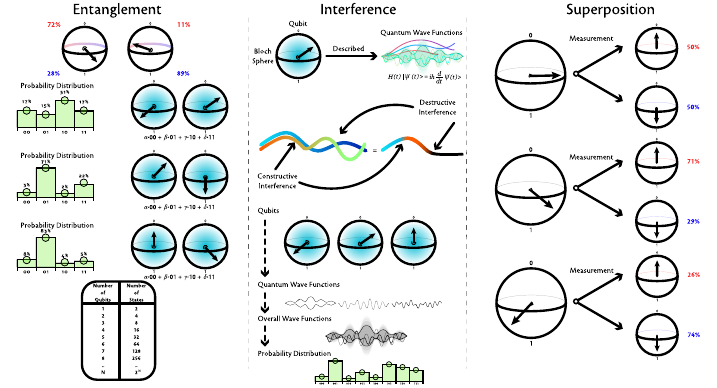}
\caption{Fundamental Concepts for Quantum Computers}
\label{fig_Quantum_Computing_Ent_Sup}
\end{figure}

The core operational mechanics of QC rely on distinct physical phenomena, conceptually outlined in Fig. \ref{fig_Quantum_Computing_Ent_Sup}. Superposition allows qubits to maintain simultaneous representations of multiple states, forming the baseline for quantum parallelism where systems evaluate vast configurations concurrently. Entanglement establishes non-local correlations between distinct qubits, ensuring that the state of one instantly dictates the state of another regardless of physical separation. Operations on these states are executed via quantum gates, which directly manipulate the underlying wavefunctions. Finally, quantum interference governs these operations by exploiting the wave-like properties of particles, utilizing constructive and destructive patterns to systematically modulate probability amplitudes across the computational state space.

The exceptional performance scaling of quantum systems originates from the systemic integration of these physical principles into algorithmic design. Superposition allows an $n$-qubit architecture to evaluate $2^n$ states in parallel, generating exponential speedups for specialized optimization, factorization, and simulation tasks. Entanglement amplifies this efficiency by enabling highly coordinated multi-qubit transformations, which are essential for complex information transfer and distributed quantum protocols. Crucially, quantum interference acts as an accuracy-filtering mechanism; by engineering destructive interference to suppress the probability amplitudes of erroneous pathways, algorithms systematically maximize the probability of measuring the correct final state. This targeted resource allocation enables quantum algorithms to efficiently resolve classically infeasible problem spaces.

\subsection{Post-Quantum Cryptography} \label{ref_label_post_quantum_cryptography}

\begin{figure} [h]
\includegraphics[width=1\linewidth]{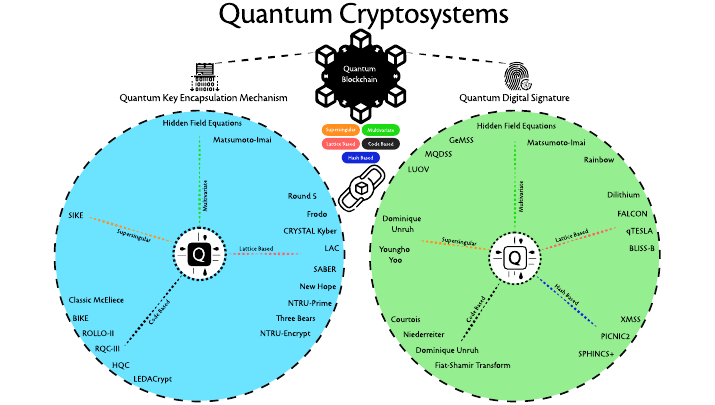}
\caption{Cryptography Systems for Blockchain Architecture}
\label{fig_Post_Quantum_Cryptosystems}
\end{figure}

PQC encompasses classical cryptographic algorithms designed to remain secure against adversaries equipped with both classical and quantum computation. Relevant families include lattice-based, code-based, hash-based, multivariate, and historically isogeny-based constructions, as summarized in Fig.~\ref{fig_Post_Quantum_Cryptosystems} \cite{b8-2020Fernandez}. Their security rests on mathematical problems for which no efficient classical or quantum attack is currently known under the selected parameters. PQC should be understood in terms of cryptographic functions rather than as a single category of ``quantum-resistant encryption.'' Key-Encapsulation Mechanisms (KEMs) establish shared secret material; digital-signature schemes provide origin authentication and integrity; Key Derivation Functions (KDFs) derive independent traffic keys; and symmetric AEAD protects high-rate data. Hybrid cryptographic systems combine classical and post-quantum mechanisms so that security does not depend exclusively on one family during migration. This integration can increase handshake size, certificate size, computation, memory use, and fragmentation, particularly in protocols such as Transport Layer Security \cite{b8-2020Fernandez}.

\begin{figure} [!htb]
\includegraphics[width=1\linewidth]{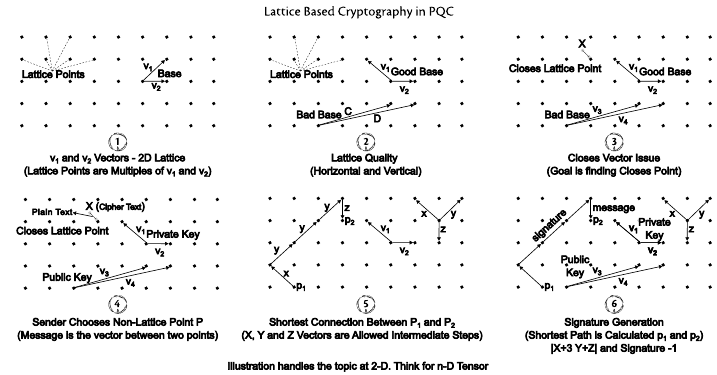}
\caption{Lattice Based Cryptography Overview} 
\label{fig_PQC_Lattice}
\end{figure}

Lattice-Based Crypto Systems (see Fig. \ref{fig_PQC_Lattice} \cite{PQC_Lattice_CryptoVision}) leverage lattices, multi-dimensional mathematical structures. Security hinges on finding the nearest point within a lattice, where each lattice point corresponds to an asymmetric key. While these algorithms demand significant key memory and usage, they offer faster transactions and efficient execution for Blockchain applications by offering both enhanced security and quicker transaction processing \cite{b5-2021Asif,b9-2021Balamurugan}. In lattice-based cryptography, key generation lays the foundation for secure communication channels. Initially, parameters are meticulously chosen to define the dimensionality of the lattice, along with security parameters to withstand potential attacks. Following this, a lattice basis is generated, often through randomized or structured methods, ensuring its orthogonality or a close approximation to facilitate efficient lattice operations. Subsequently, the public and private keys are derived. The public key is crafted by introducing randomness and noise to a randomly selected lattice point, while the private key allows for efficient lattice computations. Encryption, the next crucial step, commences with the encoding of plaintext messages into lattice-compatible representations, augmented by the addition of noise to bolster security. Through a randomized process, a lattice point is chosen and combined with the encoded message, followed by a rounding operation to produce a ciphertext closely resembling a lattice point. During decryption, the lattice point closest to the received ciphertext is recovered, employing lattice reduction algorithms or similar techniques. This recovered lattice point undergoes decoding to retrieve the original encoded message, subsequently mapping it back to plaintext. These sequential processes, from key generation through encryption and decryption, constitute the orchestrated dance of lattice-based cryptography, fortifying communication channels against classical and quantum adversaries alike.

\begin{figure} [!htb]
\includegraphics[width=1\linewidth]{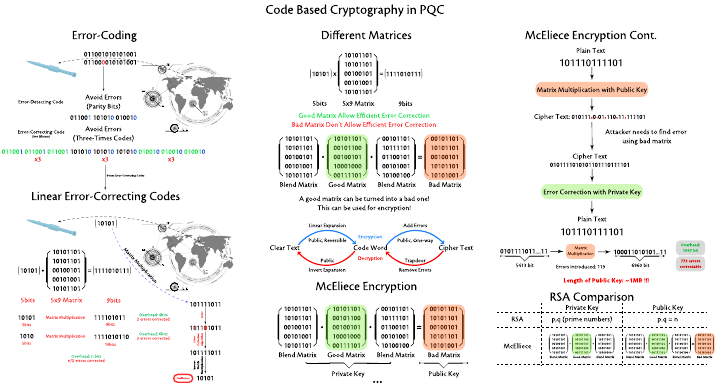}
\caption{Code Based Cryptography Overview}
\label{fig_PQC_Code}
\end{figure}

On the other hand, Code-Based Crypto Systems (see Fig. \ref{fig_PQC_Code} \cite{PQC_Understanding_Overview_with_Cartoons}) use the relationship between Asymmetric and Symmetric Keys through linear codes. The security relies on the complexity of decoding generic linear codes. These cryptosystems represent the plaintext message as a binary vector of length and add random errors to the plaintext vector to create a codeword. Then, transmit the codeword, which is the encrypted message in the encryption phase. After that, receive the codeword and attempt to decode it using the decoding algorithm and correct any errors in the received codeword to recover the plaintext in the decryption phase.

\begin{figure} [!htb]
\includegraphics[width=1\linewidth]{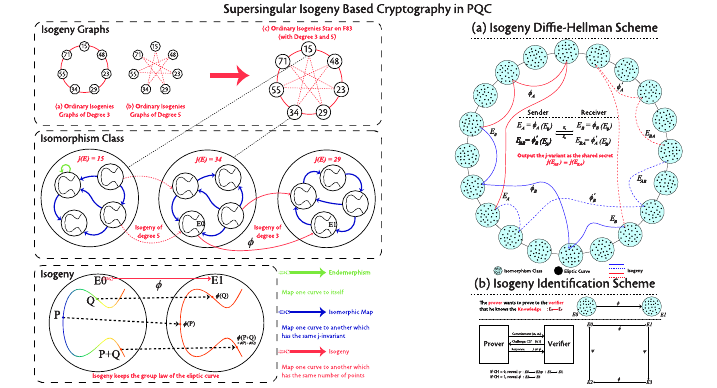}
\caption{Supersingular Isogeny Based Cryptography Overview}
\label{fig_Supersingular}
\end{figure}

Supersingular ECC Isogeny Crypto Systems (see Fig. \ref{fig_Supersingular} \cite{PQC_SUPERSINGULAR_ISOGENY}) are advanced ECC variants engineered to resist quantum attackers (QTRs). Supersingular elliptic curves are used to achieve compact key sizes. Supersingular Isogeny-based cryptography (SIKE) establishes a robust foundation for secure communication through a systematic sequence of steps. Beginning with key generation, parameters such as security levels and prime field sizes are meticulously selected, alongside the choice of supersingular elliptic curves, setting the stage for subsequent operations. Public and private keys are then generated, with each party deriving its private key and computing corresponding public keys through isogeny computations, ensuring cryptographic security. The key exchange, constituting the encryption phase, unfolds through intricate isogeny computations, where each party applies a secret isogeny to their own and the other party's public keys, facilitating the generation of a shared secret. From this shared secret, both parties derive a symmetric key for encrypting and decrypting further communications, ensuring confidentiality and integrity. SIKE's security posture is fortified by the assumed computational hardness of the isogeny problem, bolstering its resistance against classical and quantum adversaries. Efficiency considerations remain paramount, driving optimizations aimed at balancing security with computational overhead, thereby ensuring practical applicability in real-world scenarios. SIKE emerges as a reliable framework, safeguarding digital communications in an increasingly interconnected world. It is a notable scheme that enhance quantum security through random movements within isogeny graphs \cite{b4-2018Suomalainen,b8-2020Fernandez,b10-2019Fernandez}.

\begin{figure} [!htb]
\includegraphics[width=1\linewidth]{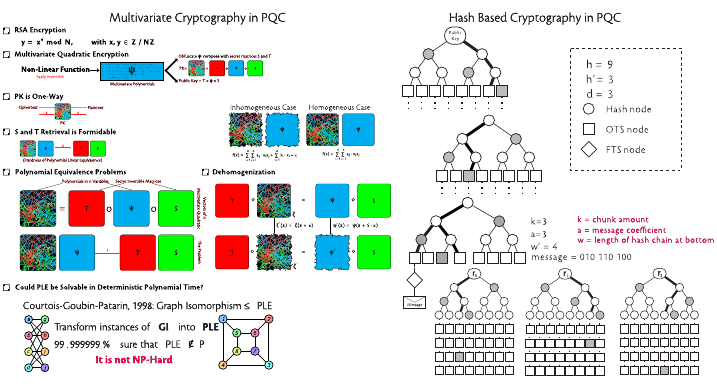}
\caption{Multivariate and Hash Based Cryptography Overview}
\label{fig_Multivariate_Hash}
\end{figure}

Multivariate-Based Crypto Systems (see Fig. \ref{fig_Multivariate_Hash} \cite{PQC_MULTIVARIATE, PQC_SPHINCS}) are designed to tackle the challenge of solving quadratic polynomial equations, making them resistant to quantum attacks \cite{b8-2020Fernandez}. Multivariate-based cryptography follows a systematic process that begins with key generation. Firstly, parameters such as the number of variables and polynomial degrees are selected to define the cryptographic scheme's structure. Then, the public and private keys are generated. The public key involves creating a system of multivariate polynomial equations, while the private key contains additional information allowing for efficient equation solving. Subsequently, during encryption, the plaintext message is encoded into variables satisfying the polynomial equations derived from the public key. These equations are then manipulated to produce ciphertexts obscuring the original message, constituting the encryption process. Decryption, on the other hand, utilizes the private key to efficiently solve the polynomial equations derived from the ciphertext, enabling the recovery of the original plaintext message. The security of multivariate-based cryptography rests on the computational hardness of solving these equations without knowledge of the private key. Efforts are made to optimize performance by balancing security, key sizes, and computational efficiency, ensuring practical applicability in various real-world scenarios. The family is historically important but should not be presented as uniformly resistant to quantum or classical cryptanalysis. The cryptanalysis of schemes such as Rainbow demonstrates that security depends on the details of the construction and parameter set rather than on the broad problem family alone \cite{b8-2020Fernandez}. Multivariate schemes should therefore appear in the taxonomy as a historical and research category unless a specific scheme has current, independently supported deployment status.

Hash-Based Crypto Systems in the Quantum Era rely on hash operations rather than mathematical complexity. It requires expanded hash function output sizes to counteract Grover's algorithm. These systems utilize one-time signatures to securely sign messages, preventing key pair reuse \cite{b4-2018Suomalainen,b5-2021Asif, b8-2020Fernandez,b10-2019Fernandez}. Attack mitigation capabilities \cite{b9-2021Balamurugan} of these schemes are also presented in Table \ref{tab_PerformanceAnalysis}. Hash-based cryptography relies on the properties and functionalities of hash functions, standardized algorithms designed to produce fixed-size hash values or digests from input data. Firstly, hash functions are generated with specific properties such as determinism, collision resistance, and preimage resistance, ensuring their reliability in cryptographic applications. Key derivation, a common use case, involves hashing input data, such as passwords, to generate cryptographic keys for encryption or authentication purposes. Additionally, hash functions are instrumental in ensuring message integrity by generating hash values or checksums for transmitted messages, allowing recipients to verify the integrity of received data by recalculating the hash value. Password hashing enhances security by storing hashed values of passwords instead of plaintext, facilitating secure authentication processes. Digital signatures, another application, utilize hash functions to hash messages before applying asymmetric encryption with the sender's private key, enabling recipients to verify message authenticity by comparing hash values. Finally, commitment schemes employ hash functions to commit to a value without revealing it, preserving confidentiality until later disclosure. Through this ordered approach, hash-based cryptography serves as a foundational tool, offering robust solutions for various cryptographic tasks while ensuring integrity, authenticity, and confidentiality in digital communications.

More generally, cryptographic hash functions provide deterministic compression to fixed-length digests and support message integrity, Merkle trees, digital signatures, commitments, transcript binding, and KDF constructions. Password hashing is a separate application requiring dedicated memory-hard or password-based mechanisms rather than direct use of a general-purpose hash. Likewise, a digital signature does not normally ``encrypt a hash with the private key''; it applies a scheme-specific signing operation to a message or message digest and is verified using the corresponding public key. These distinctions are important because the ESL relies on hashes, signatures, and authenticated state transitions for different functions and should not conflate them into a single generic security primitive.

Across all PQC families, the primary Edge AI design question is not which primitive is universally strongest, but which combination of standardized maturity, object size, execution time, memory consumption, implementation assurance, hardware support, and algorithmic diversity best satisfies a given platform and network profile. In the architecture studied here, PQC is used primarily for authenticated key establishment, signatures, and security-state transitions; symmetric AEAD remains responsible for continuous data-plane protection. Rotational secrecy and the PARS framework operate above these primitives by controlling when and how cryptographic epochs evolve, while the optional ESL provides distributed coordination when the system's trust and audit requirements justify it.
\section{Related Work: Quantum-Safe Edge Systems} \label{Related Works} 

\begin{table*}[htb]
\tiny
\centering
\caption{Quantum Security Architectures - Quantum Mechanics (S-QM) Schemes}
\label{tab_LiteratureReview_SQM}
\begin{tabular}{|c|l|l|l|l|c|}
\hline
\textbf{Type} & \textbf{Methodology} & \textbf{Pros \& Cons} & \textbf{Target} & \textbf{Contribution} & \textbf{Status} \\ \hline
{S-QM} & 
\begin{tabular}[c]{@{}l@{}} 
\textbullet\ Intertwined Architecture for IoT, BC and QCMPs \\
\textbullet\ Identity Management Mechanism \\
\textbullet\ Data Integrity Before Storage \\
\textbullet\ Heterogeneity Gateway Support Among Edges \\
\textbullet\ Data Transfer Over Quantum Distributed Ledger \\
\end{tabular} &
\begin{tabular}[c]{@{}l@{}} 
\textbf{Pros:} \\
\textbullet\ Ensuring Data Integrity Before Storage \\
\textbullet\ Facilitates Decentralization in IoT \\
\textbullet\ Focuses IoT Resource Limitation \\
\textbf{Cons:} \\
\textbullet\ Depends on Quantum Computers \\
\end{tabular} &
\begin{tabular}[c]{@{}l@{}} 
IoT \\
\cite{b11-2021Younan}
\end{tabular} &
\begin{tabular}[c]{@{}l@{}} 
\textbullet\ Propose Multilevel Distributed-Ledger \\
\textbullet\ Integration Support for Heterogeneous Edges \\
\textbullet\ Building Secure IoT Applications with BC and QCMPs \\
\end{tabular} &
Empirical \\ \cline{2-6} 

& 
\begin{tabular}[c]{@{}l@{}} 
\textbullet\ Quantum Communication Infrastructure \\
\textbullet\ Information Processing with Blockchain \\
\textbullet\ Quantum-Reflection Key Distribution Protocol \\
\end{tabular} &
\begin{tabular}[c]{@{}l@{}} 
\textbf{Pros:} \\
\textbullet\ No Need for Qubits \\
\textbullet\ Traditional Device Compliance \\
\textbullet\ Resist Various Attacks \\
\textbullet\ High Computational Efficiency \\
\textbullet\ High Level of Practicability \\
\textbf{Cons:} \\
\textbullet\ Lack of Native Quantum Resistance \\
\textbullet\ Implementation Difficulty \\
\textbullet\ Complex Infrastructure Cost \\
\end{tabular} &
\begin{tabular}[c]{@{}l@{}} 
IoV\&IoT \\
\cite{b12-2020Zhu} \\
\end{tabular} &
\begin{tabular}[c]{@{}l@{}} 
\textbullet\ Two Novel Lightweight Quantum-Reflection Protocols \\
\textbullet\ Reduced Communication Rounds \\
\textbullet\ Overcome Rare Q-Bits \\
\textbullet\ Offering Compatibility for Traditional Device \\
\end{tabular} &
\begin{tabular}[c]{@{}l@{}} 
Theoretic \\
Notional \\
\end{tabular} \\ \cline{2-6} 

& 
\begin{tabular}[c]{@{}l@{}} 
\textbullet\ Quantum Key Distribution \\
\textbullet\ Secure Authentication Algorithm \\
\end{tabular} &
\begin{tabular}[c]{@{}l@{}} 
\textbf{Pros:} \\
\textbullet\ Anonymity of Network Peers \\
\textbullet\ Supports Large Scale Blockchains \\
\textbf{Cons:} \\
\textbullet\ Lack of Attack Mitigation Analysis \\
\textbullet\ Lack of Technical Details \\
\textbullet\ Lack of Functional Supremacy \\
\end{tabular} &
\begin{tabular}[c]{@{}l@{}} 
Generic \\
\cite{b13-2018Kiktenko}
\end{tabular} &
\begin{tabular}[c]{@{}l@{}} 
\textbullet\ Efficient Blockchain for Quantum Attacks \\
\textbullet\ Experimental Realization \\
\end{tabular} &
\begin{tabular}[c]{@{}l@{}} 
Notional \\
Empirical \\
\end{tabular} \\ \cline{2-6} 

& 
\begin{tabular}[c]{@{}l@{}} 
\textbullet\ Usage of QDS and QM \\
\textbullet\ Consensus Algorithm \\
\textbullet\ Entanglement Operations \\
\end{tabular} &
\begin{tabular}[c]{@{}l@{}} 
\textbf{Pros:} \\
\textbullet\ Weighted Graph States \\
\textbullet\ Undeniability of Nodes \\
\textbullet\ Resistant to Hash Rate Attacks \\
\textbullet\ Solves Heterogeneity Problem \\
\textbullet\ Resource Friendly \\
\textbullet\ Lower Computation Volume \\
\textbullet\ Wide-Spectrum of Applicability \\
\textbf{Cons:} \\
\textbullet\ Technically Challenging \\
\textbullet\ Complex Architecture \\
\end{tabular} &
\begin{tabular}[c]{@{}l@{}} 
Generic \\
\cite{b14-2022Li}
\end{tabular} &
\begin{tabular}[c]{@{}l@{}} 
\textbullet\ Present Higher Efficiency \\
\textbullet\ Novel Consensus Mechanism \\
\end{tabular} &
\begin{tabular}[c]{@{}l@{}} 
Theoretic \\
Empirical \\
\end{tabular} \\ \cline{2-6} 

&
\begin{tabular}[c]{@{}l@{}}
\textbullet\ Quantum-native chain via \\
temporal GHZ states \\
\textbullet\ Entanglement-in-time block links \\
\end{tabular} &
\begin{tabular}[c]{@{}l@{}}
\textbf{Pros:} \\
\textbullet\ Past blocks physically tamper-proof \\
\textbullet\ Subcomponents experimentally realizable \\
\textbf{Cons:} \\
\textbullet\ No implementation needs quantum network \\
\textbullet\ No key rotation \\
\end{tabular} &
\begin{tabular}[c]{@{}l@{}}
Generic \\
\cite{rajanVisser2019}
\end{tabular} &
\begin{tabular}[c]{@{}l@{}}
\textbullet\ Pioneering quantum-native \\
blockchain concept \\
\end{tabular} &
Conceptual \\ \cline{2-6}

&
\begin{tabular}[c]{@{}l@{}}
\textbullet\ Quantum-blockchain framework for EMRs \\
\textbullet\ Quantum-authenticated transactions \\
\end{tabular} &
\begin{tabular}[c]{@{}l@{}}
\textbf{Pros:} \\
\textbullet\ Privacy-preserving EMR sharing \\
\textbullet\ Quantum-secured integrity \\
\textbf{Cons:} \\
\textbullet\ Assumes quantum channels \\
\textbullet\ No key rotation \\
\end{tabular} &
\begin{tabular}[c]{@{}l@{}}
IoMT \\
\cite{quQB2022}
\end{tabular} &
\begin{tabular}[c]{@{}l@{}}
\textbullet\ Quantum-BC framework for \\
private EMRs in IoMT \\
\end{tabular} &
Theoretical \\ \hline 

\end{tabular}
\end{table*} 
\begin{table*}[htb]
\tiny
\centering
\caption{Quantum Security Architectures - Lattice-Based Post-Quantum (S-PQCL) Schemes}
\label{tab_LiteratureReview_SPQCL}
\begin{tabular}{|c|l|l|l|l|c|}
\hline
\textbf{Type} & \textbf{Methodology} & \textbf{Pros \& Cons} & \textbf{Target} & \textbf{Contribution} & \textbf{Status} \\ \hline
{S-PQCL} & 
\begin{tabular}[c]{@{}l@{}} 
\textbullet\  Lattice Based Delegation Algorithm \\
\textbullet\  Double Signature Design \\
\textbullet\  Preimage Sampling Algorithm \\
\end{tabular} &
\begin{tabular}[c]{@{}l@{}} 
\textbf{Pros:} \\
\textbullet\  Reduced Correlation Between Message and Signature \\
\textbullet\  Shorter Signature and Key Size \\
\textbullet\  Decreased Computational Complexity \\
\textbullet\  Higher Security and Efficiency \\
\textbullet\  Resist Various Attacks \\
\textbf{Cons:} \\
\textbullet\  Implementation Complexity and Time \\
\textbullet\  Lack of Security Analysis \\
\textbullet\  Lack of Quantum Effects on Financial Backbone \\
\end{tabular} &
\begin{tabular}[c]{@{}l@{}} 
Finance \\
\cite{b24-2018Gao}
\end{tabular} &
\begin{tabular}[c]{@{}l@{}} 
\textbullet\  Novel Signature Schema \\
\textbullet\  Quantum Secure Cryptocurrency Architecture \\
\end{tabular} &
Theoretic \\ \cline{2-6} 

&

\begin{tabular}[c]{@{}l@{}} 
\textbullet\  Lattice Based Encryption Architecture \\
\end{tabular} &
\begin{tabular}[c]{@{}l@{}} 
\textbf{Pros:} \\
\textbullet\  Reduced Communication Cost \\
\textbullet\  Ensure Forward Security \\
\textbullet\  Resist Private Key Attacks \\
\textbullet\  Higher Efficiency \\
\textbullet\  Lower Execution Time \\
\textbf{Cons:} \\
\textbullet\  Reduced Trapdoor \\
\textbullet\  Complex Notation Flow for Security Analysis \\
\textbullet\  Possible Backdoors \\
\end{tabular} &
\begin{tabular}[c]{@{}l@{}} 
Health \\
\cite{b25-2022Xu}
\end{tabular} &
\begin{tabular}[c]{@{}l@{}} 
\textbullet\  BC Based Searchable Encryption Schema for QAs \\
\textbullet\  Solving Key Leakage Problem \\
\textbullet\  Robust Ontological Backbone for Healthcare \\
\end{tabular} &
\begin{tabular}[c]{@{}l@{}} 
Theoretic \\
Empirical \\
\end{tabular} \\ \cline{2-6}

&
\begin{tabular}[c]{@{}l@{}}
\textbullet\ Lattice RingCT (Module-SIS/LWE) \\
\textbullet\ Efficient PQ private payments \\
\end{tabular} &
\begin{tabular}[c]{@{}l@{}}
\textbf{Pros:} \\
\textbullet\ Compact proofs vs prior lattice RingCT \\
\textbullet\ Confidential transactions \\
\textbf{Cons:} \\
\textbullet\ Heavier than classical (ECC) RingCT \\
\textbullet\ No key rotation \\
\end{tabular} &
\begin{tabular}[c]{@{}l@{}}
Finance \\
\cite{esginMatRiCT2022}
\end{tabular} &
\begin{tabular}[c]{@{}l@{}}
\textbullet\ Efficient post-quantum private \\
blockchain payments (IEEE S\&P) \\
\end{tabular} &
Empirical \\ \hline

\end{tabular}
\end{table*} 

The integration of PQC and quantum mechanics (QM) primitives across decentralized architectures has a long way to become consolidated due to recent developments in attack landscapes and asymmetric cyber threats as a result of rising AI and QC functionalities. The literature on quantum-safe distributed systems spans two broad architectural directions: designs that depend on QM communication or computation primitives, and software-oriented approaches that integrate Post-Quantum Cryptography (PQC) into otherwise classical network and ledger infrastructures. Recent surveys \cite{gharavi2024PQBlockchainIoT,yang2024PQBlockchainSurvey,naik2023portfoliooptimizationquantumblockchain,chhetri2025PQCSurvey} show that this area is expanding rapidly, but they also reveal a persistent disconnect between cryptographic proposals and the operational constraints of physical Edge AI and Cyber-Physical Systems (CPS). In particular, resource-constrained agents must reconcile cryptographic security with limited memory, processing capability, energy, link capacity, and tolerance for intermittent connectivity. The relevant literature is therefore reviewed here according to architectural dependence, cryptographic maturity, edge-system suitability, dynamic key-management support, and empirical validation.

QM-Based Architecture Set (S-QM) includes proposals that incorporate quantum communication, quantum states, or quantum-native validation into distributed architectures. At the conceptual end of this spectrum, quantum-native blockchain designs encode ledger relationships into temporal Greenberger--Horne--Zeilinger (GHZ) states so that modification of previously linked states becomes physically detectable \cite{rajanVisser2019}. Such designs are important for understanding the long-term convergence of quantum networking and distributed ledgers, but they assume communication and processing capabilities that are not generally available to present-day embedded Edge AI platforms.
Closer to practical implementation, several hybrid architectures embed QM primitives into otherwise classical systems. Younan \cite{b11-2021Younan} proposes a multilevel distributed ledger for heterogeneous IoT applications, with data-integrity validation performed before information is stored on a quantum-supported ledger. Its main limitation for near-term edge deployment is the dependence on specialized quantum infrastructure. Zhu \cite{b12-2020Zhu} introduces a quantum-reflection key-distribution protocol for IoT and Internet-of-Vehicles (IoV) environments, reducing communication rounds and addressing selected transport-layer attacks. However, the approach still requires specialized infrastructure and does not by itself mitigate physical capture, malicious firmware, false sensor data, or endpoint compromise. Kiktenko et al. \cite{b13-2018Kiktenko} experimentally investigate QKD-supported blockchain operation and peer anonymity, demonstrating the feasibility of combining quantum key distribution with distributed consensus. Nevertheless, long-distance deployment, infrastructure cost, denial-of-service exposure, and comprehensive end-to-end attack analysis remain open concerns. Li \cite{b14-2022Li} combines Quantum Digital Signatures (QDS), weighted graph states, and a quantum-oriented consensus mechanism to reduce dependence on conventional hash-rate competition and accommodate node heterogeneity. The resulting architecture is technically significant, although its hardware and operational complexity limits direct applicability to conventional resource-constrained edge devices.

PQC-Based Architecture Set (S-PQC) consist a series of approaches respectively a software-driven post-quantum approaches offer greater deployment feasibility. Zhang \cite{b15-2021Zhang} develops a lattice-based blockchain utilizing the qTESLA signature algorithm to shrink transaction and block sizes, but lacks parallel transaction pipelines or resource analytical models for edge deployment. Chen \cite{b16-2021Chen} proposes a quantum Proof-of-Work (PoW) mechanism that forces nodes to solve random quadratic equations via memory mining to mitigate quantum advantage. However, because its core relies on the Rainbow multivariate scheme, it remains structurally compromised following Rainbow's cryptanalysis. For high-assurance settings, Gupta \cite{b17-2022Gupta} offers a lattice-based, privacy-preserving data authentication scheme for IoT/IoV consortium ledgers that minimizes computational, storage, and energy costs. Due to the complex model, it lacks concrete cost proofs for constrained microcontrollers.

Other domain specific designs, the Quantum Bitcoin scheme \cite{b20-2018Anhao} leverages the hash-based eXtended Merkle Signature Scheme (XMSS) \cite{b21-2011Buchmann} and the W-OTS+ one-time scheme \cite{b22-2013Hulsing} to prevent key reuse and resist Shor's algorithm. Complementary research by Ilie \cite{b7-2020Ilie} and Aggarwal \cite{b23-2018Aggarwal} details quantum signature and mining attacks while proposing alternative PoW structures, though the substantial signature migrations generate storage overheads incompatible with tactical hardware. To minimize mathematical constraints, Gao et al. \cite{b24-2018Gao} formulate a lattice-based financial delegation algorithm that decouples messages from signatures to shrink key sizes and computational complexity. Similarly, lattice-based confidential-payment designs such as MatRiCT+ \cite{esginMatRiCT2022} deliver post-quantum private ring-confidential transactions with compact proofs. Beyond finance, Xu et al. \cite{b25-2022Xu} present a healthcare-tailored lattice blockchain utilizing searchable encryption to guarantee forward security and private key recovery resistance, while related quantum-blockchain designs secure records within the medical things \cite{quQB2022}. Notably, these historical lattice frameworks \cite{b15-2021Zhang,b24-2018Gao,b25-2022Xu} predate and do not adopt finalized NIST standardized parameter sets, leaving an active compliance gap.

The deployment context of edge networks introduces distinct architectural paradigms. The lightweight distributed ledgers encompasses optimized for resource-constrained environments. IOTA \cite{popov2018Tangle} avoids block-building overheads by replacing linear chains with a directed acyclic graph enabling feeless, parallel transactions. It has higher security than Bitcoin, even if not capable to fully avoid QAs. Similarly, IoTChain layers authentication, ledger, and application tiers to provide lightweight identity and fault tolerance for IoT networks \cite{alphand2018IoTChain}. However, these systems focus primarily on footprint optimization, remain vulnerable in the QAs, and omit key rotation. The other critical area that remains in the center of this application is military and defense applications. Programs such as DARPA's Internet of Battlefield Things (IoBT) and NATO initiatives have long explored decentralized, tamper-evident messaging and logistics, with recent work combining blockchain with attribute-based encryption for coalition-capable Military IoT \cite{pioro2024MIoT}. Furthermore, blockchain has been deployed to secure autonomous multi-robot and swarm systems that have been utilized to isolate Byzantine robots \cite{strobel2023ByzantineRobots}, and comprehensive frameworks have been proposed to coordinate mobile multi-robot operations \cite{dorigo2024BlockchainRobots}. Despite their operational benefits, these swarm architectures still rely on quantum-vulnerable signatures (e.g. ECDSA) and lack quantum protections or dynamic key rotation.

Bridging the gap between physical execution and edge intelligence will drive the next wave of applied researches and industry applications at the intersection of collaborative Edge AI and quantum-safe systems. Gharavi et al. \cite{Gharavi2025PQBFL} proposed PQBFL, a post-quantum blockchain-federated learning protocol that mitigates adversarial gradient poisoning attacks via lattice-based signatures, though it adds computational latency on non-accelerated edge clients. Concurrently, Zareei et al. \cite{Zareei2025Blockchain} designed a zero-trust IoT framework featuring a domain-adaptive anomaly detector (TinyDA) requiring less than 100 KB of RAM on ESP32 microcontrollers, logging events off-chain via IPFS while securing validations through Dilithium to yield significant throughput and latency gains.

For high-frequency telemetry, Hao et al. \cite{Hao2025PostQuantum} proposed a simulation-based, post-quantum-inspired alliance blockchain for medical data ingestion, using Kyber-inspired hybrid encryption along with Raft and HotStuff consensus engines to cut encryption overheads and achieve up to 14\% memory savings. Physical edge acceleration has also been optimized by \cite{Kaushik2026Karana} developed Karana, a multimodal AI surveillance system for autonomous threat detection operating on NVIDIA Jetson hardware. Karana fuses inputs from radar and imaging sensors (RF, IR, RGB) through a YOLOv8 engine, Swin Transformers, and a quantum variational classifier, employing Kyber key encapsulation and Dilithium signatures to protect remote links in contested physical environments.

\begin{table*}[!htb]
\tiny
\centering
\caption{Quantum Security Architectures - Post-Quantum Cryptography (S-PQC) Schemes}
\label{tab_LiteratureReview_SPQC}
\begin{tabular}{|c|l|l|l|l|c|}
\hline
\textbf{Type} & \textbf{Methodology} & \textbf{Pros \& Cons} & \textbf{Target} & \textbf{Contribution} & \textbf{Status} \\ \hline
{S-PQC} & 
\begin{tabular}[c]{@{}l@{}} 
\textbullet\ PARS Protocol (Rotational Secrecy) \\
\textbullet\ PoE Consensus Mechanism \\
\textbullet\ Layer-1 Private Quantum Blockchain \\
\textbullet\ Light Node Executive Architecture \\
\end{tabular} &
\begin{tabular}[c]{@{}l@{}} 
\textbf{Pros:} \\
\textbullet\ Algorithm-agnostic (modular structure) \\
\textbullet\ High energy efficiency \\
\textbullet\ Low storage footprint \\
\textbullet\ Robust quantum attack mitigation \\
\textbullet\ Optimized for edge hardware \\
\textbullet\ Strong empirical baseline \\
\textbf{Cons:} \\
\textbullet\ Lacks holistic empirical \\
\textbullet\ Performance metrics \\ 
\textbullet\ (deferred to this study) \\
\textbullet\ Coordination overhead for pattern consensus \\
\end{tabular} &
\begin{tabular}[c]{@{}l@{}} 
IoMT\&\\
Military\&\\
Generic \\
\cite{b15-2024kcso} \\
\end{tabular} &
\begin{tabular}[c]{@{}l@{}} 
\textbullet\ Proposes Layer-1 Blockchain for military systems \\
\textbullet\ Dynamic key reconciliation for tactical swarms \\
\textbullet\ Introduces novel PoE consensus \\
\textbullet\ Comprehensive benchmark of NIST PQC candidates \\
\end{tabular} &
Empirical  \\ \cline{2-6} 

&
\begin{tabular}[c]{@{}l@{}} 
\textbullet\ Lattice Based DSA Blockchain \\
\textbullet\ IPFS \\
\end{tabular} &
\begin{tabular}[c]{@{}l@{}} 
\textbf{Pros:} \\
\textbullet\ Lower Transaction Size \\
\textbullet\ Increased Performance \\
\textbullet\ Optimized Block Size in Traditional Architectures \\
\textbf{Cons:} \\
\textbullet\ Limitations for Parallel Transactions \\
\textbullet\ Lack of Performance and Scalability Analysis \\
\end{tabular} &
\begin{tabular}[c]{@{}l@{}} 
Generic \\
\cite{b15-2021Zhang}
\end{tabular} &
\begin{tabular}[c]{@{}l@{}} 
\textbullet\ Propose Quantum-Resistant DSA Based Blockchain \\
\textbullet\ Plan to Resist Quantum Attacks by qTesla \\
\textbullet\ Ensure Higher Number of Transactions \\
\end{tabular} &
Empirical  \\ \cline{2-6} 

&
\begin{tabular}[c]{@{}l@{}} 
\textbullet\ Novel Quantum PoW Consensus Mechanism  \\
\end{tabular} &
\begin{tabular}[c]{@{}l@{}} 
\textbf{Pros:} \\
\textbullet\  Solve Random Quadratic Equations \\
\textbullet\  Enhanced Security \\
\textbullet\  Higher Transaction Amount \\
\textbf{Cons:} \\
\textbullet\  Holds Possible Back Doors of Rainbow \\
\end{tabular} &
\begin{tabular}[c]{@{}l@{}} 
Generic \\
\cite{b16-2021Chen}
\end{tabular} &
\begin{tabular}[c]{@{}l@{}} 
\textbullet\  Memory Mining Capabilities\\
\textbullet\  Reduce Validity Concerns on Transactions\\
\textbullet\  Efficient Block Expansion \\
\end{tabular} &
\begin{tabular}[c]{@{}l@{}} 
Theoretic \\
Empirical \\
\end{tabular} \\ \cline{2-6} 

&
\begin{tabular}[c]{@{}l@{}} 
\textbullet\ Conditional Privacy-Preserving Authentication \\
\textbullet\ Lattice Based Architecture \\
\textbullet\ Authentication Algorithm \\
\end{tabular} &
\begin{tabular}[c]{@{}l@{}} 
\textbf{Pros:} \\
\textbullet\  Energy and Storage Efficiency \\
\textbullet\  Lower Computation Cost \\
\textbullet\  Rich Set Amount Attacks Mitigation Capability \\
\textbullet\  Rich Set Amount Attacks Mitigation Capability \\
\textbullet\  Sophisticated Security Functionalities \\
\textbullet\  Strong Analytical Analysis \\
\textbf{Cons:} \\
\textbullet\  Loosely Correlated Cost Results\\
\textbullet\  Loosely Correlated Proofs \\
\textbullet\  Complex Model for Implementation Phase \\
\end{tabular} &
\begin{tabular}[c]{@{}l@{}} 
IoV\&\\
IoT \\
\cite{b17-2022Gupta} \\
\end{tabular} &
\begin{tabular}[c]{@{}l@{}} 
\textbullet\  Propose Quantum-Resistant Authentication Schema \\
\textbullet\  Reliable Consortium Blockchain for Data Management \\
\textbullet\  Enhance Security Functionalities of Literature \\
\end{tabular} &
\begin{tabular}[c]{@{}l@{}} 
Notional \\
Theoretic \\
Empirical \\
\end{tabular} \\ \cline{2-6}

&
\begin{tabular}[c]{@{}l@{}} 
\textbullet\ Quantum-Safe XMSS \\
\textbullet\ Novel Mining Algorithm \\
\textbullet\ Key Migration from Old to XMSS Based Keys \\
\textbullet\ W-OTS+ Schema for Avoiding One-Time \\
\textbullet\ Non-Interactive ZKP for Privacy  \\
\end{tabular} &
\begin{tabular}[c]{@{}l@{}} 
\textbf{Pros:} \\
\textbullet\  Lack of Transaction Malleability \\
\textbullet\  Suitable for Lightning Network \\
\textbullet\  Protected Network Efficiency \\
\textbullet\  Burn Operation for Non-Migrated Coins \\
\textbf{Cons:} \\
\textbullet\  Extended XMSS Signature Migration is Required \\
\textbullet\  Mix of Classical Approaches like Equihash, SHA3 \\
\textbullet\  Higher Witness Scale Factor \\
\end{tabular} &
\begin{tabular}[c]{@{}l@{}} 
Finance \\
\cite{b21-2011Buchmann}
\end{tabular} &
\begin{tabular}[c]{@{}l@{}} 
\textbullet\  Quantum-Safe Cryptocurrency Schema \\
\textbullet\  Serves for Avoiding Possible Defects of Bitcoin for QAs \\
\end{tabular} &
\begin{tabular}[c]{@{}l@{}} 
Theoretic \\
Notional \\
\end{tabular} \\ \cline{2-6}

&
\begin{tabular}[c]{@{}l@{}} 
\textbullet\ Multi-sensor input fusion (RF/IR/radar/RGB) \\
\textbullet\ Post-Quantum Cryptography (Kyber/Dilithium2) \\
\textbullet\ YOLOv8, Swin-T, and Quantum Variational Classifier \\
\end{tabular} &
\begin{tabular}[c]{@{}l@{}} 
\textbf{Pros:} \\
\textbullet\ On-device threat detection and active response \\
\textbullet\ Low average inference latency (52.9 ms) \\
\textbullet\ High accuracy (F1-score of 90.0) \\
\textbf{Cons:} \\
\textbullet\ High local computational/thermal demand \\
\textbullet\ Heavy hardware requirements for multimodal execution \\
\end{tabular} &
\begin{tabular}[c]{@{}l@{}} 
Edge AI \\
Surveillance \\
\cite{Kaushik2026Karana}
\end{tabular} &
\begin{tabular}[c]{@{}l@{}} 
\textbullet\ Proposes first fully integrated, PQC-secure \\
\quad Edge AI threat mitigation platform (Karana) \\
\textbullet\ Hardware-validated on NVIDIA Jetson architectures \\
\end{tabular} &
Empirical \\ \cline{2-6}

&
\begin{tabular}[c]{@{}l@{}} 
\textbullet\ Lattice-based signature verification \\
\textbullet\ Decentralized model gradient protection \\
\textbullet\ Post-Quantum Federated Learning \\
\end{tabular} &
\begin{tabular}[c]{@{}l@{}} 
\textbf{Pros:} \\
\textbullet\ Robust protection against local model poisoning \\
\textbullet\ Secure, privacy-preserving client collaboration \\
\textbf{Cons:} \\
\textbullet\ Substantial signature verification latency \\
\textbullet\ High network overhead on non-accelerated edges \\
\end{tabular} &
\begin{tabular}[c]{@{}l@{}} 
Edge AI \& \\
Federated Learning \\
\cite{Gharavi2025PQBFL}
\end{tabular} &
\begin{tabular}[c]{@{}l@{}} 
\textbullet\ Evaluates lattice signatures for model sharing \\
\textbullet\ Maps cryptographic validation costs to FL metrics \\
\end{tabular} &
Theoretic \\ \cline{2-6}

&
\begin{tabular}[c]{@{}l@{}} 
\textbullet\ Compressed anomaly detector (TinyDA) \\
\textbullet\ CRYSTALS-Dilithium transaction signing \\
\textbullet\ Off-chain IPFS storage and permissioned ledger \\
\end{tabular} &
\begin{tabular}[c]{@{}l@{}} 
\textbf{Pros:} \\
\textbullet\ Ultra-low memory usage (under 100 KB RAM) \\
\textbullet\ ESP32 microcontroller compatibility \\
\textbullet\ 35\% reduction in end-to-end latency \\
\textbf{Cons:} \\
\textbullet\ IPFS chunk retrieval latency under network splits \\
\textbullet\ Prone to intermittent routing delays \\
\end{tabular} &
\begin{tabular}[c]{@{}l@{}} 
Zero-Trust \\
IoT Edge \\
\cite{Zareei2025Blockchain}
\end{tabular} &
\begin{tabular}[c]{@{}l@{}} 
\textbullet\ Proposes a low-power, zero-trust IoT framework \\
\textbullet\ Implements hardware-constrained anomaly detection \\
\textbullet\ Integrates permissioned post-quantum blockchain \\
\end{tabular} &
\begin{tabular}[c]{@{}l@{}} 
Theoretic \\
Empirical \\
\end{tabular} \\ \cline{2-6}

&
\begin{tabular}[c]{@{}l@{}} 
\textbullet\ Kyber-inspired hybrid encryption simulation \\
\textbullet\ Patient-centric authorization proofs \\
\textbullet\ Raft, HotStuff, and Tendermint consensus \\
\end{tabular} &
\begin{tabular}[c]{@{}l@{}} 
\textbf{Pros:} \\
\textbullet\ 72.3\% reduction in encryption latency vs RSA-2048 \\
\textbullet\ Fivefold throughput increase (1.26M TPS under Raft) \\
\textbullet\ 14\% lower memory consumption \\
\textbf{Cons:} \\
\textbullet\ Emulated security model lacks full polynomial rings \\
\textbullet\ Requires synchronized network models \\
\end{tabular} &
\begin{tabular}[c]{@{}l@{}} 
Healthcare \\
IoMT \\
\cite{Hao2025PostQuantum}
\end{tabular} &
\begin{tabular}[c]{@{}l@{}} 
\textbullet\ Proposes modular post-quantum hospital alliance ledger \\
\textbullet\ Benchmarks and compares modern consensus protocols \\
\textbullet\ Implements lightweight non-interactive access control \\
\end{tabular} &
\begin{tabular}[c]{@{}l@{}} 
Theoretic \\
Empirical \\
\end{tabular} \\ \cline{2-6}

&
\begin{tabular}[c]{@{}l@{}} 
\textbullet\ Post-Quantum Cryptographic Review \\
\textbullet\ Trilemma Analysis (Security, Decentr., Perf.) \\
\end{tabular} &
\begin{tabular}[c]{@{}l@{}} 
\textbf{Pros:} \\
\textbullet\ Establishes clear baseline for ledger PQC migration \\
\textbullet\ Maps cryptographic cost to network metrics \\
\textbf{Cons:} \\
\textbullet\ High-level review lacking localized edge code \\
\end{tabular} &
\begin{tabular}[c]{@{}l@{}} 
Generic \\
\cite{b8-2020Fernandez}
\end{tabular} &
\begin{tabular}[c]{@{}l@{}} 
\textbullet\ First structured evaluation of PQC impact on \\
\quad blockchain databases and block expansion limits \\
\end{tabular} &
Theoretic \\ \cline{2-6}

&
\begin{tabular}[c]{@{}l@{}} 
\textbullet\ DAG-based Tangle consensus (IOTA) \\
\textbullet\ IoTChain gateway integrations \\
\end{tabular} &
\begin{tabular}[c]{@{}l@{}} 
\textbf{Pros:} \\
\textbullet\ Low transaction fees with fast localized consensus \\
\textbullet\ Ideal for high-frequency low-payload sensors \\
\textbf{Cons:} \\
\textbullet\ High vulnerability to transaction grinding attacks \\
\textbullet\ Relies on centralized coordinators under splits \\
\end{tabular} &
\begin{tabular}[c]{@{}l@{}} 
IoT \& \\
Lightweight \\
\cite{popov2018Tangle}
\end{tabular} &
\begin{tabular}[c]{@{}l@{}} 
\textbullet\ Proposes high-speed, non-linear ledger structures \\
\quad for high-frequency edge IoT workloads \\
\end{tabular} &
Empirical \\ \cline{2-6}

&

\begin{tabular}[c]{@{}l@{}} 
\textbullet\ Layer-1 Private Quantum Blockchain (ESL,MiliChain) \\
\textbullet\ Integrated PARS Protocol (Rotational Secrecy) \\
\textbullet\ PoE Consensus Engine with Multi-Domain ZKPs \\
\textbullet\ Shifting to Light Node Executive Blockchain \\
\textbullet\ 7 Layered Defense Attack Mitigation Doctrines \\
\end{tabular} &
\begin{tabular}[c]{@{}l@{}} 
\textbf{Pros:} \\
\textbullet\ Organic crypto-agility for dynamic algorithm \\
\textbullet\ High resource efficiency (peak memory under 3.5 MB) \\
\textbullet\ Zero dependency on specialized quantum hardware \\
\textbullet\ Low time complexity\\
\textbullet\ Low storage footprint \\
\textbullet\ Robust quantum attack mitigation \\
\textbullet\ Optimized for edge hardware \\
\textbullet\ Strong empirical baseline \\
\textbf{Cons:} \\
\textbullet\ Minor consensus denial vulnerability (0.1\% to 0.01\% \\
 error rates) under high ZKP complexity \\
\textbullet\ Variable bandwidth demands (1.67 Kbps to 78.09 Kbps) \\
depending on PQC candidate choice \\ 
\textbullet\ Coordination overhead for pattern consensus \\
\end{tabular} &
\begin{tabular}[c]{@{}l@{}} 
IoMT\&\\
Military\&\\
Generic\\
$\left[\text{This Study}\right]$
\end{tabular} &
\begin{tabular}[c]{@{}l@{}} 
\textbullet\ Introduces first Layer-1 Military QB architecture \\
\textbullet\ Dynamic key reconciliation for tactical swarms \\
\textbullet\ Introduces novel PoE consensus \\
\textbullet\ Comprehensive benchmark of NIST PQC candidates \\
\textbullet\ Outlines 7 novel offensive/defensive \\
cyber-attack doctrines for kinetic swarms \\
\textbullet\ Implements localized ZKP authentication \\
models tailored for edge sensor networks \\
\textbullet\ Fully evaluates protocol overhead, \\
transaction sizes, and structural failure rates \\
\end{tabular} &
Empirical  \\ \hline

\end{tabular}
\end{table*} 

Hardware--software co-design offers a complementary path for reducing PQC overhead. Waris et al. \cite{Waris2025AreaTime} develop an area--time-efficient pipelined Number Theoretic Transform (NTT) architecture for Artix-7 FPGAs, accelerating the polynomial-multiplication operations used by Kyber by 68\%. Liao et al. \cite{Liao2026Accelerating} use Large Language Model (LLM)-assisted co-design to generate FPGA acceleration structures for the FALCON signature scheme, reporting up to a 2.6$\times$ kernel speedup. These results show that PQC cost is not fixed solely by the algorithm; it also depends on arithmetic architecture, memory organization, implementation quality, and available accelerators.

At the architectural level, Gupta and Mittal \cite{Gupta2026Secure} propose a secure-by-design industrial-control model in which computationally intensive post-quantum handshakes terminate at capable edge gateways while internal control paths continue to use symmetric cryptography. This separation closely aligns with the distinction between the PARS-ESL security-control plane and the operational data plane. Dynamic cryptographic orchestration has also been investigated through reinforcement-learning-based decision layers that select among classical, ML-KEM-based, and QKD-assisted mechanisms according to network conditions and assessed threat levels \cite{RL2026Cryptographic}. Such adaptive selection motivates crypto-agility, but secure deployment additionally requires authenticated policy updates, downgrade prevention, deterministic fallback behavior, and protection against adversarial manipulation of the decision inputs.

The reviewed literature reveals several recurring gaps. First, many proposals analyze one component---for example, a signature scheme, consensus mechanism, anomaly detector, or hardware accelerator---without connecting it to the complete lifecycle of identity provisioning, authenticated key establishment, key rotation, membership change, revocation, partition recovery, and algorithm replacement. Second, evaluations often report isolated latency or memory values while omitting packet fragmentation, retransmission, energy, persistent storage, failure classification, and confidence intervals. Third, lightweight or edge-oriented ledgers typically optimize transaction processing but do not explicitly manage authenticated cryptographic epochs. Fourth, designs that use the terms zero knowledge, quantum security, or forward security do not always define the exact adversary model, statement--witness relation, compromise timeline, or implementation assumptions required for those properties.

Tables~\ref{tab_LiteratureReview_SQM}, \ref{tab_LiteratureReview_SPQCL}, and \ref{tab_LiteratureReview_SPQC} summarize representative studies according to these architectural and evaluation dimensions. The tables are not intended to imply that every deployment requires a ledger or that PARS is the only method of obtaining key evolution. Centralized key-management services, non-ledger group key agreement, and gateway-mediated security can be preferable when a continuously available authority is acceptable or when the cost of replicated consensus exceeds its auditability and fault-tolerance benefits. PARS-ESL is positioned for deployments in which multiple authorized nodes require a shared, tamper-evident view of security state without continuous reliance to online services.

This study addresses the identified gaps through three interconnected levels of abstraction. First, \emph{rotational secrecy} is introduced as a general design objective for limiting temporal key exposure through authenticated and policy-controlled cryptographic-epoch transitions. Second, PARS provides a reference framework for time-based (TB), action-based (AB), and group-based (GB) key evolution. Third, an optional Edge Security Ledger (ESL) provides distributed coordination of key epochs, membership, revocation, algorithm policy, validation outcomes, and recovery state when a deployment requires a shared and auditable view of security control.

The PARS-enabled ESL is not presented as a universal replacement for centralized key management, PKI, replicated directories, or conventional Byzantine Fault-Tolerant (BFT) state machines. Its intended deployment space consists of multi-agent systems in which authorized nodes require distributed security-state agreement, operate under intermittent connectivity, and cannot continuously depend on a single online authority. The architecture is derived from the authors' earlier work \cite{b15-2024kcso}, but is expanded here into a reference design and empirical evaluation framework.

The reference implementation supports classical, hybrid, and post-quantum algorithm suites, including standardized ML-KEM, ML-DSA, and SLH-DSA where applicable. Its broader benchmark corpus includes more than 60 historical and contemporary PQC candidates, while the deployment recommendations distinguish standardized algorithms from selected, historical, withdrawn, or cryptanalytically broken schemes. The evaluation framework complements existing IoT benchmarking studies \cite{fitzgibbon2024PQCBenchmark,liuPQCIoT2024} by examining protocol-level latency, memory, storage, bandwidth, scalability, and operational reliability rather than primitive execution time.

The current prototype has been evaluated with up to 16 nodes and includes an 18-hour run in which the reported peak memory metric remained below 3.5~MB. These measurements provide evidence for the evaluated configuration, but they do not by themselves establish universal suitability for all resource-constrained CPS platforms. Similarly, the adaptive challenge and consensus mechanism is analyzed in terms of challenge complexity, verification cost, and operational context without assuming that tunable computational difficulty automatically constitutes a formal Zero-Knowledge Proof (ZKP). The article therefore uses PARS-enabled ESL as a case study to connect post-quantum primitives, key-epoch evolution, distributed security state, implementation constraints, and migration planning across Edge AI and CPS.

\section{PROPOSED METHOD} \label{Proposed Method}
\begin{figure} [!htb]
\includegraphics[width=1\linewidth]{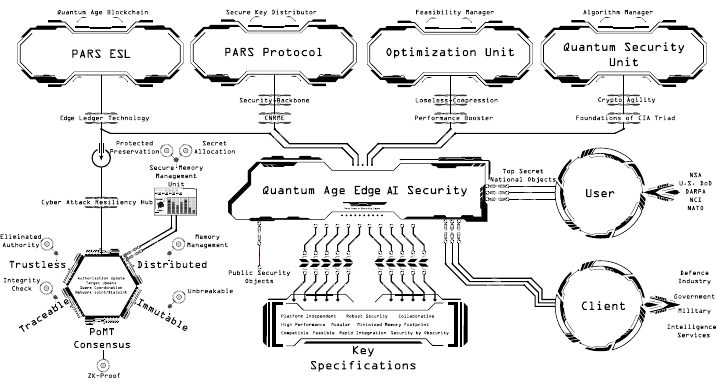}
\caption{Modular Overview of Proposed Study}
\label{fig_ESL_Overview}
\end{figure}

The acquisition of QC for large-scale attacks is considerably more challenging for traditional users compared to obtaining them for space and defense applications by administrations. Therefore, proposing cutting-edge architectures for space and defense systems is the priority to achieve mitigation of quantum attacks. Hence, the development of Quantum Blockchain (QBC) for Decentralized Autonomous Applications by focusing on a Secure Consensus Mechanism in defense environments (see Fig. \ref{fig_ESL_Overview}). This research is of paramount importance and plays a critical role in mission planning, coordination, swarm behaviors, and the implementation of decentralized smart consensus systems in the realm of asymmetric warfare and the evolution of space and defense coordination.

This section instantiates PARS and PARS-ESL as a security-control architecture for distributed Edge AI and CPS. The design is motivated by deployments in which multiple authorized agents must maintain a consistent view of cryptographic epochs, group membership, revocation, and algorithm policy without continuous dependence on a single online authority. The application surface include industrial automation, autonomous mobility, critical infrastructure, robotics, aerospace, and defense environments in which multiple agents must coordinate despite constrained processors, limited memory, intermittent connectivity, and long operational lifetimes. Space and defense systems provide a demanding representative environment as high consequence examples due to their long service lives, intermittent connectivity, heterogeneous hardware, high consequences of authentication, integrity failure, combination of long-lived cryptographic dependencies, distributed autonomy, heterogeneous communication links, and stringent availability requirements. PARS coordinates authenticated transitions between key epochs, whereas PARS-ESL records the corresponding security state through an optional permissioned ledger. High-rate operational telemetry and physical-control messages remain outside the ledger path and are protected through the symmetric keys associated with the active epoch. Validator-capable gateways or edge nodes may maintain the replicated ledger state, while more constrained agents may participate as lightweight clients. The architecture therefore does not assume that every endpoint executes full consensus or stores a complete blockchain.

The PQC Algorithm Standardization undertaken by NIST \cite{b18-2023NIST} represents a significant milestone deserving of scrutiny. This study asses over 60 algorithms, providing an extensive analysis of potential candidates for space and defense system security. The scaled analysis encompasses 9 Key Encapsulation Mechanisms (KEM) and 9 Digital Signature Algorithms (DSA). The empirical assessment is conducted using a machine equipped with an Intel i5-12450H 2.00GHz, 3.20GHz 64GB DDR4. It is running Ubuntu 18.04 with GCC optimization set to level 3 and 100 epochs are executed to observe potential performance variations across iterations processes for each algorithm, including key generation, encryption/decryption, signing, and verification. Performance details of KEM and DSA algorithms given in Table \ref{tab_PerformanceAnalysis}. 

Due to the extensive range of data streams, including Voice, Electro-Optic, Imagery, Telemetry, Lidar, Textual, Geospatial, Weather, Sonar, Radar, and GPS, even the lightest storage strategy can surpass the storage limitations of available space and defense systems with traditional blockchain approaches. Recording these high-rate data streams directly on a replicated ledger would impose unnecessary storage, bandwidth, privacy, and synchronization overhead. To address this, the proposed architecture separates system operation into three complementary planes: the data plane which transports high-rate sensor, telemetry, coordination, and actuation traffic using symmetric authenticated encryption; the security-control plane which manages identity validation, key establishment and rotation, membership updates, revocation, and security-state agreement; and the management plane which provisions credentials, cryptographic algorithm policies, software, and recovery information.Instead, given study focus on mission critical C4ISR related datas for Edge AI and CPS deployments. It addresses and investigates the potential control process of data signals that could be acquired and monitored by external forces.  This shifts the main focus from transaction-oriented blockchain to executive blockchain where the decentralization is a main characteristic.

PARS-ESL adopts a security-state-oriented model rather than a transaction-oriented data-retention. Hence, the PARS does not store each data transaction or message exchange on blockchain that creates insane storage for decentralization. It focus on executing data security over key chains. The ledger stores key-epoch commitments, membership and revocation events, algorithm-suite identifiers, validation outcomes, and recovery checkpoints. Operational data remain off-chain and are protected by the active traffic keys. In this article, the earlier term \emph{executive blockchain} is replaced by the more precise term \emph{Edge Security Ledger}. 

The traditional key exchange and management strategies are highly vulnerable to QC due to their vulnerabilities to Shor’s algorithm. PQC Algorithms often employ deterministic strategies for key exchange through KEM. However, this approach raises concerns due to its potential risks when used in defense systems requiring inevitable authentication among nodes and possible back doors. An alternative robust solution can be seen as QKD in this case but these approaches come with its limitations like requirements of special purpose equipment, physical layer hardware costs, additional network costs, and increased risks for denial of services pointed by NSA \cite{NSA_QKD_QC_Decide}. Therefore, a major necessity arises for a feasible and cost-effective secure key exchange method that can resist quantum attacks for next-generation space and defense systems. The study pioneers the literature by handling these requirements with blockchain-based rotational approach.

\begin{figure} [!htb]
\includegraphics[width=1\linewidth]{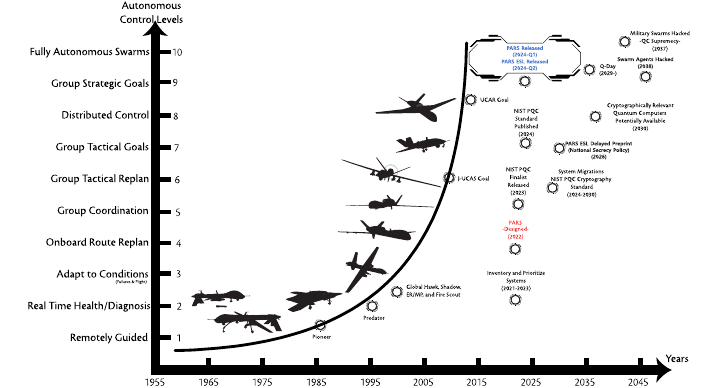}
\caption{Defense Roadmap Security Paradigm Shift}
\label{fig_USA_UAV_Roadmap}
\end{figure}

In addition to the information contained by the report of US Secretery of Defense, Fig. \ref{fig_USA_UAV_Roadmap} \cite{b15-2024kcso,bib_UAS_Roadmap_2005}) represent the importance of the PARS Protocol and ESL studies proposed in this field because there is highly important point which intrigues the attack cases given as doctrines. Although QCMPs cannot yet obtain suspicious deterministic outputs within the framework of the published information, security level update studies should be started for these very critical systems. The PARS and ESL architecture lead the literature thanks to its lightweight and secure infrastructure. It stands at a remarkable point among the 2005-2030 targets mentioned in the published report for fully autonomous swarms are included in the operational flow of the inventory. While there is no problem in the functioning of a fully autonomous system (provided that in case of no defects in its autonomous features), there are major problems and risks in the task flow of autonomous swarms due to the effects of the quantum computing power factor on the cyber world. These problems can generally be called risks and cyber threats that come with the potential processing power of QCMPs. Fully autonomous systems can detect, diagnose, and make decisions themselves. With these decisions, it is possible to create a great space and defense power potential by working collectively and collaboratively. If precautions are not taken against attacks carried out by QC in the communication of these systems, QCMPs can turn into very dangerous elements. ESL and PARS offer a solution to this critical problem. The key doctrines offered by the proposed study framework are discussed over possible attack cases in the next section. These attack cases should be handled and evaluated at the most critical category level for space and defense systems. All these scenarios can be extended in a way that affects the strategic flow of operation instead of interaction-level data. 

\subsection{Attack Mitigation Cases and Cyber Attack Surface} \label{crosRefAttackMitigationCases}
Space and Defense oriented CPS should prioritize key distribution and management due to the critical nature of the information they handle and the potential consequences of unauthorized access or tampering, because of the sensitive and classified information. These systems are prime targets for cyber attacks by state actors, terrorists, or hackers seeking to disrupt space and defense operations or gain strategic advantages. Proper key distribution and management are essential to ensure the CIA triad of this data. Without secure key management, adversaries could intercept, manipulate, or eavesdrop on communications, leading to compromised missions, loss of life, or strategic disadvantages. Effective key distribution and management mechanisms help defend against these threats by providing robust encryption, authentication, and access control measures. Due to the immense power of QCMPs, creating a system whose security level can be updated is a vital step to involve challenges. %In its true nature of public key is explicitly shareable. Even if the risks posed by this reality were acceptable before, in the quantum age this is indigestible for space and defense standards.

\begin{figure} [!htb]
\includegraphics[width=1\linewidth]{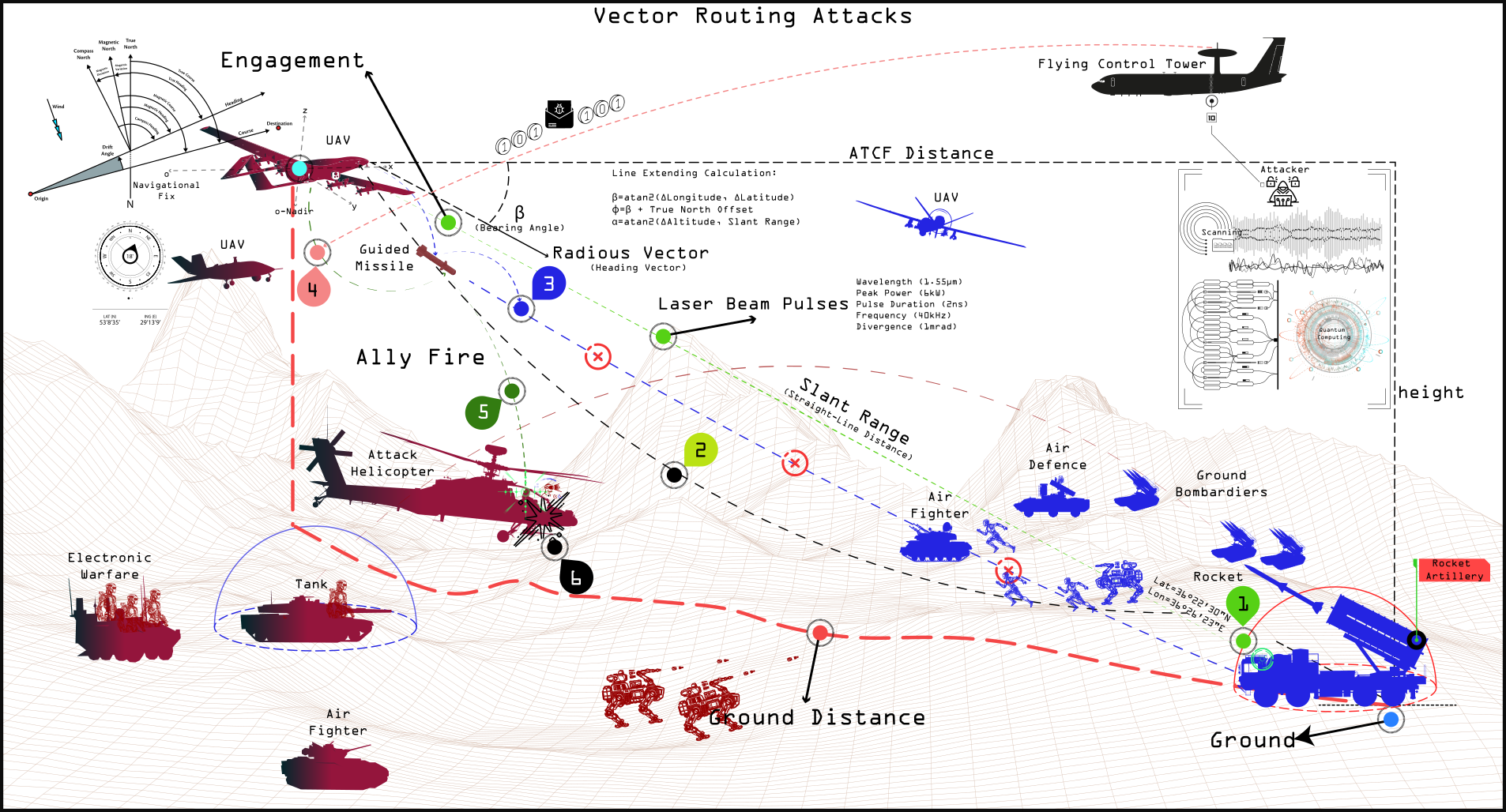}
\caption{Quantum Attacks - Denial of Engagement 1}
\label{fig_QuantumAttacks_Denial_of_Engagement_1}
\end{figure}

In scenario Vector Routing Attacks [\ref{fig_QuantumAttacks_Denial_of_Engagement_1}], the attacker node (Side-H) can measure the air speed of a specific target and relative coordinates. The relative coordinates and air-ground speed of the engager node (Side-C) can be used to produce bearing vectors of the engager node to its target. Furthermore, attackers can validate parameters through transmitted message packages over telemetry due to their massive computation power and decryption ability. If the battlefield consists of a compatible ally related to the engager node, the attacker can alter packages and convert suitable parameters to the missile to route. Firstly, the engager observes the positioning parameters for the target and releases the missile through the line extending to the target after the transmission of calculated outputs to the missile, respectively through S1, S2, and S3. During data feeds to the missile, the QTRs intercepts and manipulates positioning angles to allow the engager to damage his allies in S4. Thus the missile deviated from the actual line extending in S5 and damaged ally forces in S6. So the task is completed by Zero Cost Doctrine (ZCD). In the ZCD, firstly, the engager node designates the target and acquires the coordinates of the destination. After the acquisition of coordinates missile released in the navigation line with continuous feeds. During the stationary feed, the QTRs alters the critical exchange parameters and the missile guidance line changed to ally.

\begin{figure} [!htb]
\includegraphics[width=1\linewidth]{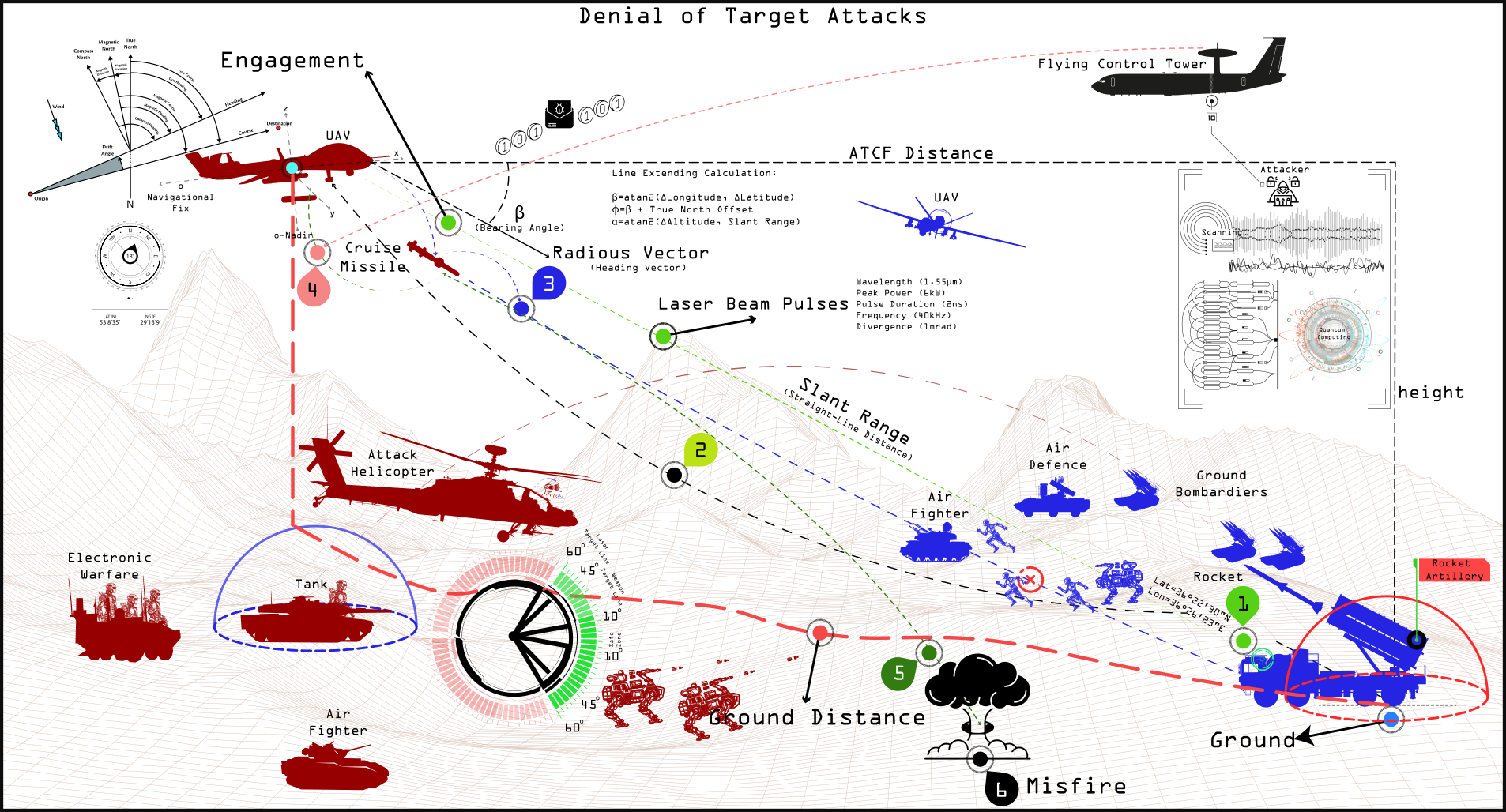}
\caption{Quantum Attacks - Denial of Engagement 2}
\label{fig_QuantumAttacks_Denial_of_Engagement_2}
\end{figure}

In scenario Denial of Target Attacks [\ref{fig_QuantumAttacks_Denial_of_Engagement_2}], the attacker node (Side-H) can measure the air speed of specific targets and relative coordinates. The relative coordinates and air-ground speed of the engager node (Side-C) can be used to produce bearing vectors of the engager node to its target. Furthermore, attackers can validate parameters through transmitted telemetry message packages due to their massive computation power and decryption ability. In this case, since there is no compatible allied force connected to the engager node, it makes no sense for an attacker to alter packages and convert parameters to route the missile through. So the attacker selected to cause little bias for the guidance of the missile in terms of completing tasks by Misfire Doctrine (MD). In the MD, firstly, the engager node designates the target and acquires the coordinates of the destination in S1 and S2. After the acquisition of coordinates missile released in the navigation line with continuous feeds in S3. During the stationary feed, the QTRs alters the critical exchange parameters and the guidance line changes with a small bias out of deviation range as if the missile successfully engaged in S4. Finally, the missile deviated from the actual line extending in S5 and caused to misfire doctrine in S6.

\begin{figure} [!htb]
\includegraphics[width=1\linewidth]{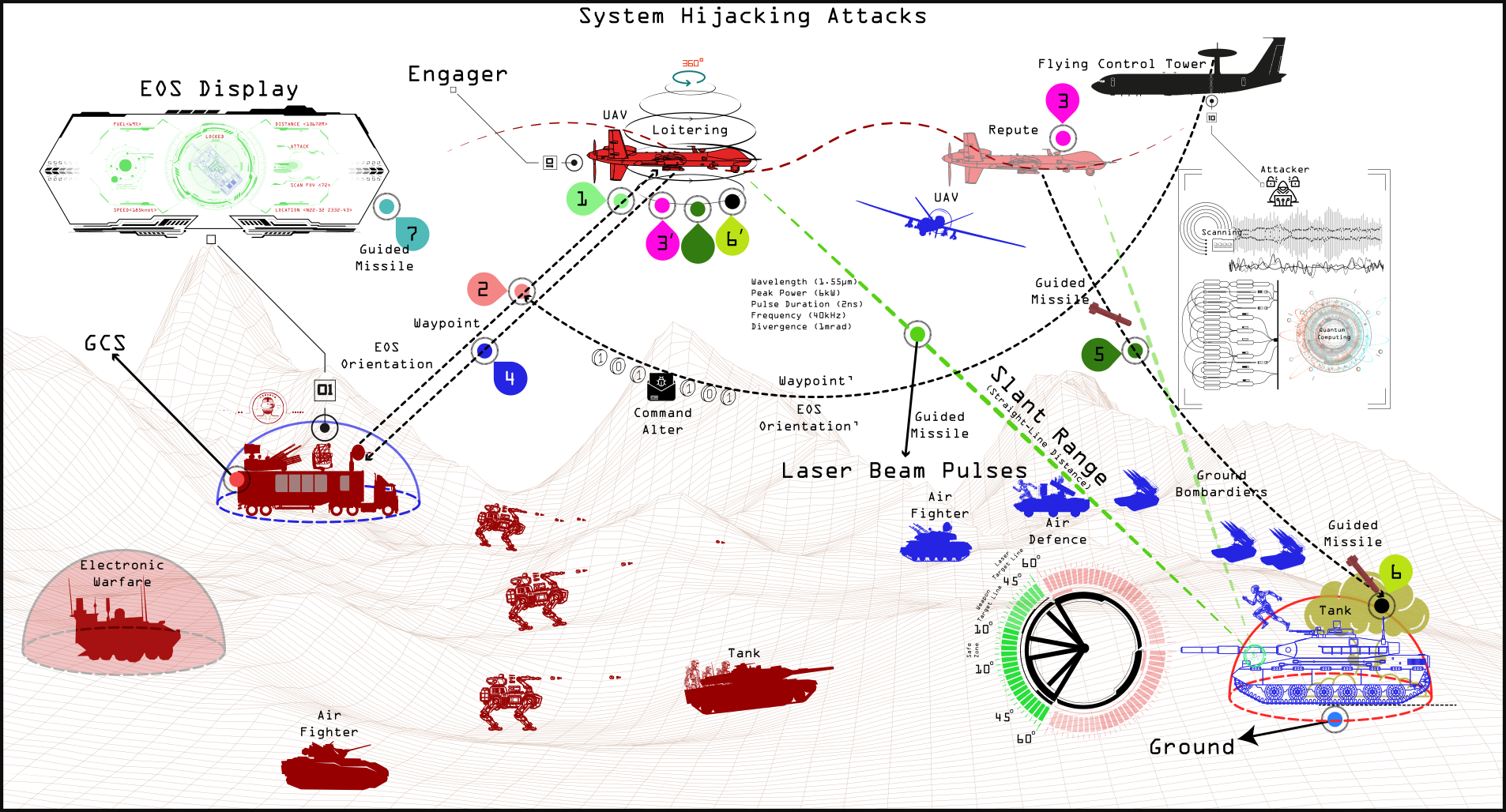}
\caption{Quantum Attacks - Denial of Command Control}
\label{fig_QuantumAttacks_Denial_of_Command_Control}
\end{figure}

In scenario Denial of Command Control Attacks [\ref{fig_QuantumAttacks_Denial_of_Command_Control}], the attacker node (Side-H) can resolve C4ISR patterns and identifies by insider infiltrator. The engager node (Side-C) can be deceived by QTRs as if it is a Ground Control Station (GCS). This attack type constitutes a high risk for both autonomous and non-autonomous driving. In this case, initially, the firing sequence for the engager node is initiated by LOS coordinate retrieval by GCS to push the engager through the planned path. The attacker interferes with the transmission, as in S4, by making the GCS assume that the engagement phase is in action. This intervention covers sending suitable monitoring messages to GCS despite the engager in a state of rotational inertia that stands still. 

The quantum attack rises where the decryption and manipulation of suitable command control exchanges are realized in the operational period. After the ACK was received by GCS, the operational flow was monitored as S3'-S5'-S6' while the fact is S3-S5-S6. The intrusion is noticeable, as in S7, when the capture is done. This process is called the Domineer Doctrine (DD). In the DD, the main objective is capturing the system functionality. The attacker can manipulate target node actions in a way that assumes operational flow is proper while the system is tried to be destroyed by moving it closer to air defense ranges. Moreover, it can land safely on its runways. This is not always the result, but the main principle tendency is manipulating control owners in a way that creates illusive minor effects that affect the operation. The basis of the entire process is to perform successful manipulation by quantum attacks in operational.

\begin{figure} [!htb]
\includegraphics[width=1\linewidth]{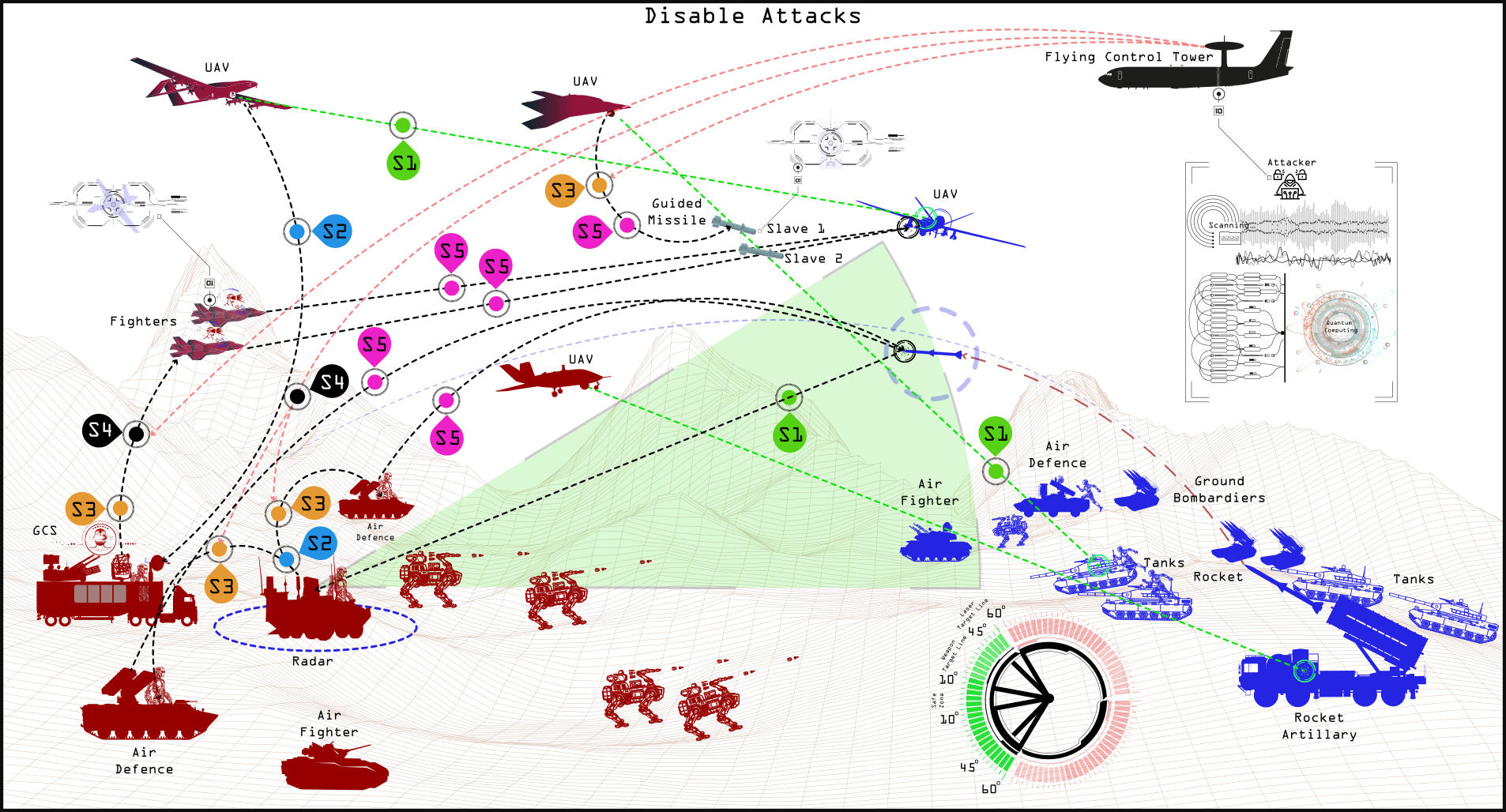}
\caption{Quantum Attacks - Denial of System}
\label{fig_QuantumAttacks_Denial_of_System}
\end{figure}

In scenario Denial of System [\ref{fig_QuantumAttacks_Denial_of_System}], the attacker node (Side-H) is capable of presenting herself as a friend in case of any engagement or assault attempt on its slaves. The intruder can manipulate its existence over the exchange between enemy communication in a way that causes enemies to perceive its identification as something other than what it is on the battlefield. The imaging and radar nodes detect environment objects and announce connected action nodes as seen in S1 and S2. The attacker intervenes in the transmission, as in S3, and lunge message stack which is determined responsible for reporting in a way that pretends there is no abnormality in S4. Then, the manipulated nodes do not fulfill the mission scope. If the attacker can destroy antagonist forces by cyber spoofing the task is completed by Chameleon Doctrine (CHD). If the intruder can prevent a potential attack on its swarm cluster as well as not being able to destroy it, the task is completed by Cuttlefish Doctrine (CTD).

\begin{figure} [!htb]
\includegraphics[width=1\linewidth]{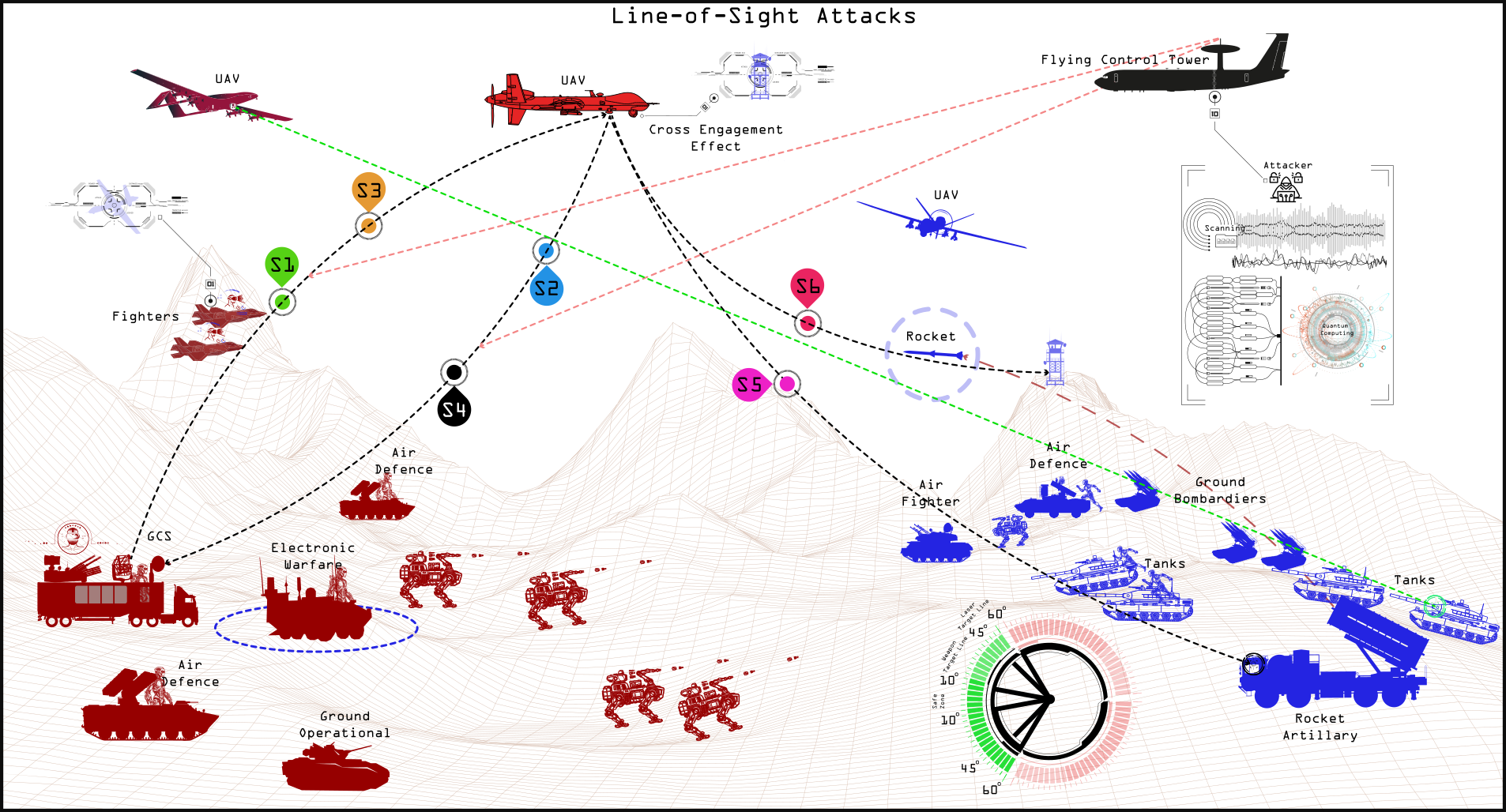}
\caption{Quantum Attacks - Denial of Line-of-Sight}
\label{fig_QuantumAttacks_Denial_of_LOS}
\end{figure}

In scenario Denial of Line-of-Sight [\ref{fig_QuantumAttacks_Denial_of_LOS}], it demonstrates how the EOS Controller, either AI-supported autopilot or payload operator, can be manipulated through the QTRs node (Side-H). It can alter microcontrol actions such as orientation, pointing, designation, slewing, panning, tilting, and zooming over transmission. In addition to these values, it is also possible to create a shift by positioning data in the computational base. The data exchange-oriented control is available in the nature of the job due to surveillance tasks. This comes with high cyber security risks in case of target diversion by intruders. It is possible to achieve this diversion without disturbing the control mechanism. The attack case starts by S1 when the transmission is detected between GCS and Engager. Then the engager returns responses to GCS, as in S2, about the confirmation of micro control actions stated above. The attacker intervenes in communication lines for consecutive data transmission as in S3 and S4 respectively for command and response. After the quantum attack, the engager system is successfully dominated for expected target destruction as S6 instead of S5. This attack strategy is the Naked Ape Doctrine which can be also called the Eclipse Doctrine.

\begin{figure} [!htb]
\includegraphics[width=1\linewidth]{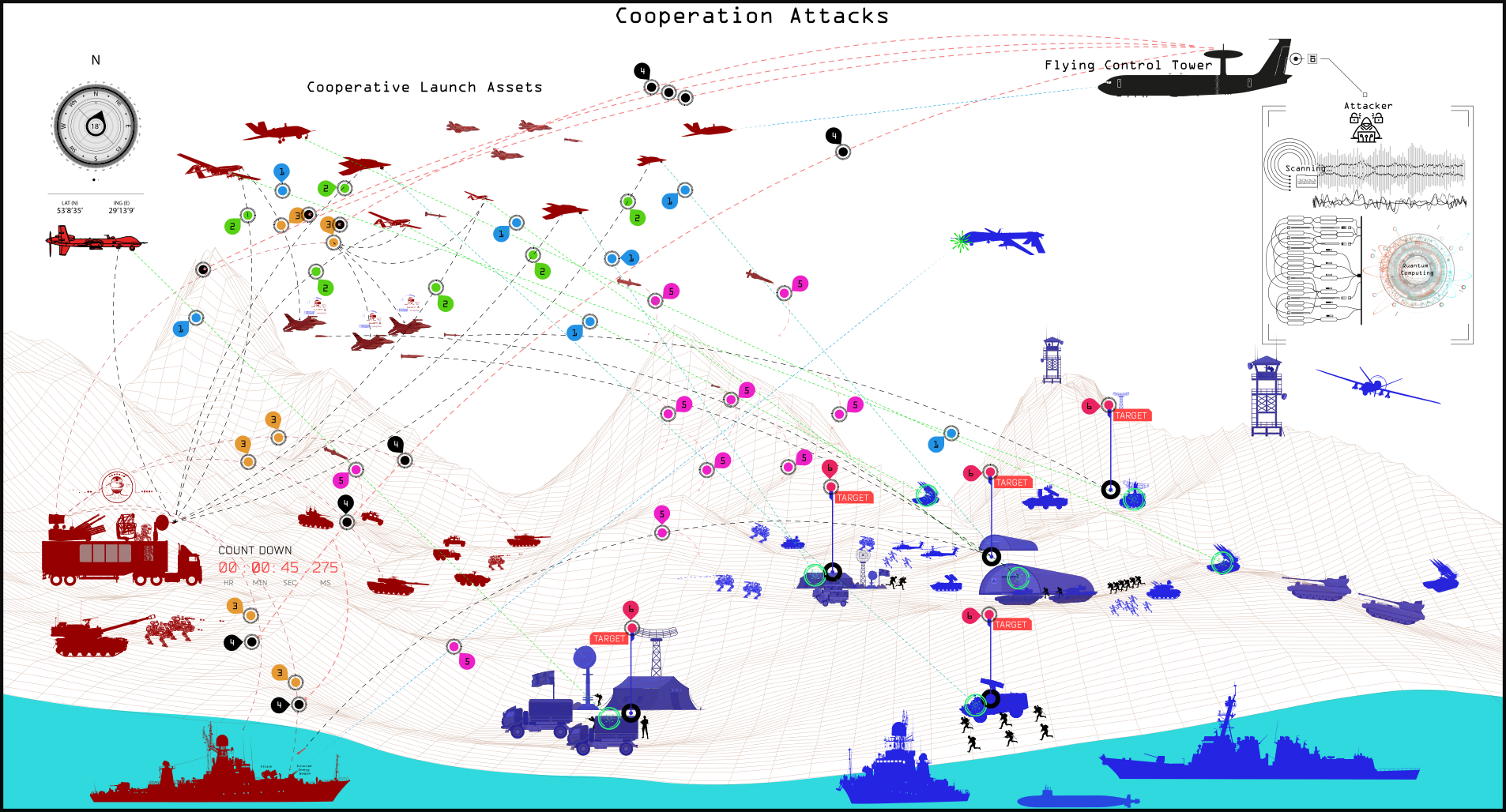}
\caption{Quantum Attacks - Denial of Cooperation}
\label{fig_QuantumAttacks_Denial_of_Cooperation}
\end{figure}

In scenario Denial of Cooperation[\ref{fig_QuantumAttacks_Denial_of_Cooperation}], it is possible to manipulate and alter swarm communication by the QTRs node (Side-H). It can process massive data transmission between swarm systems by exploring exchange patterns and altering them. This kind of attack are Octopus Doctrine (OD) which aims for holistic intervention on the battlefield. This holistic approach is a cutting-edge devastating attack strategy after the Directed-Energy Weapons with great cost savings. In the OD, firstly, observer nodes designate, reveal, or point the target acquire the coordinates of destinations in S1, and notify related edge nodes in S2 and S3. During the stationary feed, the QTRs alters the critical exchange parameters, and the missile guidance line changes with an important bias out of deviation range as if the missile successfully engaging. Finally, the missile deviated from the actual line extending in S5 and avoided destroying targets in S6.

\subsection{Private Authentication for Rotational Secrecy (PARS)} \label{crosRefPARSProtocol}
The fundamental need for a Collective Space and Defense System is secure message exchange between peers. The proposal focus on ensuring critical data security by refreshing the main key chain among peers which serves as the cornerstone of secure transmission. It aims to transform traditional security approaches for message exchange in mission critical environments, introducing a decentralized approach. The proposed quantum key management protocol aims to handle the security of strategic message transmission cycles during tactical swarm operations which NSA, DARPA, NATO, and U.S. DoD targets in long-term periods.

\begin{figure} [!htb]
\includegraphics[width=1\linewidth]{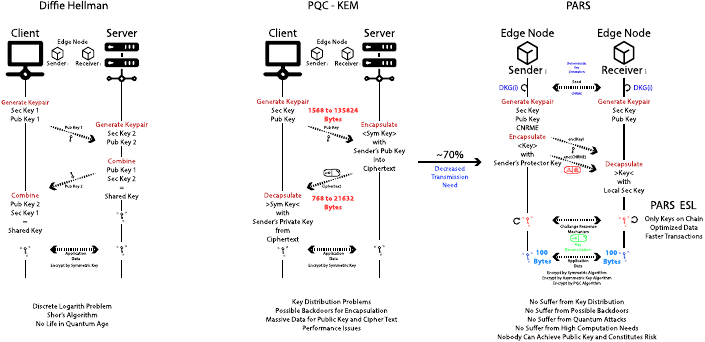}
\caption{Comparison of Key Exchange Methods}
\label{Figure_DH_PQC_PARS_Compact}
\end{figure}

\begin{figure*}[!htb]
\includegraphics[width=1\linewidth]{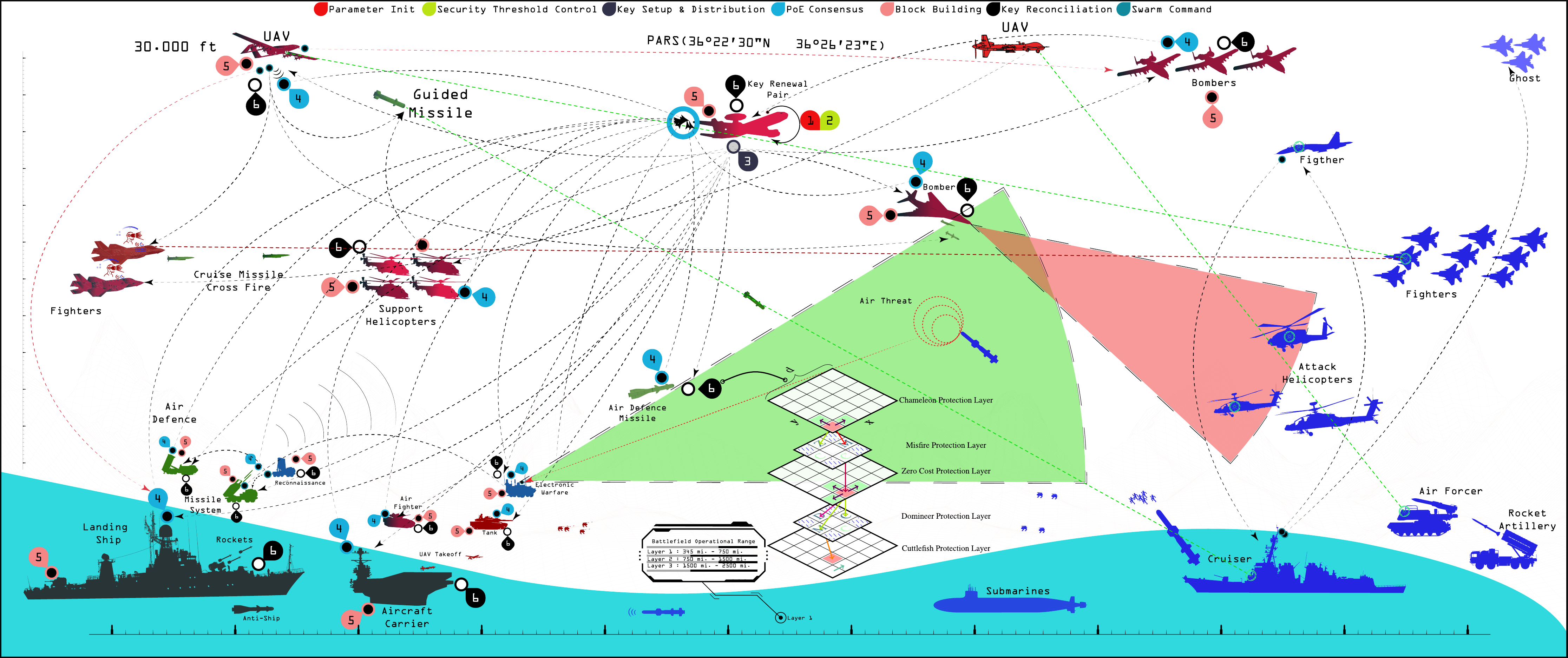}
\caption{PARS Protocol Operational Flow.}
\label{fig_PARS_Operational_Flow}
\end{figure*}

The PARS protocol is designed for dynamically rebuilding key chains (encryption, decryption, sign, and verify) through heterogeneous space and defense nodes different from available method as shown in Fig. \ref{Figure_DH_PQC_PARS_Compact}. It lets swarm systems renew security levels in periodic order for asymmetric cyber warfare shown in Fig. \ref{fig_PARS_Operational_Flow}. In traditional systems, cryptography keys serve as indispensable components in ensuring the secure storage and seamless transmission of information. This protocol focus on creating an infinitely secure system for subsequent data exchange flows by making the time insignificant required to decipher these keys, even under massive computational capabilities. It uses deterministic periods for renewing keys over a blockchain architecture. The selected PQC algorithm can vary among node devices to enforce the system against heterogeneous attack classes and serves as an additional security booster. To broaden the accepted operational envelope, PARS realises its key-renewal core, the Combinatorial Non-Reversible Mixing Effect (CNRME), in two interoperable instantiations selected through the protocol's crypto-agility parameters: a post-quantum encapsulation mode that protects and conveys a public key under module-lattice hardness, traditional symmetric mode and seed-decoupled mode that transmits no key material and instead reconstructs the shared key chain locally from end-to-end seeds decoupled from the key. The latter is, by construction, resistant to the Harvest-Now-Decrypt-Later (HNDL) threat---since no quantity bound to the key by a public-key relation ever appears on the wire, an adversary recording all traffic and later acquiring a quantum computer still cannot reconstruct past keys---and it avoids transporting large post-quantum key material; the former offers explicit lattice hardness and interoperability with public-key infrastructure. Both modes share the same six core steps, swarm consensus, and ledger, differing only in the key-setup step. The protocol's main pseudo-code is given in two parts, and each CNRME mode is specified as a further parts.

Each edge node will initialize $n$ auxiliary key protectors, for pre-defined parameter initialization, during system start-up apart from their mission type. These parameters are Protector Keys ($PK$) during key setup transmission, Machine ID ($M_{id}$), and Pattern Design Parameters ($P_{dp}$). Selected PQC algorithm will common for all nodes to achieve collective synchronization and initiated as off-chain parameters. Finally, the Differentiation Effect Table ($DET$) is initiated for each node to prevent potential vulnerabilities for $M_{id}$ and $Alg_{id}$. DETs are used for masking identities of on-chain consecutive activities during protocol flow.

The Sender Node ($SN$) periodically checks the remaining safe-usage budget of the current key, during periodic security threshold control, material shared with the Receiver Node ($RN$) before message transmission. Then, rotational key exchange activity starts if the available usage exceeds the Security Threshold ($ST$). The $ST$ will be the safe time zone before risks rise. The $ST$ can be determined in three ways --- Time-Based ($TB$), Action-Based ($AB$), and Group-Based ($GB$) --- in PARS. $TB$ means a pre-defined time threshold to hold security. $AB$ means detected anomalies and operational concept change. $GB$ stands for entire peer key change instead of twins. Each pair will continue message exchange with available keys until the consensus is ensured and block synchronization is achieved for exchange activity.

The system parameters governing post-quantum key updates within the proposed framework are dynamically orchestrated by the Periodic Security Threshold Control (PSTC) function. The PSTC mathematical model continuously adapts the operational security threshold based on (\ref{pstc_model}) real-time threat vector assessments and tactical communication link dynamics.

\begin{equation}
ST(t) = f(\mathcal{Q}_A, \Delta \tau) = \prod_{i=1}^{n} \left( \frac{\lambda_i \cdot \alpha_i}{\log_2(\mathcal{M}_t)} \right) - \gamma \Delta \tau
\label{pstc_model}
\end{equation}

\begin{itemize}
    \item $ST(t)$: The dynamic security threshold at time $t$. Lower scalar values indicate an elevated risk profile or degraded link quality, triggering accelerated post-quantum rekeying intervals.
    \item $f(\mathcal{Q}_A, \Delta \tau)$: The governing PSTC orchestration function mapping the threat assessment vector $\mathcal{Q}_A$ and link latency differential $\Delta \tau$ into the unified scalar threshold.
    \item $\mathcal{Q}_A$: The multi-dimensional risk evaluation vector aggregating threat inputs across physical, cryptographic, and network operational domains.
    \item $n$: The total number of evaluated security risk categories ($i \in \{1, \dots, n\}$).
    \item $i$: Index corresponding to a specific security risk domain (e.g., side-channel power leakage, packet injection rate, or quantum cryptanalysis speedup metrics).
    \item $\lambda_i$: The threat intensity coefficient for category $i$, quantifying active attack severity or signal jamming intensity.
    \item $\alpha_i$: The hardware vulnerability weighting factor for category $i$, reflecting the intrinsic physical susceptibility of the underlying Edge AI hardware execution substrate (e.g., FPGA PL fabric, GPU co-processor, or RISC-V vector core).
    \item $\mathcal{M}_t$: The effective cryptographic state space cardinality (key space entropy) at operational time $t$.
    \item $\log_2(\mathcal{M}_t)$: The effective bit-security level of the active key state (e.g., 128, 192, or 256 bits).
    \item $\Delta \tau$: The measured channel delay variation or differential transmission latency across the tactical communication link (operating under Denied, Disrupted, Intermittent, and Limited [DDIL] conditions).
    \item $\gamma$: The latency penalty weighting coefficient that converts raw transmission delay spikes into proportional threshold adjustments.
    \item $\gamma \Delta \tau$: The dynamic link latency penalty term, preventing network congestion collapse during key exchanges over bandwidth-constrained channels.
\end{itemize}

The multiplicative compounding term $\prod_{i=1}^{n} \left( \frac{\lambda_i \cdot \alpha_i}{\log_2(\mathcal{M}_t)} \right)$ models the combined risk across all $n$ evaluated dimensions. An increase in active threat intensity ($\lambda_i$) or platform vulnerability ($\alpha_i$) expands the numerator and raises the risk threshold, whereas maintaining high effective key entropy ($\log_2(\mathcal{M}_t)$) counteracts threat amplification.

The subtractive term $-\gamma \Delta \tau$ introduces physical communication awareness. Because post-quantum cryptographic primitives (such as FIPS 203 ML-KEM-1024 or FIPS 204 ML-DSA-87) impose significant key and signature byte overheads (reaching up to 9.53~KB per transaction under CNSA 2.0 standards), naive continuous rekeying risks saturating low-bandwidth tactical links (1.67--78~kbit/s). By dynamically penalizing threshold delays as $\Delta \tau$ increases, the PSTC function avoids packet fragmentation and network collapse while enforcing strict security guarantees.

\begin{figure} [!htb]
\includegraphics[width=1\linewidth]{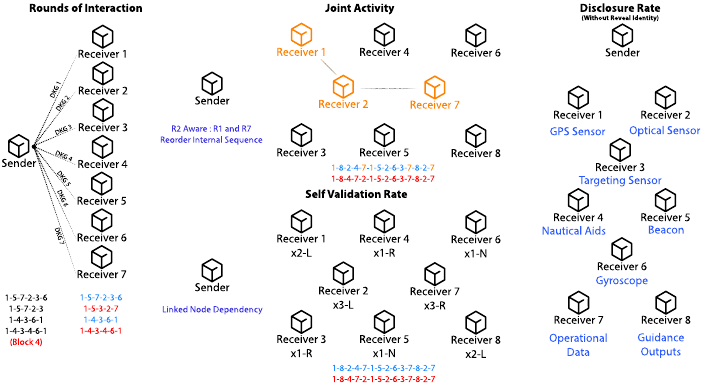}
\caption{ZKP Hardness for Challange Response Mechanism}
\label{fig_ZK_Proof_Hardness_Challange_Response}
\end{figure}

\begin{figure}[!htb]
\includegraphics[width=1\linewidth]{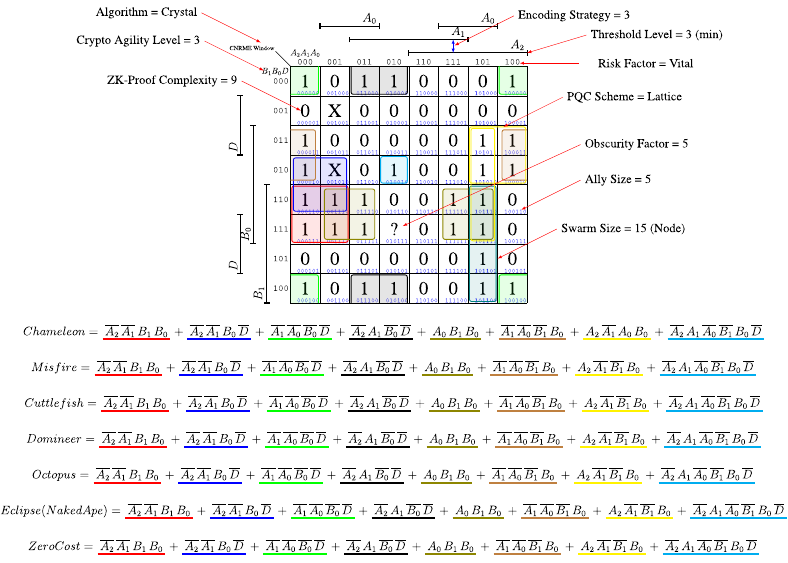}
\caption{PARS Protocol Generation and Protection Metrics}
\label{fig_PARS_CNRME_Fundatemental_GP_Metrics}
\end{figure}

Generation of a fresh mutual key chain, during key setup and distribution, for a selected peer pair. Each peer may select its scheme independently, provided that the variable Algorithm ID ($Alg_{id}$) is conveyed to the chain under local obscuration; transferring this identifier creates no vulnerability because its heterogeneous projection means the same $Alg_{id}$ can be represented by many on-chain values, which preserves node authenticity and prevents infiltrators from exposing the messaging architecture or rendering command interaction vulnerable. The obscurity primitive at the heart of the renewal is the CNRME. To enlarge the operational envelope of the protocol, PARS instantiates CNRME in two interoperable modes that a deployment selects through the crypto-agility component of the parameter vector (the PQC~Scheme selector): a post-quantum encapsulation mode, $\mathsf{CNRME.Encapsulate}$ (Part~\ref{alg:cnrme_enc}), which protects and conveys a public key under module-lattice hardness; and a traditional symmetric mode and a seed-decoupled mode, $\mathsf{CNRME.Derive}$ (Part~\ref{alg:cnrme_derive}), which never transmits key material and instead reconstructs the shared key chain locally from end-to-end seeds that are decoupled from the key. The two modes are deliberately symmetric in their interface: each is keyed by the same pre-shared Protector Keys ($PK$), each is governed by the same major parameters, Algorithm, Crypto Agility Level, ZKP Complexity (see Fig. \ref{fig_ZK_Proof_Hardness_Challange_Response}), Encoding Strategy, Threshold Level, Risk Factor, PQC Scheme, Obscurity Factor, Ally Size, and Swarm Size, as indicated in Fig. \ref{fig_PARS_CNRME_Fundatemental_GP_Metrics}, and each emits a public, one-way value (encapsulation, commitment) that is broadcast to the edge interactors with hash and timestamp, so that the swarm can verify agreement on-chain without disclosing the key. They differ only in what is placed on the wire and in which hardness assumption. The ESL manages modes to mitigate extended spectrum of attacks by layered heterogeneous approach (see Fig. \ref{fig_PARS_Operational_Flow}). The exact construction of each mode as a single deterministic (post-quantum and local-derivation) are specified in Parts~\ref{alg:cnrme_enc} and~\ref{alg:cnrme_derive}.

% 2022/10/09 : Algoritma Denklemi tek bir parca buyuk olursa hizalamada problem yasaniyor. Bu sebeple ayrildi. 
\vspace{2mm}
\setlength{\intextsep}{1pt} % Adjust the value as needed
\setlength{\textfloatsep}{1pt} % Adjust the value as needed
\floatname{algorithm}{PARS Protocol (Alignment) - Part}
\begin{algorithm}[!htb]
\small % Set the font size to small
\caption{PARS Protocol (Alignment)}
\label{alg:PARSProtocol}
\begin{tabular}{p{0.95\linewidth}}
\textbf{Parameter Initialization} \\
\end{tabular}
\begin{tabular}{p{0.12\linewidth}p{0.78\linewidth}}
%\textbf{Inputs:} & Superscripts: $b^2$, $Y^{\text{super}}$, $B^{\mathrm{super}}$ \\
\textbf{Inputs:} & $Node_i$, $Count_{\text{Node}}$ \\
\end{tabular}
\begin{tabular}{p{0.14\linewidth}p{0.76\linewidth}}
\textbf{Outputs:} & $PK_i$, $M_{\text{id}}$, $P_{\mathrm{dp}}$ \\
\end{tabular}

\begin{tabular}{p{0.04\linewidth}p{0.86\linewidth}}
1: & \textbf{function} initParams($Node_i$, $Count_{\text{Node}}$) \\
2: & \hspace{10pt} $PK_i, M_{\text{id}}, P_{\mathrm{dp}} \gets 0$ \\
3: & \hspace{10pt} \textbf{for each} $Node_i \in \text{Military Space}$ \\
4: & \hspace{20pt} $PK_i \gets \text{ProtectorKeysForNewSession}$ \\
5: & \hspace{20pt} $M_{\text{id}_{i}} \gets \text{GenerateMachineID}$ \\
6: & \hspace{20pt} $P_{\mathrm{dp}} \gets \text{GeneratePatternDesignParams}$ \\
7: & \hspace{10pt} \textbf{return} $PK_i$, $M_{\text{id}}$, $P_{\mathrm{dp}}$ \\
\end{tabular}

\begin{tabular}{p{0.95\linewidth}}
\textbf{Periodic Security Threshold Control} \\
\end{tabular}
\begin{tabular}{p{0.12\linewidth}p{0.78\linewidth}}
\textbf{Inputs:} & $Node_i$, $M_{\text{id}_{i}}$, $Telemetry_{i}$ \\
\end{tabular}
\begin{tabular}{p{0.14\linewidth}p{0.76\linewidth}}
\textbf{Outputs:} & $Vulnerability_{\text{i}}$ \\
\end{tabular}
\begin{tabular}{p{0.04\linewidth}p{0.86\linewidth}}
1: & \textbf{function} checkThreshold($Node_i$, $M_{\text{id}_{i}}$, $Telemetry_{i}$) \\
2: & \hspace{10pt} $validity_i \gets 0$ \\
3: & \hspace{10pt} \textbf{for each} $Node_i \in \text{Military Space}$ \\
4: & \hspace{20pt} $validity_i \gets \text{ControlKeyValidity}(Node_{SN})$ \\
5: & \hspace{20pt} \textbf{if} $validity_i > \text{SecurityThreshold}$ \\
6: & \hspace{20pt} $PARS$.continueExchange($Node_{SN}$, $Node_{RN}$) \\
7: & \hspace{10pt} \textbf{while} $validity_i$ is not valid \\
8: & \hspace{20pt} $validity_i \gets PARS$.KeyReconciliation() \\
9: & \hspace{10pt} \textbf{return} $validity_i$ \\
\end{tabular}

\begin{tabular}{p{0.95\linewidth}}
\textbf{Key Setup and Distribution} \\
\end{tabular}
\begin{tabular}{p{0.12\linewidth}p{0.78\linewidth}}
\textbf{Inputs:} & $Node_i$, $M_{\text{id}_{i}}$ \\
\end{tabular}
\begin{tabular}{p{0.14\linewidth}p{0.76\linewidth}}
\textbf{Outputs:} & $PUB_{\text{i}}$ \\
\end{tabular}
\begin{tabular}{p{0.04\linewidth}p{0.86\linewidth}}
1: & \textbf{function} keySetup($Node_i$, $M_{\text{id}_{i}}$, $Telemetry_{i}$) \\
2: & \hspace{10pt} $PUB_i, PR_i, Alg_{id} \gets 0$ \\
3: & \hspace{10pt} \textbf{for each} $Node_i \in \text{Military Space}$ \\
4: & \hspace{20pt} $Alg_{id} \gets \text{selectPQCAlg}(Node_{SN})$ \\
5: & \hspace{20pt} $PUB_i, PR_i \gets \text{genKeyPair}(Node_{SN}, Alg_{id})$ \\
6: & \hspace{20pt} $PUB_{SN} \gets \text{encrypt}(PK_i, PUB_{SN})$ \\
7: & \hspace{20pt} $PUB_{SN} \gets \text{generateCNR-ME}(PUB_{SN})$ \\
8: & \hspace{20pt} Broadcast($PUB_{SN}$, $Node_{RN}$, $Alg_{id}$) \\
9: & \hspace{10pt} \textbf{return} $validity_i$ \\
\end{tabular}
\end{algorithm}
\vspace{2mm}

The two modes share the notation, the primitive set and the parameter space introduced above, and diverge only where the post-quantum mode (Mode~A) consumes a public key and the symmetric mode (Mode~B) consumes an end-to-end seed. Throughout, $\lambda$ denotes the security parameter (target $\lambda=256$, i.e.\ a $128$-bit margin against quantum adversaries), $\mathbin{\|}$ denotes bit-string concatenation, $\oplus$ bitwise \textsc{xor}, $\lfloor\cdot\rceil$ rounding to the nearest integer, and $\lfloor x\rfloor_{\lambda}$ truncation to the leading $\lambda$ bits of $x$. Both modes draw from a single standardised, quantum-resistant toolset; what changes between them is the role of the lattice layer. A keyed pseudo-random function $\mathsf{F}\colon\{0,1\}^{\lambda}\times\{0,1\}^{*}\!\to\!\{0,1\}^{\lambda}$, instantiated as AES-256 operated as a PRF/PRG (CMAC and CTR modes) \cite{FIPS197}, whose best known quantum key recovery is Grover search in $\Theta(2^{\lambda/2})$. A Keccak sponge providing an extendable-output function $\Xi\colon\{0,1\}^{*}\!\to\!\{0,1\}^{*}$ (\textsc{shake}-256) and a fixed-length hash $\mathsf{H}\colon\{0,1\}^{*}\!\to\!\{0,1\}^{\lambda}$ (\textsc{sha3}-256) \cite{FIPS202}, whose preimage resistance is $\Theta(2^{\lambda})$ classically and $\Theta(2^{\lambda/2})$ quantumly. A module-lattice ring $R_q=\mathbb{Z}_q[X]/(X^{d}+1)$ with $d$ a power of two and $q$ prime, together with a deterministic number-theoretic-transform (NTT) diffusion $\mathsf{L}(\mathbf v)=\mathrm{NTT}^{-1}\!\big(\hat{\mathbf A}\odot\mathrm{NTT}(\mathbf v)\big)$, where $\hat{\mathbf A}\in R_q^{k\times k}$ is a key-scheduled matrix and $\odot$ the point-wise product, whose hardness is inherited from Module-LWE \cite{NIST2024FIPS203,Regev2009}. In Mode~A this lattice diffusion is a core and load-bearing pillar---the encapsulated public key is bound to a Module-LWE-structured element; in Mode~B the identical primitive is retained only as an optional crypto-agility booster that is not required for any of the symmetric-mode guarantees. Canonical, prefix-free encoders $\mathsf{Encode}(\cdot)$ and $\mathsf{Combine}(\cdot)$ together with the octet-string conversion $\mathrm{I2OSP}(\cdot)$. Every CNRME invocation, in either mode, is keyed by the secret protector bundle $\mathcal P=(PK_1,\dots,PK_\kappa)$ established off-chain at initialization and which never traverses the network during operation, and is governed by the public parameter vector.

\vspace{2mm}
\floatname{algorithm}{PARS Protocol (Consensus Mechanism) - Part 2}
\begin{algorithm}[!htb]
\small % Set the font size to small
\caption{PARS Protocol (Consensus Mechanism)}
\label{alg:PARSProtocol_Part2}
\begin{tabular}{p{0.95\linewidth}}
\textbf{PoE Consensus} \\
\end{tabular}
\begin{tabular}{p{0.12\linewidth}p{0.78\linewidth}}
\textbf{Inputs:} & $Node_i$, $P_{\mathrm{dp}}$, $PK_i$, $PUB_i$ \\
\end{tabular}
\begin{tabular}{p{0.14\linewidth}p{0.76\linewidth}}
\textbf{Outputs:} & $ValidationCount_{\text{PUB}_{i}}$ \\
\end{tabular}
\begin{tabular}{p{0.04\linewidth}p{0.86\linewidth}}
1: & \textbf{function} PoEConsensus($Node_i$, $P_{\mathrm{dp}}$, $PK_i$, $PUB_i$) \\
2: & \hspace{10pt} $f(x \gets PUB_i) :$ StartConsensus($x$) \\%\rightarrow
3: & \hspace{10pt} $f(y \gets PK_i) :$ DetectPKsIdentityInEdgeNodes($y$) \\
4: & \hspace{10pt} $f(z, t \gets Node_i, P_{\mathrm{dp}}) :$ ProcessPatterns($z, t$) \\ 
5: & \hspace{10pt} $f(t, u \gets Node_i, VC_{\mathrm{i}}) :$ ExecutePUBConsensus($t, u$) \\ %\rightarrow
6: & \hspace{10pt} \textbf{return} $ValidationCount_{\text{PUB}_{i}}$ \\
\end{tabular}
\begin{tabular}{p{0.95\linewidth}}
\textbf{Block Building} \\
\end{tabular}
\begin{tabular}{p{0.12\linewidth}p{0.78\linewidth}}
\textbf{Inputs:} & $H_i$, $EV_i$, $GT_i$, $CT_i$, $O_i$ \\
\end{tabular}
\begin{tabular}{p{0.04\linewidth}p{0.86\linewidth}}
1: & \textbf{function} BlockBuilding($H_i$, $EV_i$, $GT_i$, $CT_i$, $O_i$) \\
2: & \hspace{10pt} $block_i \gets \text{CreateBlock}(H_i, EV_i, GT_i, CT_i, O_i)$ \\
3: & \hspace{10pt} $RAM \gets \text{StoreBlock}(block_i, addr_{ram})$ \\ %UntilSync
4: & \hspace{10pt} $localChain \gets \text{AddChain}(block_i)$ \\
\end{tabular}
\begin{tabular}{p{0.95\linewidth}}
\textbf{Key Reconciliation} \\
\end{tabular}
\begin{tabular}{p{0.12\linewidth}p{0.78\linewidth}}
\textbf{Inputs:} & $Node_i$, $PUB_i$ \\
\end{tabular}
\begin{tabular}{p{0.14\linewidth}p{0.76\linewidth}}
\textbf{Outputs:} & $OpSignal_{\text{i}}$ \\ %Operational Signal
\end{tabular}
\begin{tabular}{p{0.04\linewidth}p{0.86\linewidth}}
1: & \textbf{function} KeyReconciliation($Node_i$, $PUB_i$) \\
2: & \hspace{10pt} $result \gets \text{CheckIntegrity}(Node_i, PUB_i, addr_{flash})$ \\ %OfLocalChain RAMADDRESS
3: & \hspace{10pt} $match_{\text{signal}_i} \gets \text{GenerateSignals}(block_i, addr_{flash})$ \\ %UntilSync
4: & \hspace{10pt} $chain_{local} \gets \text{MakeOperational}(PUB_i)$ \\
5: & \hspace{10pt} \textbf{return} $validity_i$ \\
\end{tabular}
\end{algorithm}
\vspace{2mm}

In addition to the parameters of Fig.~\ref{fig_PARS_CNRME_Fundatemental_GP_Metrics}, per-epoch nonce $\nu$ (the generation time $GT$) is included. The remaining, mode-specific input is the quantity to be protected, a public key $PUB$ in Mode~A, or a fresh end-to-end seed $\sigma=\sigma_{SN}\mathbin{\|}\sigma_{RN}$ to which Sender and Receiver each contribute in Mode~B. Writing $\theta_i$ for the admissible domain of the $i$-th component of \eqref{eq:cnrme_theta}, the configuration space cardinality given as \eqref{eq:cnrme_cfg}.
\begin{equation}
\theta=\big(\mathit{A},\,\mathit{CAL},\,\mathit{ZKc},\,\mathit{E},\,\mathit{T},\,\mathit{R},\,\mathit{PQC},\,\mathit{Obf},\,\mathit{AS},\,\mathit{SS}\big)
\label{eq:cnrme_theta}
\end{equation}
\begin{equation}
N_{\mathrm{cfg}}=\prod_{i}\lvert\theta_i\rvert ,
\label{eq:cnrme_cfg}
\end{equation}
Each $\theta$ deterministically selects a distinct key schedule, (optional or core) diffusion matrix and domain context, contributing $\log_2 N_{\mathrm{cfg}}$ bits of uncertainty over the instance of the function itself shared by both modes.

The Mode A is post-quantum encapsulation phase and the protected quantity here is a public key and the lattice layer is load-bearing given as (\ref{eq:cnrme_fn_enc}). The mode is the deterministic, $\mathcal P$-keyed map where $C\in\{0,1\}^{\ell_{pub}}$ is the recoverable masked key, $\tau\in\{0,1\}^{\lambda}$ is the non-reversible binding tag, $n=\ell_{pub}+\lvert\mathsf{Encode}(\theta)\rvert+\lvert\nu\rvert$ and $m=\ell_{pub}+\lambda$. Its evaluation is given in Part~\ref{alg:cnrme_enc}; the broadcast emitted during key setup is $(EV,\,H{=}\tau,\,\nu)$.
\vspace{-1mm}
\begin{equation}
EV=\mathsf{f}^{\mathrm{enc}}_{\mathcal P}\big(PUB\mathbin{\|}\mathsf{Encode}(\theta)\mathbin{\|}\nu\big)=(C,\tau),
\label{eq:cnrme_fn_enc}
\end{equation}

A peer holding $(\mathcal P,\theta,\nu)$, authorised inversion, recovers $PUB$ by re-deriving the master seed and the schedule of Part~\ref{alg:cnrme_enc}, verifying $\tau\stackrel{?}{=}\lfloor\mathsf{H}(k_{\mathrm{tag}}\mathbin{\|}C\mathbin{\|}\mathit{ctx}\mathbin{\|}\nu)\rfloor_{\lambda}$ (aborting on mismatch), removing the keystream as $S=C\oplus\mathsf{AES\text{-}CTR}_{k_{\mathrm{enc}}}(\nu)$, inverting the $\rho$ keyed Feistel rounds, and applying $\hat{\mathbf A}^{-1}$ in the NTT domain before decoding $\mathbf v$ to $PUB$. Every layer except the truncated tag is a bijection under the shared key, so recovery is exact; the tag $\tau$ is one-way and is reused as the on-chain hash $H$. This resolves the apparent paradox in the name: $PUB$ is recoverable by peers that know the function dynamics, whereas the obscurity layer that protects it is non-reversible to everyone else.

The construction reduces the claim to named assumptions. (i)~Determinism and consensus correctness. $\mathsf{f}^{\mathrm{enc}}_{\mathcal P}$ is a fixed composition of deterministic maps, so honest peers sharing $(\mathcal P,\theta,\nu)$ compute identical $EV$; this enables PoE agreement and the block hash $H=\tau$. (ii)~Confidentiality. For fixed $(\mathcal P,\theta)$ and distinct $\nu$, $C$ is $PUB$ masked by the keystream $\mathsf{AES\text{-}CTR}_{k_{\mathrm{enc}}}(\nu)$ whose key is sponge-derived; if AES-256 is a quantum-secure PRF and \textsc{shake}-256 a quantum-secure generator, the keystream is indistinguishable from uniform to any quantum polynomial-time (QPT) adversary lacking $\mathcal P$, so $\mathsf{f}^{\mathrm{enc}}_{\mathcal P}$ is IND-CPA and generic key recovery costs $\Omega(2^{\lambda/2})=\Omega(2^{128})$ Grover queries. (iii)~Non-reversibility. The tag $\tau=\lfloor\mathsf{H}(\cdot)\rfloor_{\lambda}$ is a truncated Keccak digest, preimage-resistant at $\Omega(2^{\lambda})$ classically and $\Omega(2^{\lambda/2})$ quantumly. (iv)~Unlinkability. Distinct nonces $\nu$ yield independent pseudo-random $(C,\tau)$, defeating traffic-pattern exploitation and complementing the $DET$ masking. (v)~Layered lattice hardness. The diffusion $\mathsf{L}$ binds the masked state to a Module-LWE-structured element of $R_q^{k}$, so recovering $\mathbf v$ without $\hat{\mathbf A}$ is governed by Module-LWE hardness \cite{NIST2024FIPS203,Regev2009}---the independent obstacle that realises the ``layered'' qualifier of this mode. (vi)~Combinatorial uncertainty. By Part \eqref{eq:cnrme_cfg}, an adversary ignorant of $\theta$ additionally faces $\log_2 N_{\mathrm{cfg}}$ bits of uncertainty. None of (ii)--(v) exposes the abelian hidden-subgroup or periodic structure that Shor require, residual quantum advantage is the square-root Grover speed-up, neutralised at $\lambda\ge 256$.

\vspace{2mm}
\floatname{algorithm}{PARS Protocol (CNMRE - Traditional)}
\begin{algorithm}[!htb]
\small % Set the font size to small
\caption{PARS Protocol (CNMRE - Traditional) - Deterministic Post-Quantum Obscurity Encapsulation}
\label{alg:cnrme_enc}

\begin{tabular}{p{1\linewidth}p{0.6\linewidth}}
\textbf{CNRME Encapsulation} &  \\
\end{tabular}

\begin{tabular}{p{0.09\linewidth}p{0.85\linewidth}}
\textbf{Inputs:} & public key $PUB$; secret protector bundle $\mathcal P=(PK_1,\dots,PK_\kappa)$; parameter vector $\theta$; epoch nonce $\nu$; restricted constants $(\rho,D_0,D_1,D_2,q,d,k)$. \\
\end{tabular}

\begin{tabular}{p{0.12\linewidth}p{0.82\linewidth}}
\textbf{Outputs:} & encapsulation $EV=(C,\tau)$. \\
\end{tabular}

\begin{tabular}{p{0.06\linewidth}p{0.89\linewidth}}
1: & $\mathit{idx}\gets\mathsf{Combine}(\theta)$,\quad $\mathit{ctx}\gets\mathsf{Encode}(\theta)\mathbin{\|}\mathrm{I2OSP}(\mathit{idx})$ \\
2: & $\sigma\gets\Xi\!\big(D_0\mathbin{\|}\mathit{ctx}\mathbin{\|}PK_1\mathbin{\|}\cdots\mathbin{\|}PK_\kappa\mathbin{\|}\nu\big)$ \\
3: & $\big(k_{\mathrm{enc}}\mathbin{\|}k_{\mathrm{lat}}\mathbin{\|}k_{\mathrm{tag}}\mathbin{\|}k_1\mathbin{\|}\cdots\mathbin{\|}k_\rho\big)\gets\Xi\!\big(D_1\mathbin{\|}\sigma\big)$ \\
4: & $\hat{\mathbf A}\gets\mathsf{SampleInvertible}(k_{\mathrm{lat}})\in R_q^{k\times k}$ \\
5: & $\mathbf v\gets\mathsf{Decode}_{R_q}(PUB)$, \quad $\bar w\gets\mathsf{Serialize}\!\big(\mathrm{NTT}^{-1}(\hat{\mathbf A}\odot\mathrm{NTT}(\mathbf v))\big)$ \\
6: & $S\gets\bar w$ \\
7: & \textbf{for} $j=1$ \textbf{to} $\rho$ \textbf{do} \\
8: & \hspace{10pt} $(S_L,S_R)\gets\mathsf{Split}(S)$ \\
9: & \hspace{10pt} $S\gets\big(\,S_R,\;\; S_L\oplus\Xi(D_2\mathbin{\|}k_j\mathbin{\|}S_R)\,\big)$ \\
10: & \textbf{end for} \\
11: & $C\gets S\oplus\mathsf{AES\text{-}CTR}_{k_{\mathrm{enc}}}(\nu)$ \\
12: & $\tau\gets\big\lfloor\mathsf{H}\!\big(k_{\mathrm{tag}}\mathbin{\|}C\mathbin{\|}\mathit{ctx}\mathbin{\|}\nu\big)\big\rfloor_{\lambda}$ \\
13: & \textbf{return} $EV\gets(C,\tau)$ \\
\end{tabular}
\end{algorithm}
\vspace{2mm}

The Mode B is traditional symmetric, seed-decoupled derivation phase and no key material is placed here on the wire, the protected quantity is reconstructed locally from the exchanged seed given as (\ref{eq:cnrme_fn_derive}). The mode is the deterministic, $\mathcal P$-keyed map where $K_{\mathrm{chain}}$ is the locally reconstructed symmetric key chain (encryption, decryption, signing and verification keys) that is never transmitted, $EV\in\{0,1\}^{\lambda}$ is the one-way public commitment that is broadcast and written on-chain, $n=\lvert\sigma\rvert+\lvert\mathsf{Encode}(\theta)\rvert+\lvert\nu\rvert$ and $m=\lvert K_{\mathrm{chain}}\rvert+\lambda$. Only the $\lambda$-bit component $EV$ is released; the remainder stays inside the node. The evaluation is given in Part~\ref{alg:cnrme_derive}.
\vspace{-1mm}
\begin{equation}
\big(K_{\mathrm{chain}},\,EV\big)=\mathsf{f}^{\mathrm{der}}_{\mathcal P}\big(\sigma\mathbin{\|}\mathsf{Encode}(\theta)\mathbin{\|}\nu\big),
\label{eq:cnrme_fn_derive}
\end{equation}

Because $\mathcal P$, $\sigma$, $\theta$ and $\nu$ are shared by the pair during local agreement and verification, both endpoints run $\mathsf{CNRME.Derive}$ independently and obtain the same $K_{\mathrm{chain}}$ and the same $EV$ without any key material crossing the link. The Sender broadcasts only $EV$ with its hash and timestamp; the swarm then confirms through PoE that the committed values agree across peers, certifying consistent reconstruction without revealing $K_{\mathrm{chain}}$. On agreement the key chain becomes operational and $EV$ is written to the block as the on-chain anchor $H$. The key chain is thus reconstructible by peers holding the protector keys and the function dynamics, yet the seed-to-key map is non-reversible for everyone else.

The claim reduces to named assumptions and to the secrecy of $\mathcal P$, which never traverses the network. (i)~Determinism and agreement. As in Mode~A, peers sharing $(\mathcal P,\sigma,\theta,\nu)$ reconstruct identical $K_{\mathrm{chain}}$ and $EV$. (ii)~HNDL resistance. The only key-related quantities on the wire are the seed $\sigma$ and the commitment $EV$. Since $K_{\mathrm{chain}}=\Xi(D_2\mathbin{\|}k_{\mathrm{kdf}}\mathbin{\|}t)$ with $k_{\mathrm{kdf}}$ and $t$ both derived under $\mathcal P$, $K_{\mathrm{chain}}$ is pseudo-random---hence computationally independent of $(\sigma,EV)$---for any QPT adversary lacking $\mathcal P$. No transmitted quantity is bound to $K_{\mathrm{chain}}$ by a public-key relation, so unlike a KEM (and unlike Mode~A) there is no ciphertext/public-key pair from which the session key is algebraically recoverable; an adversary that records all traffic and later applies a quantum computer gains no advantage. This is the decoupling the seed provides: $\sigma$ is an input to a $\mathcal P$-keyed map, not a share of its output, so it ``is not a part of the key''. (iii)~Key-recovery bound. Recovering $K_{\mathrm{chain}}$ requires recovering the symmetric root $\mathcal P$; the best generic quantum attack is Grover search at $\Omega(2^{\lambda/2})=\Omega(2^{128})$. (iv)~Non-reversibility of the commitment. $EV=\lfloor\mathsf{H}(\cdot)\rfloor_{\lambda}$ is a truncated Keccak digest, preimage-resistant at $\Omega(2^{\lambda})$ classically and $\Omega(2^{\lambda/2})$ quantumly, disclosing neither $K_{\mathrm{chain}}$ nor $\mathcal P$. (v)~Unlinkability. Fresh seeds and nonces per epoch give independent pseudo-random $EV$. (vi)~Combinatorial uncertainty. By \eqref{eq:cnrme_cfg}, an adversary ignorant of $\theta$ faces a further $\log_2 N_{\mathrm{cfg}}$ bits. Shor's algorithm does not apply, and the optional lattice layer adds independent Module-LWE diffusion \cite{NIST2024FIPS203,Regev2009} for defense in depth without the need for (i)--(vi).

Both modes target the same $128$-bit post-quantum level and expose only Grover-style square-root speed-ups, but they trade off differently. Mode~A provides explicit, certifiable Module-LWE hardness and interoperates with public-key/KEM infrastructure, at the cost of transporting an encapsulated public key and the associated bandwidth and computation. Mode~A offers standardized KEM interoperability and an independently updatable public-key layer at the cost of larger transmitted objects. Mode~B reduces bandwidth and computation but depends more strongly on pre-shared roots and their lifecycle management. The same downstream ledger can represent either profile because both produce an authenticated public commitment $EV$ and epoch anchor $H$. Mode~B places no key material on the wire and is therefore HNDL-resistant by construction, and it removes the cost of large post-quantum key material, which makes it lightweight default for resource-constrained edge nodes; its trust root is the pre-shared $\mathcal P$. Because each mode emits a public one-way value ($EV$) and an on-chain anchor $H$, the downstream consensus and ledger are identical, and a deployment may even assign different modes to different peer pairs through the PQC~Scheme selector of $\theta$ without altering the rest of the protocol. The exact constants of both modes---the round count $\rho$, the domain-separation tags $D_0,D_1,D_2$, the sampling routine for $\hat{\mathbf A}$, the seed-contribution routine, the ring parameters $(q,d,k)$, and the canonical encoders $\mathsf{Encode}/\mathsf{Combine}$---are deployment-sensitive and are fixed in a controlled technical report \cite{b15-2024kcso} in accordance with defense-sector obligations. The specifications above are nonetheless complete up to these constants and suffice to reproduce each function and verify its security properties.

\vspace{2mm}
\floatname{algorithm}{PARS Protocol (CNRME - PQC)}
\begin{algorithm}[!htb]
\small % Set the font size to small
\caption{PARS Protocol (CNMRE - PQC) - Deterministic Post-Quantum Obscurity Encapsulation}
\label{alg:cnrme_derive}

\begin{tabular}{p{1\linewidth}p{0.6\linewidth}}
\textbf{CNRME Key Derivation} &  \\
\end{tabular}

\begin{tabular}{p{0.09\linewidth}p{0.85\linewidth}}
\textbf{Inputs:} & pre-shared secret protector bundle $\mathcal P=(PK_1,\dots,PK_\kappa)$; end-to-end seed $\sigma=\sigma_{SN}\mathbin{\|}\sigma_{RN}$; parameter vector $\theta$; epoch nonce $\nu$; restricted constants $(\rho,D_0,D_1,D_2)$ and optional ring parameters $(q,d,k)$. \\
\end{tabular}

\begin{tabular}{p{0.12\linewidth}p{0.82\linewidth}}
\textbf{Outputs:} & local key chain $K_{\mathrm{chain}}$ (never transmitted); public commitment $EV$. \\
\end{tabular}

\begin{tabular}{p{0.06\linewidth}p{0.89\linewidth}}
1: & $\mathit{idx}\gets\mathsf{Combine}(\theta)$,\quad $\mathit{ctx}\gets\mathsf{Encode}(\theta)\mathbin{\|}\mathrm{I2OSP}(\mathit{idx})$ \\
2: & $\kappa_0\gets\Xi\!\big(D_0\mathbin{\|}PK_1\mathbin{\|}\cdots\mathbin{\|}PK_\kappa\mathbin{\|}\mathit{ctx}\mathbin{\|}\nu\big)$ \\
3: & $\big(k_{\mathrm{mix}}\mathbin{\|}k_{\mathrm{kdf}}\mathbin{\|}k_{\mathrm{com}}\mathbin{\|}k_{\mathrm{opt}}\big)\gets\Xi\!\big(D_1\mathbin{\|}\kappa_0\big)$ \\
4: & $t\gets\sigma\oplus\mathsf{AES\text{-}CTR}_{k_{\mathrm{mix}}}\!\big(\mathit{ctx}\mathbin{\|}\nu\big)$ \\
5: & \textbf{if} $\theta$ enables the lattice booster \textbf{then} \\
6: & \hspace{10pt} $\hat{\mathbf A}\gets\mathsf{Sample}(k_{\mathrm{opt}})\in R_q^{k\times k}$;\quad $t\gets\mathsf{Serialize}\!\big(\mathrm{NTT}^{-1}(\hat{\mathbf A}\odot\mathrm{NTT}(\mathsf{Decode}_{R_q}(t)))\big)$ \\
7: & \textbf{end if} \\
8: & \textbf{for} $j=1$ \textbf{to} $\rho$ \textbf{do} \\
9: & \hspace{10pt} $(t_L,t_R)\gets\mathsf{Split}(t)$;\quad $t\gets\big(t_R,\; t_L\oplus\Xi(D_1\mathbin{\|}k_{\mathrm{kdf}}\mathbin{\|}\mathrm{I2OSP}(j)\mathbin{\|}t_R)\big)$ \\
10: & \textbf{end for} \\
11: & $K_{\mathrm{chain}}\gets\Xi\!\big(D_2\mathbin{\|}k_{\mathrm{kdf}}\mathbin{\|}t\big)$ \\
12: & $EV\gets\big\lfloor\mathsf{H}\!\big(k_{\mathrm{com}}\mathbin{\|}K_{\mathrm{chain}}\mathbin{\|}\mathit{ctx}\mathbin{\|}\nu\big)\big\rfloor_{\lambda}$ \\
13: & \textbf{return} $(K_{\mathrm{chain}},\,EV)$ \\
\end{tabular}
\end{algorithm}
\vspace{2mm}

When the key-setup broadcast finishes, Proof-of-Edge (PoE) Consensus continue, PARS lets the network start a pattern-based consensus hierarchy. After the Sender Node's encapsulation or key commitment ($EV_{SN}$) is received, all nodes transmit a core-sized signal-level message, as part of a specific pattern, to the Key Renewal Peer ($KRP$ or $SN$) by preserving the pattern-step values in the local node. Until this pattern is achieved, other nodes must send a signal to SN and a return must be made for the packet that requires it to stop sending the signal. Later, SN initiates a rotational communication loop by aligning the received patterns. If all nodes successfully stick to the flow specified in the aligned pattern by achieving target Validation Count ($VC$), consensus is established which confirms that all participating peers hold a consistent updated key. The protocol is designed in such a way that actual message exchange is not affected during key consensus progress. The first step is the detection of the identity of PKs in the edge nodes. Because the master node used different PKs for different edges. Then, the initiated $P_{dp}$ is used by the master to achieve consensus on keys that will be used in swarm communication securely. Each slave tries to find their pattern order by communicating with the master node. Consensus verification cannot be completed for the new node unless the node is in the correct order before it has been verified with the $KRP$ node. Master conveys the consensus signal to peers when the verification over pattern is completed.

Peers, for block building, will create the blocks locally by using Hash ($H$), the key commitment or encapsulated version ($EV$) that anchors the renewed key chain without revealing it, Generation Time ($GT$), Consensus Time ($CT$), and Owner ($O$) when the consensus is achieved. The created blocks will be stored in RAM until the sync is completed over peers. When the ongoing progress is completed, the blocks will link to the local chain by prioritizing $CTs$ and transfer into Flash. Thus, the PARS Protocol lets swarm to ensure simultaneous key exchange for multiple progress. Genesis block will be standalone for possible forks.

Finally, for key reconciliation, key updates will be fulfilled by checking the integrity of local chains. After the arrival of the approval and match signals over the peers to the master, a new key will be operational. Thus, a secure rotational key exchange mechanism will be held and all peers will renew security levels to be ready for asymmetric quantum cyber threats. Light Nodes ($LN$) are preferred over Full Nodes ($FN$) to achieve a reasonable and secure paradigm in the space and defense systems. The PARS mostly focus on storing executive data in a particular order to achieve distributed architecture. The sensitive data will be stored encrypted version correlated with rotational reconciliations.

\subsection{PARS-Enabled ESL} \label{crosRefESL}

In traditional IoT security infrastructures, the confidentiality of information is ensured through accepted algorithms and their static keys. Additionally, electronic security strategies are also implemented at the signal level. These methods can be used independently or in combination. Measures taken at the signal level alone are not always sufficient. Regarding the methods implemented using static algorithms and keys, when QCMPs are ready, the security vulnerabilities of the systems will be at a very high level. For this reason, it is vital to provide concepts such as authorization, confidentiality, integrity of critical information, and key management in the communication infrastructure of the systems through a distributed architecture.

ESL is a blockchain infrastructure built to meet available gaps mentioned. The main focus is to secure the information exchange between nodes at high speed and with minimum memory requirements. ESL is a space and defense ledger technology that allows all controls regarding the operational flow in herd systems to be executed in an unchangeable and decentralized manner. Using the ZKP infrastructure, it ensures agreement between the validator node and the prover node without revealing confidential information. The infrastructure utilizes various proof domains, including Spatial Space, Sensor Readings, Event Logs, Physical Components, Surveillance Outputs, and Tactical Decision-Making, as key parameters for ZKP implementation. For example, the process can be advanced by selecting the ZKP technique Surveillance Outputs category for nodes that are in the same region space and want to perform ZKP because the line of sight of the enemy element can already be clearly distinguished. Thus, space and defense systems can execute a series of functions such as joining nodes to the existing network, conducting identity checks and new node authentication, managing nodes leaving the network, handling authorization and its escalation, evaluating node security scores, and ensuring communication security.

\begin{figure}[!htb]
\includegraphics[width=1\linewidth]{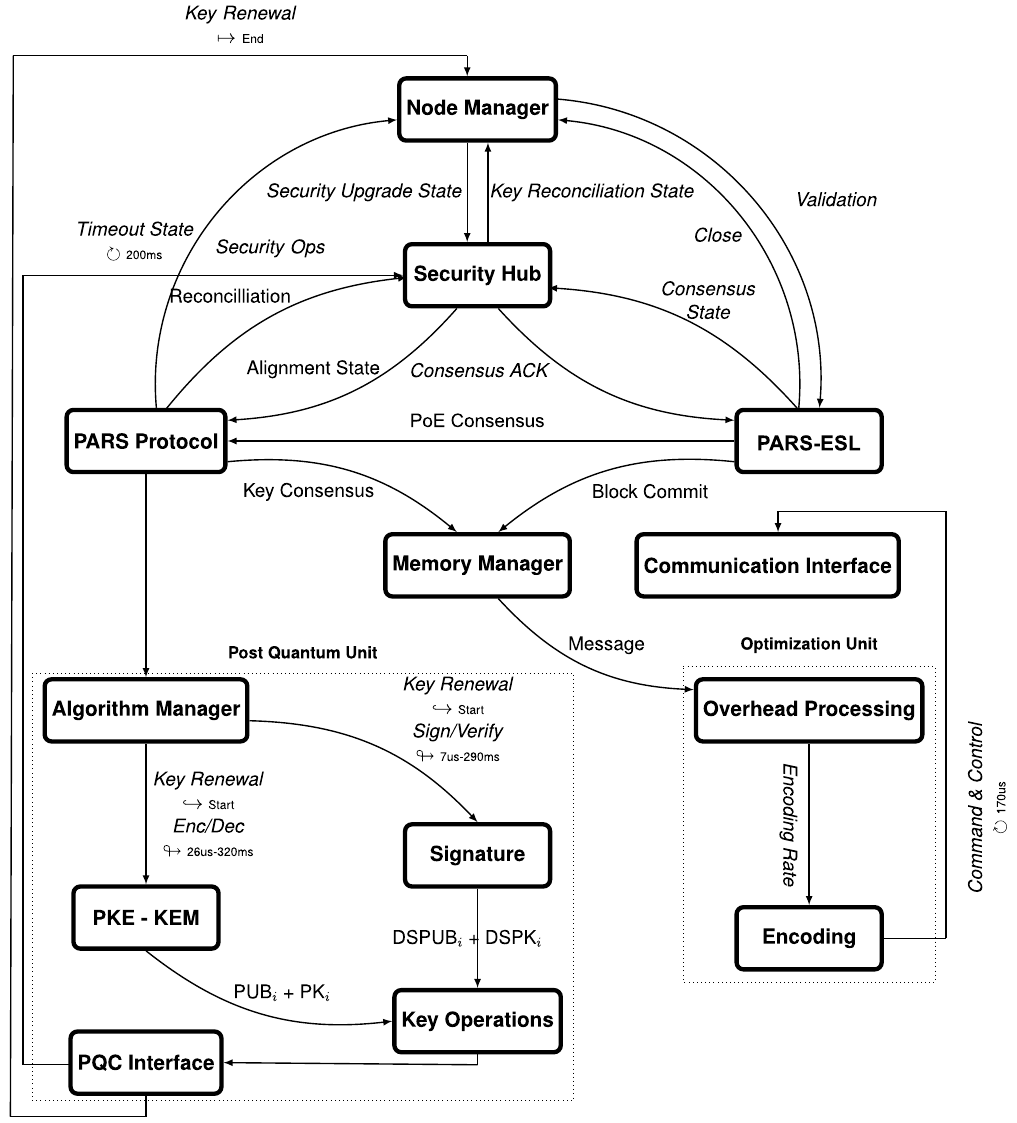}
\caption{Edge AI Nodes Quantum Security Flow}
\label{label_FSM_ESL}
\end{figure}

\begin{figure}[!htb]
\includegraphics[width=1\linewidth]{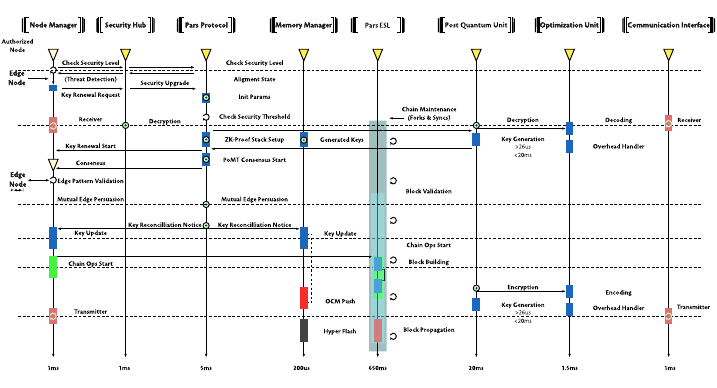}
\caption{Sequence Diagram Proposed Security Stack}
\label{fig_Sequence_Diagram}
\end{figure}

\begin{figure}[!htb]
\includegraphics[width=1\linewidth]{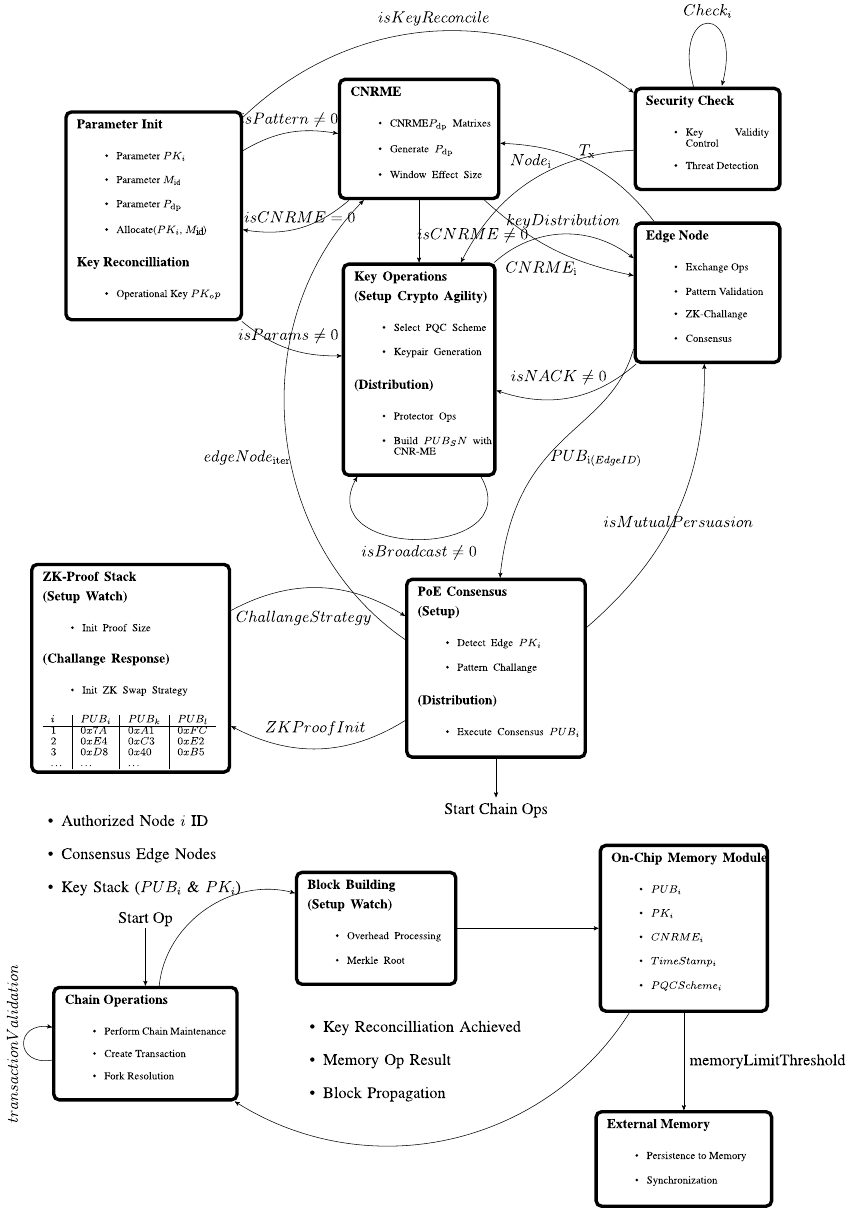}
\caption{Private Authentication for Rotational Secrecy}
\label{label_FSM_PARS}
\end{figure}

ESL stands out from existing solutions in the literature due to its advanced security mechanisms, capacity to defend against heterogeneous attacks, effective memory usage, scalability, and exceptional performance tailored for mission-critical systems. It offers a robust secrecy mechanism using the PARS Protocol. Using its dynamic key management mechanism, ESL efficiently mitigates significant memory costs by storing keys and authentication data on-chain, instead of burdening the blockchain with all data storage responsibilities. Thus, it ensures data can be stored in limited hardware environments rather than central stations. It maintains the traceability, decentralization, immutability, and trustless of the system by maintaining the connection between data and keys and protecting against active attacks. The mission critical space and defense systems empowered by this technology attain the strategic advantage of conducting independent data and key management, facilitating seamless strategic analyses within integration centers. In addition, it provides a real-time mutual proof mechanism without revealing any information about the systems and offers enhanced confidentiality by applying ZKP technology to IoMT. Similar to blockchain solutions utilized in the financial sector, which safeguard monetary value and facilitate secure transactions, offering decentralization, traceability, and immutability, ESL provides a service focus on securing critical processes within space and defense systems. This service encompasses managing, distributing, and ensuring the security of data, while also offering robust key management to support decision-making workflows. The experimental results are obtained  through flows in Fig. \ref{label_FSM_PARS} and Fig. \ref{label_FSM_ESL} as well as conceptual (see Fig. \ref{fig_Sequence_Diagram}) operational flow.

The PARS-enabled ESL thus provides a reference implementation for studying how authenticated key evolution, distributed security-state agreement, crypto-agility, and constrained communication interact. Its security and performance claims remain bounded by the disclosed protocol, cryptographic assumptions, provisioning model, validator threshold, network conditions, and experimental platforms. The following evaluation sections examine these costs and limitations empirically.
\lstdefinestyle{qhtpseudo}{
  language=Python,
  basicstyle=\ttfamily\fontsize{6.25}{6.8}\selectfont,
  keywordstyle=\bfseries,
  commentstyle=\itshape,
  numbers=left,
  numberstyle=\tiny,
  numbersep=5pt,
  xleftmargin=1.5em,
  frame=single,
  breaklines=true,
  columns=fullflexible,
  keepspaces=true,
  showstringspaces=false,
  aboveskip=4pt,
  belowskip=4pt
}

\subsection{Quantum Hyperlattice Tensor Key Agreement}
\label{subsec_qht_key_agreement}

Quantum Hyperlattice Tensor Key Agreement (QHT-KA) is a two-phase construction for authenticated edge nodes $G=\{P_1,\ldots,P_n\}$, $n\geq2$. During \emph{preparation}, direct multipartite quantum conference key agreement (QCKA) gives the nodes the same private seed, from which they reconstruct a virtual module tensor and commit to a bank of path representations. During \emph{reveal}, a fresh, independently executed QCKA supplies entropy that did not exist at the endpoints during preparation; it selects one committed path
and completes the application key. Neither seed nor final key is carried in a classical message.

The construction deliberately separates security from representation. QCKA supplies correlated secret entropy under its device and adversary model \cite{MurtaEtAl2020QCKA,PortmannRenner2022Security}. A keyed virtual tensor supplies a multidimensional namespace, nested paths, many-to-one display symbols, synchronized touch updates, and a commit-before-selection workflow. It adds neither entropy nor a lattice hardness assumption. Indeed, direct $\mathsf{KDF}(\mathsf{ENC}(S,R),\text{context})$ would provide the same key-secrecy level; the tensor is justified only by these scheduling and representation services.

QHT-KA is not communication-free. It distributes quantum systems and exchanges authenticated public data for parameter estimation, reconciliation, privacy amplification, state certification, and key confirmation. It also requires an initial authentication basis and cannot authenticate previously unknown nodes by itself. Four-party QCKA has been demonstrated over fibre \cite{ProiettiEtAl2021Experimental}, while a network experiment shows that direct GHZ-based QCKA can consume fewer quantum-network resources than building
a conference key from pairwise links \cite{PickstonEtAl2023Network}. To retain the literal claim that no common secret is classically transported, the core profile uses direct multipartite key distillation rather than a dealer choosing a group key and distributing it under pairwise keys.

\subsubsection{Two independently distilled quantum secrets}
\label{subsubsec_qht_quantum_seed}

For each QCKA execution, an untrusted source distributes many instances of an $n$-party GHZ or protocol-compatible graph state. In the qubit case,
\begin{equation}
 \lvert\mathrm{GHZ}_n\rangle
 =\frac{\lvert0\rangle^{\otimes n}+\lvert1\rangle^{\otimes n}}{\sqrt2}
 \in\mathcal H_G,\qquad
 \mathcal H_G=\bigotimes_{i=1}^{n}\mathbb C^2 .
 \label{eq_qht_ghz}
\end{equation}
The chosen QCKA protocol assigns systems to test and key rounds, authenticates all discussion, and aborts unless its phase-error, bit-error, loss, source, and finite-size conditions hold. Measurement-device-independent QCKA can reduce detector trust and has been studied for an arbitrary finite number of parties; state-generation rate, loss, finite-key cost, and the worst participant link still bound practical group size \cite{LiEtAl2023MDIQCKA}.

For execution $x\in\{\mathsf{prep},\mathsf{rev}\}$, let $X_i^{x}$ be the verified reconciled string at $P_i$, $E$ the adversary's quantum side information, and $\Gamma_x$ a versioned, type-tagged, length-prefixed session descriptor. It binds ordered membership and credential digests, a fresh nonce, purpose, protocol and parameter identifiers, test results, and the authenticated transcript digest. Each accepted node computes
\begin{equation}
 \begin{aligned}
 Q_i^x&=\mathsf{Ext}(X_i^x;s_x),\\
 Z_i^x&=\mathsf{KDF}(Q_i^x,
       \mathsf{ENC}(\texttt{QHT},x,\Gamma_x)),\\
 \ell_x&\leq H_{\min}^{\epsilon_x}(X_i^x\mid E,\Gamma_x)
 -\mathsf{leak}_{\rm EC}^x\\
 &\quad-\lambda_{\rm ver}^x
 -2\log_2\!\frac{1}{2\epsilon_{\rm PA}^x}.
 \end{aligned}
 \label{eq_qht_extraction}
\end{equation}
The public extractor seed $s_x$ is included in $\Gamma_x$. It must be uniform and independent of $X_i^x$, for example through an authenticated, precommitted public-coin procedure with at least one honest contribution; an exact QCKA proof may impose a different ceremony. Correctness gives a common $Z^x$ except with the stated error, and privacy amplification makes it close to uniform conditioned on the public transcript and $E$ \cite{PortmannRenner2022Security}. Set $S=Z^{\mathsf{prep}}$ and, only after an
authorized reveal, $R=Z^{\mathsf{rev}}$. The two executions use independent nonces, quantum rounds, extractor seeds, and transcript domains; otherwise a joint entropy proof is required. ``Complete randomness'' is replaced by these quantified finite-key guarantees. Device-independent expansion can strengthen certification when its experimental cost is acceptable \cite{LiEtAl2021DIRandomness}.

\subsubsection{Virtual module tensor and fixed-budget sampling}
\label{subsubsec_qht_tensor_creation}

Let $d$ be a power of two,
$\mathcal R=\mathbb Z[X]/(X^d+1)$,
$\mathcal R_q=\mathcal R/q\mathcal R$ for odd $q$, and let
$\Lambda=\mathcal R^k$ be a module lattice under the canonical embedding. Its
finite quotient is
$\Lambda_q=\Lambda/q\Lambda\cong\mathcal R_q^k$. For order $\nu$,
shape $(s_1,\ldots,s_\nu)$, and
$\mathcal I=\mathbb Z_{s_1}\times\cdots\times\mathbb Z_{s_\nu}$, define
\begin{equation}
 \mathcal T_q=\Lambda_q\otimes_{\mathcal R_q}\mathcal R_q^{s_1}
 \otimes_{\mathcal R_q}\cdots\otimes_{\mathcal R_q}
 \mathcal R_q^{s_\nu}\cong(\Lambda_q)^{\mathcal I}.
 \label{eq_qht_tensor_module}
\end{equation}
This is a tensor-indexed finite module derived from an underlying module lattice; it is not an LWE instance. The term \emph{Hyper Lattice Space} denotes the author-defined product
$\mathfrak X_G=\mathcal H_G\times\mathcal T_q\times\mathcal D_\Gamma$ and must not be presented as a recognized hard problem \cite{RaviEtAl2022LatticeSurvey}.

No node allocates the $N=\prod_{u=1}^{\nu}s_u$ cells. It evaluates requested
cells online:
\begin{equation}
 \mathsf{Cell}_{S,\Gamma}(\boldsymbol a)=
 \mathsf{Parse}_{\Lambda_q}^{\rm ct}\!\left(
 \mathsf{XOF}(S,\mathsf{ENC}(\texttt{CELL},\Gamma,\boldsymbol a))\right).
 \label{eq_qht_virtual_cell}
\end{equation}
$\mathsf{Parse}^{\rm ct}$ and every bounded sampler consume a fixed public budget of XOF blocks, select the first in-range value by constant-time masking, and abort if none exists. Conditional on acceptance, rejection sampling is unbiased; the fixed budget converts variable timing into a parameterized, negligible abort probability. Direct modulo reduction is forbidden when it is biased. This design gives $O(1)$ tensor-cell storage; materializing the tensor would cost $Nkd\lceil\log_2q\rceil$ bits before software overhead.

\subsubsection{Certified touches and explicit tensor views}
\label{subsubsec_qht_touch}

A touch is a public state-transition instruction, not a rekey. At sequence $h$, nodes accept $(h,c_h)$ only with a certificate over the previous state:
\begin{equation}
 \begin{aligned}
 \mathsf{Cert}_h&=\mathsf{CertSign}(
   \mathsf{ENC}(\texttt{TOUCH},\Gamma_{\mathsf{prep}},
   \tau_{h-1},h,c_h)),\\
 \tau_h&=\mathsf H(\mathsf{ENC}(\texttt{TOUCH-STATE},
   \tau_{h-1},h,c_h,\mathsf{Cert}_h)).
 \end{aligned}
 \label{eq_qht_touch_chain}
\end{equation}
The profile either names one trusted stateful issuer or uses a threshold certificate. With at most $f$ Byzantine signers, two conflicting certificates are excluded when quorum intersection contains an honest signer; for the common $n\geq3f+1$ profile, $2f+1$ signers suffice if every honest signer signs at most one command for a given $(\Gamma_{\mathsf{prep}},\tau_{h-1},h)$. Nodes reject gaps, duplicates, stale predecessors, invalid certificates, and equivocation evidence.

The view transform is concrete rather than an unspecified permutation. Let
$\mathsf{rank}:\mathcal I\rightarrow\mathbb Z_N$ be mixed-radix ranking.
From $\mathsf{XOF}(S,\mathsf{ENC}(\texttt{VIEW},
\Gamma_{\mathsf{prep}},\tau_h,c_h))$, sample $u_h\in\mathbb Z_N^*$,
$b_h\in\mathbb Z_N$, a component permutation $\varrho_h$, signs
$e_l\in\{-1,1\}$, and negacyclic rotations $a_l\in\mathbb Z_d$. Define
\begin{equation}
 \begin{aligned}
 \pi_h(\boldsymbol a)&=\mathsf{unrank}(
   u_h\mathsf{rank}(\boldsymbol a)+b_h\bmod N),\\
 \sigma_h(p_1,\ldots,p_k)_l&=e_lX^{a_l}p_{\varrho_h(l)},\\
 \boldsymbol c_h(\boldsymbol a)&=
   \mathsf{Cell}_{S,\Gamma_{\mathsf{prep}}}(\pi_h(\boldsymbol a)),\\
 \mathcal L^{(h)}[\boldsymbol a]&=\sigma_h(\boldsymbol c_h(\boldsymbol a))
   +\Delta_h(\boldsymbol a)\pmod q .
 \end{aligned}
 \label{eq_qht_view}
\end{equation}
Here $\gcd(u_h,N)=1$ makes $\pi_h$ bijective; signed monomial multiplication and component permutation make $\sigma_h$ invertible; and bounded $\Delta_h$ is evaluated lazily. A small instruction therefore changes every virtual coordinate consistently. It adds no entropy, forward secrecy, or compromise recovery. Any accepted touch after bank commitment invalidates that commitment and requires rebuilding it.

\subsubsection{Streaming nested paths and bank commitment}
\label{subsubsec_qht_paths}

For candidate $j\in\{0,\ldots,M-1\}$, fixed-budget samplers derive its start and moves. For compactness, let
$\mathsf F_{j,\ell}(\boldsymbol x)=\mathsf{XOF}(S,
\mathsf{ENC}(\ell,\Gamma_{\mathsf{prep}},\tau_H,j,\boldsymbol x))$, with a
type-tagged tuple $\boldsymbol x$. A rolling digest avoids storing the complete
path:
\begin{equation}
 \begin{aligned}
 \boldsymbol v_{j,0}&=\mathsf{Sample}^{\rm ct}_{\mathcal I}(
   \mathsf F_{j,\texttt{START}}(\varnothing)),\\
 \boldsymbol z_{j,0}&=\boldsymbol0,\qquad
   \eta_{j,0}=\mathsf H(\mathsf{ENC}(\texttt{P0},j,\boldsymbol v_{j,0})),\\
 d_{j,t}&=\mathsf{Sample}^{\rm ct}_{\mathcal N}(
   \mathsf F_{j,\texttt{MOVE}}(t,\mathcal L^{(H)}[\boldsymbol v_{j,t}],
   \boldsymbol z_{j,t})),\\
 \boldsymbol v_{j,t+1}&=\mathsf{Neighbor}(\boldsymbol v_{j,t},d_{j,t}),\\
 \boldsymbol z_{j,t+1}&=\mathsf{Parse}^{\rm ct}_{\Lambda_q}(
   \mathsf F_{j,\texttt{ACC}}(t,\boldsymbol z_{j,t},
   \mathcal L^{(H)}[\boldsymbol v_{j,t+1}])),\\
 \eta_{j,t+1}&=\mathsf H(\mathsf{ENC}(\texttt{PS},
   \eta_{j,t},\boldsymbol v_{j,t+1},\boldsymbol z_{j,t+1})).
 \end{aligned}
 \label{eq_qht_nested_path}
\end{equation}
After $r$ steps, $D_j=\eta_{j,r}$. A deterministic representation function returns $(w_j,\chi_j)=\mathsf{Rep}^{\rm ct}(D_j,j,\rho)$, where the public
policy $\rho$ either sets $w_j=0$ and maps the digest to an alphabet, or uses a fixed-budget counter search for a requested display symbol. Failure to find a
counter is a synchronized abort, never an unbounded search. Many paths may display the same $\chi_j$, while their digests remain distinct except with the hash-collision probability. This is a representation property, not extra entropy.

Preparation commits to the entire bank before the selected index exists:
\begin{equation}
 \begin{aligned}
 L_j&=\mathsf H(\mathsf{ENC}(\texttt{QLOCK-LEAF},
   \Gamma_{\mathsf{prep}},\tau_H,j,w_j,\chi_j,D_j)),\\
 B&=\mathsf{MerkleRoot}(L_0,\ldots,L_{M-1}),\\
 C_B&=\mathsf H(\mathsf{ENC}(\texttt{BANK-COMMIT},
   \Gamma_{\mathsf{prep}},\tau_H,M,\mathsf H(\rho),B)).
 \end{aligned}
 \label{eq_qht_bank_commitment}
\end{equation}
The Merkle profile fixes binary arity, left-to-right index order, separate leaf and internal-node tags, padding, and digest suite. Every node requires a complete, identity-signed set of identical $C_B$ votes before freezing the bank. A public audit service, including a blockchain, learns the bank commitment but not which Q-Lock will be selected. The selected opening later contains $(j^*,w_{j^*},\chi_{j^*},D_{j^*},\mathsf{AuthPath}_{j^*})$ and proves membership in the precommitted bank. It does not prove that the secret tensor computation itself was performed correctly; that stronger property needs a separately analyzed zero-knowledge proof or attestation mechanism. An implementation may
store the $O(M)$-digest Merkle tree or recompute it at reveal; candidate keys and full paths need not be retained.

\subsubsection{Authorized just-in-time reveal and confirmation}
\label{subsubsec_qht_selection}

The recommended delayed profile does not place long-lived quantum shares at the endpoints. After $C_B$ is frozen, nodes wait for a reveal certificate $A_{\rm rev}$ that binds $C_B$, ordered membership, a fresh session nonce, QCKA profile, one-shot sequence, and expiry. Only then do they execute an independent QCKA and derive $R$. Separate subkeys prevent the public index from reusing the exact input that masks the final key. The extract-and-expand KDF treats both secret components as input key material:
\begin{equation}
 \begin{aligned}
 R_{\rm sel}&=\mathsf{KDF}(R,\mathsf{ENC}(\texttt{SELECT-KEY},
   \Gamma_{\mathsf{rev}},C_B)),\\
 R_{\rm mix}&=\mathsf{KDF}(R,\mathsf{ENC}(\texttt{MIX-KEY},
   \Gamma_{\mathsf{rev}},C_B)),\\
 j^*&=\mathsf{Sample}^{\rm ct}_M(\mathsf{XOF}(R_{\rm sel},
   \mathsf{ENC}(\texttt{SELECT},C_B))),\\
 \widetilde K&=\mathsf{KDF}(S,
   \mathsf{ENC}(\texttt{QP},\Gamma_{\mathsf{prep}},
   \tau_H,j^*,D_{j^*},\boldsymbol z_{j^*,r})),\\
 c_{\rm rev}&=\mathsf{ENC}(\texttt{RV},\Gamma_{\mathsf{rev}},C_B,
   \mathsf H(A_{\rm rev})),\\
 K_0&=\mathsf{KDF}(\mathsf{ENC}(\widetilde K,R_{\rm mix}),c_{\rm rev}).
 \end{aligned}
 \label{eq_qht_reveal}
\end{equation}
Thus neither the selected index nor final key exists at an endpoint before the fresh reveal execution. The quantum superposition is in the distributed QCKA resource before measurement; the committed Q-Lock bank remains classical. This delay assumes that fewer than the authorization threshold can trigger or obtain the reveal quantum systems early. If any endpoint can independently invoke the source, the time-of-release property disappears.

Nodes derive a distinct confirmation key and reliably broadcast an identity-bound tag plus post-quantum signature. Write
$g_x=\mathsf H(\Gamma_x)$:
\begin{equation}
 \begin{aligned}
 K_{\rm conf}&=\mathsf{KDF}(K_0,
   \mathsf{ENC}(\texttt{CONF-KEY},\Gamma_{\mathsf{rev}})),\\
 t_i&=\mathsf{MAC}_{K_{\rm conf}}(
   \mathsf{ENC}(\texttt{CONF},i,C_B,j^*,g_{\mathsf{prep}},g_{\mathsf{rev}})),\\
 F&=\mathsf H(\mathsf{ENC}(\texttt{CONF-SET},
   \mathsf{sort}(i,t_i,\mathsf{sig}_i)_{i=1}^{n})),\\
 K_{\rm app}&=\mathsf{KDF}(K_0,
   \mathsf{ENC}(\texttt{APP-KEY},C_B,F)).
 \end{aligned}
 \label{eq_qht_confirmation}
\end{equation}
Every node requires one consistent signed value from the complete expected identity set or aborts. The signature prevents one group member, which also knows $K_{\rm conf}$, from impersonating another. After confirmation, QHT-KA publishes the Merkle opening, erases $S$, $R$, path state, and intermediate keys, and marks the bank consumed. This one-shot default prevents later seed exposure from revealing unused candidates.

This is transcript-consistent group agreement, not atomic release. Network latency means one honest node may finish before another, and a malicious member may learn the key and then abort. Exact wall-clock simultaneity or fairness requires an additional synchrony, threshold-release, or trusted-hardware assumption; QCKA alone does not provide it.

Long-lived stored GHZ registers remain an experimental alternative, but a malicious endpoint can measure its own share early without the basic QHT-KA logic stopping it. Enforced delayed ignorance against admitted endpoints needs a proved $(t,n)$ quantum-secret-sharing or distributed-MPC release scheme, or tamper-resistant quantum memory. Threshold QSS has been demonstrated, but its dealer, dishonest-player, channel, and finite-size assumptions must be adopted explicitly \cite{LiuEtAl2023QSS}.

\subsubsection{Two-phase Qiskit-style pseudocode}
\label{subsubsec_qht_algorithm}

Listing~\ref{lst_qht_ka} makes the preparation/reveal boundary explicit. All variable fields use canonical $\mathsf{ENC}$, and every bounded draw uses the fixed-budget sampler described above.

\begin{lstlisting}[style=qhtpseudo,
caption={Two-Phase QHT-KA at $P_i$},
label={lst_qht_ka}]
from qiskit import QuantumCircuit
def ghz_resource(n):
    qc = QuantumCircuit(n); qc.h(0)
    for p in range(1, n): qc.cx(0, p)
    return qc                 # distribute one qubit to each node

def qcka_secret(P_i, session, purpose, prm):
    pa_seed = authenticated_precommitted_coin(session, purpose)
    raw, stats, T = direct_qcka_acquire(
        P_i, session, purpose, ghz_resource, prm)
    verify_authenticated_transcript(T, session, purpose)
    if not finite_key_accept(stats, T, prm): raise Abort()
    Gamma = finalize_descriptor(session, purpose, prm, stats,
                                pa_seed, H(T))
    X = reconcile_and_verify(raw, T)
    Q = strong_extract(X, pa_seed, T)
    Z = KDF(Q, ENC("QHT", purpose, Gamma))
    erase(raw, X, Q)
    return Z, Gamma

def virtual_cell(S, Gamma, tau, a, prm):
    b = affine_view_index(S, Gamma, tau, a, prm) # bijective
    x = ct_xof_to_lattice(XOF(S, ENC("CELL", Gamma, b)), prm)
    return (apply_invertible_view(x, S, Gamma, tau, prm) + \
            bounded_delta(S, Gamma, tau, a, prm)) % prm.q

def path_id(S, Gamma, tau, j, prm):
    v = ct_sample_index(XOF(S, ENC("START", Gamma, tau, j)), prm)
    z, eta = zero_lattice(prm), H(ENC("PATH-INIT", j, v))
    for t in range(prm.path_len):
        x = virtual_cell(S, Gamma, tau, v, prm)
        move = ct_sample_step(XOF(S, ENC("MOVE", Gamma, tau,
                              j, t, x, z)), prm)
        v = neighbor(v, move, prm.shape)
        y = virtual_cell(S, Gamma, tau, v, prm)
        z = ct_xof_to_lattice(XOF(S, ENC("ACC", Gamma, tau,
                              j, t, z, y)), prm)
        eta = H(ENC("PATH-STEP", eta, v, z))
    return eta, z

def qht_prepare(P_i, session, prm, touch_certs):
    validate_profile(prm); verify_initial_credentials(session)
    S, Gamma = qcka_secret(P_i, session, "PREP", prm)
    tau, expected = H(ENC("INIT", Gamma)), 1
    for cert in touch_certs:
        verify_unique_quorum_cert(cert, Gamma, tau, expected, prm)
        tau = H(ENC("TOUCH-STATE", tau, expected,
                    cert.command, cert))
        expected += 1
    tree = MerkleBuilder(prm.M)
    for j in range(prm.M):
        D, _ = path_id(S, Gamma, tau, j, prm)
        w, symbol = ct_represent(D, j, session.display_policy, prm)
        tree.add(H(ENC("QLOCK-LEAF", Gamma, tau, j, w, symbol, D)))
    Cbank = H(ENC("BANK-COMMIT", Gamma, tau, prm.M,
                  H(session.display_policy), tree.root()))
    vote = PQ_sign(P_i.signing_key,
                   ENC("BANK-VOTE", P_i.id, Cbank))
    votes = authenticated_reliable_all_gather(P_i, Cbank, vote)
    verify_complete_identical_bank(votes, Gamma, Cbank)
    return FrozenState(S, Gamma, tau, tree, Cbank,
                       session.display_policy)

def qht_reveal(P_i, state, reveal_cert, fresh_session, prm):
    verify_reveal_quorum(reveal_cert, state.Cbank, fresh_session)
    bound = bind_reveal_session(
        fresh_session, state.Cbank, H(reveal_cert),
        state.Gamma.participants)
    R, GammaR = qcka_secret(P_i, bound, "REVEAL", prm)
    Rsel = KDF(R, ENC("SELECT-KEY", GammaR, state.Cbank))
    Rmix = KDF(R, ENC("MIX-KEY", GammaR, state.Cbank))
    j = ct_sample_M(XOF(Rsel, ENC("SELECT", state.Cbank)), prm.M)
    D, z = path_id(state.S, state.Gamma, state.tau, j, prm)
    w, symbol = ct_represent(D, j, state.display_policy, prm)
    opening = (j, w, symbol, D, state.tree.proof(j))
    verify_merkle_opening(opening, state.tree.root())
    Kpre = KDF(state.S, ENC("QP", state.Gamma,
               state.tau, j, D, z))
    K0 = KDF(ENC(Kpre, Rmix), ENC("RV", GammaR,
             state.Cbank, H(reveal_cert)))
    Kconf = KDF(K0, ENC("CONF-KEY", GammaR))
    tag = MAC(Kconf, ENC("CONF", P_i.id, state.Cbank, j,
              H(state.Gamma), H(GammaR)))
    sig = PQ_sign(P_i.signing_key, ENC(P_i.id, tag, state.Cbank))
    confirms = authenticated_reliable_all_gather(P_i, tag, sig)
    verify_complete_unique_set(confirms, Kconf, state, GammaR)
    F = H(ENC("CONF-SET", sorted(confirms)))
    Kapp = KDF(K0, ENC("APP-KEY", state.Cbank, F))
    publish(opening)
    erase(state.S, R, Rsel, Rmix, z, Kpre, K0, Kconf)
    state.consume()
    return Kapp, opening
\end{lstlisting}

The GHZ constructor is Qiskit-compatible, but \texttt{direct\_qcka\_acquire} is a distributed protocol, not a simulator call. Its basis choices, network routing, entropy estimation, reconciliation, privacy amplification, authentication, and finite-key proof must be concrete. The touch quorum, reliable broadcast, Merkle implementation, fixed-budget
samplers, signatures, side-channel controls, and secure erasure likewise need reviewed implementations. The listing is implementable structure, not a security proof. In particular, ordinary Python does not guarantee constant-time execution or reliable zeroization; those operations require a suitable native or hardware-backed implementation.

QHT-KA supports several release profiles with distinct security boundaries. In the **immediate-release profile**, authorized members derive both the selection information and the key from \(S\), so the security claim is limited to excluding outsiders rather than preventing access by already authorized members \cite{PortmannRenner2022Security}. In the **just-in-time (JIT) reveal profile**, a second direct QCKA execution after \(A_{\rm rev}\) generates \(R\), avoiding the need for long-term quantum memory while postponing key establishment until the reveal stage \cite{PickstonEtAl2023Network,LiEtAl2023MDIQCKA}. Alternatively, the **stored profile** retains quantum shares at the endpoints until release; this supports honest scheduling but does not prevent an authorized malicious endpoint from measuring its share prematurely. Finally, the **threshold profile** distributes the protected information such that fewer than \(t\) authorized shares are insufficient to release the protected value, subject to the security guarantees of the selected QSS/MPC construction \cite{LiuEtAl2023QSS}. These profiles therefore represent different trade-offs between release timing, quantum-storage requirements, and resistance to premature disclosure by authorized participants.

\begin{comment}
    
\begin{center}
\begin{minipage}{\columnwidth}
\refstepcounter{table}\label{tab_qht_modes}
\centering
\footnotesize
\textbf{TABLE \thetable}\\[-1pt]
\textsc{QHT-KA Release Profiles and Exact Boundaries}\\[3pt]
\begin{tabularx}{\columnwidth}{@{}p{1.45cm}X@{}}
\toprule
\textbf{Profile} & \textbf{Defensible claim} \\
\midrule
Immediate & Members derive selection and key from $S$; only outsiders are
excluded \cite{PortmannRenner2022Security}. \\
JIT reveal & A second direct QCKA after $A_{\rm rev}$ creates $R$ without
long-term quantum memory \cite{PickstonEtAl2023Network,LiEtAl2023MDIQCKA}. \\
Stored & Quantum shares wait at endpoints; honest scheduling is supported, but
early malicious measurement is not prevented. \\
Threshold & Fewer than $t$ authorized shares cannot release the protected
value under the selected QSS/MPC proof \cite{LiuEtAl2023QSS}. \\
\bottomrule
\end{tabularx}
\end{minipage}
\end{center}

\end{comment}

A low-entropy password may gate admission only through a separately analyzed post-quantum PAKE; a two-party Module-LWE PAKE cannot be silently promoted to an $n$-party proof \cite{RenGuWang2023PAKE}. A user-chosen word may instead be a display target: the preparation stage searches representation counters until secret paths map to the requested characters, and binds each counter into the leaf. The target is display metadata, may be exposed by an opening, and is not an authentication secret. It changes the visible representation, not the key entropy. Likewise, an action vector is only a credential unless a fuzzy extractor has measured conditional min-entropy, helper leakage, liveness, replay, and spoofing bounds. Experimental QKD--PQC integration supports combining quantum entropy with post-quantum authentication \cite{GarmsEtAl2024Hybrid}. Lattice-based proofs of quantumness can challenge an outsourced quantum device but do not generate the QHT-KA group secret \cite{AlnawakhthaEtAl2024LatticeQuantum}.

\subsubsection{Conditional security claim and residual limitations}
\label{subsubsec_qht_security}

For honest accepted nodes, correctness reduces to the two QCKA correctness errors, canonical encodings, identical certified touch state, deterministic path derivation, Merkle-root agreement, and complete confirmation. Fixed-budget sampling and representation-search exhaustion are synchronized aborts, not biased fallbacks; the parameter profile must bound their aggregate probability by an explicit $\epsilon_{\rm samp}$. Against an external quantum adversary, an indicative modular bound is
\begin{equation}
 \begin{aligned}
 \mathsf{Adv}_{\rm QHT}
 &\leq\epsilon_{\rm sec}^{\mathsf{prep}}
 +\epsilon_{\rm sec}^{\mathsf{rev}}
 +\mathsf{Adv}^{Q}_{\rm XOF/KDF}\\
 &\quad+\mathsf{Adv}^{Q}_{\rm auth}
 +\mathsf{Adv}^{Q}_{\rm hash},
 \end{aligned}
 \label{eq_qht_security_bound}
\end{equation}
with query and multi-user factors instantiated for the selected suite. This is not yet a composable theorem: a proof must define adaptive corruption, selective abort, insider leakage, reveal scheduling, reliable broadcast, commitment opening, erasure, and both QCKA interfaces. The tensor size, path count, and number of representations do not reduce the bound or increase the entropy available from $(S,R)$.

The just-in-time profile improves the delayed-release claim but cannot create an unconditional time lock. It relies on the reveal certificate and controlled availability of fresh quantum systems; an authorized or sufficiently large colluding set can reveal early. Every admitted member learns the final group key at release and may disclose it. QHT-KA also does not prevent denial of service, participant abort, traffic analysis, endpoint compromise before erasure, sampler faults, signature-key compromise, or implementation leakage. The bank is one-shot: membership change, post-commit touch, failed confirmation, or suspected seed exposure requires abort and fresh preparation. Consequently, QHT-KA should be described as a research composition of direct QCKA and a virtual tensor key schedule---not as an unbreakable lattice algorithm.

A concrete profile must freeze the exact QCKA theorem and error budget; authentication, XOF, extractor, KDF, MAC, signature, and hash suites; canonical encoding and Merkle rules; $(d,q,k)$, tensor shape, $M$, $r$, and all sampler budgets; quorum and expiry rules; and storage, erasure, and recovery behavior. None of the lattice dimensions, path count, or representation counters may be counted as additional secret entropy.

\section{Resource-Constrained Edge AI Implementation}
\label{Experimental Results}

This section evaluates the performance and resource implications of the PARS framework and its PARS-ESL reference implementation. Although the physical testbed is instantiated using an Internet of Military Things (IoMT) configuration, the evaluated mechanisms address the broader problem of coordinating post-quantum key epochs, membership state, and revocation across resource-constrained Edge~AI and Cyber-Physical System (CPS) nodes. The evaluation therefore concerns a security-control plane for distributed re-keying rather than the execution of blockchain consensus within high-rate inference, telemetry, or physical-control loops. Earlier post-quantum IoT and blockchain studies provide the broader architectural frameworks that remain confined to theoretical designs or simulations in terms of cryptography algorithms \cite{b8-2020Fernandez}. To the best of our knowledge, this work presents the first quantum-safe decentralized key-management architecture validated end-to-end on an operational, physical Edge~AI cyber-physical testbed. The present results should be interpreted according to the specific hardware, topology, algorithm, and workload conditions reported below as well as evidence of universal Edge~AI suitability.

The evaluation distinguishes four evidence classes. First, \emph{primitive-level measurements} quantify the execution time of Key-Encapsulation Mechanisms (KEMs) and digital-signature algorithms on an Intel Core i5-12450H platform. Second, \emph{protocol-level measurements} characterize consensus, reconciliation, message size, accumulated state, bandwidth, and unsuccessful rounds in the multi-node prototype. Third, \emph{derived metrics} calculate object-size and data-flow overhead from standardized cryptographic parameters. Fourth, \emph{platform projections} estimate deployment behavior on heterogeneous edge processors and accelerators using published cycle counts, device specifications, and implementation studies. These categories must not be interpreted as equivalent, projected platform values are design guidance and do not constitute physical measurements.

The evaluation covers six performance dimensions: (i)~execution time of cryptographic primitives, (ii)~PARS-ESL consensus latency, (iii)~volatile-memory and ledger-state growth, (iv)~message, security-record, and bandwidth overhead, (v)~incomplete or rejected consensus rounds, and (vi)~end-to-end reconciliation failures. The analysis explicitly distinguishes physical measurements from analytical projections. Results obtained from the implemented testbed are reported as empirical observations, whereas estimates for additional processors, adaptive SoCs, FPGAs, and space/aerospace/robotics-grade platforms are identified as projections derived from cited hardware characteristics and implementation studies. The cryptographic microbenchmarks use an Intel i5-12450H gateway-class platform to establish a comparative upper-tier baseline. The benchmark workflow follows the operation categories and reporting conventions commonly used in NIST-oriented and Open~Quantum~Safe evaluations \cite{stebila2016oqs}, although complete reproducibility additionally requires disclosure of the exact library version, compiler flags, CPU-frequency policy, random-number source, warm-up procedure, and trial-level data.

\begin{figure}[!htb]
\includegraphics[width=1\linewidth]{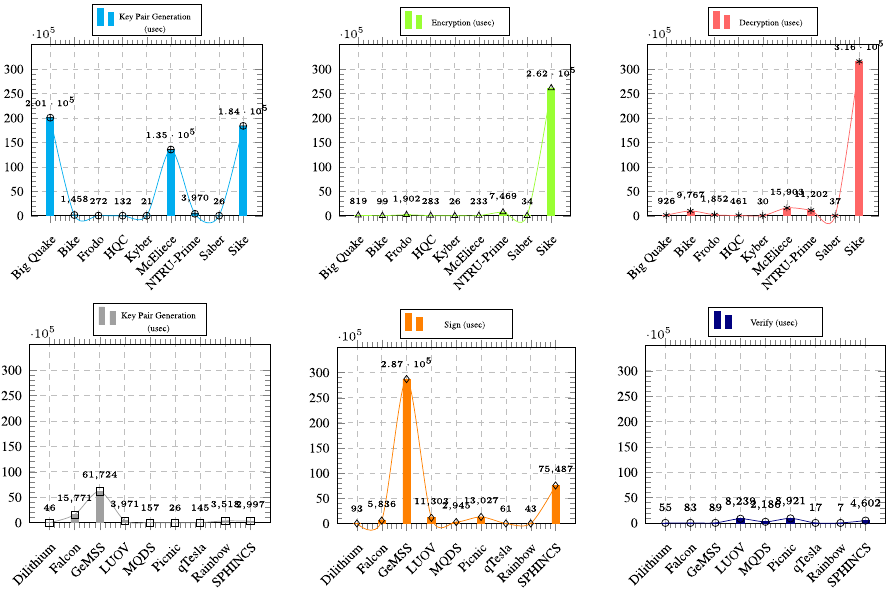}
\includegraphics[width=1\linewidth]{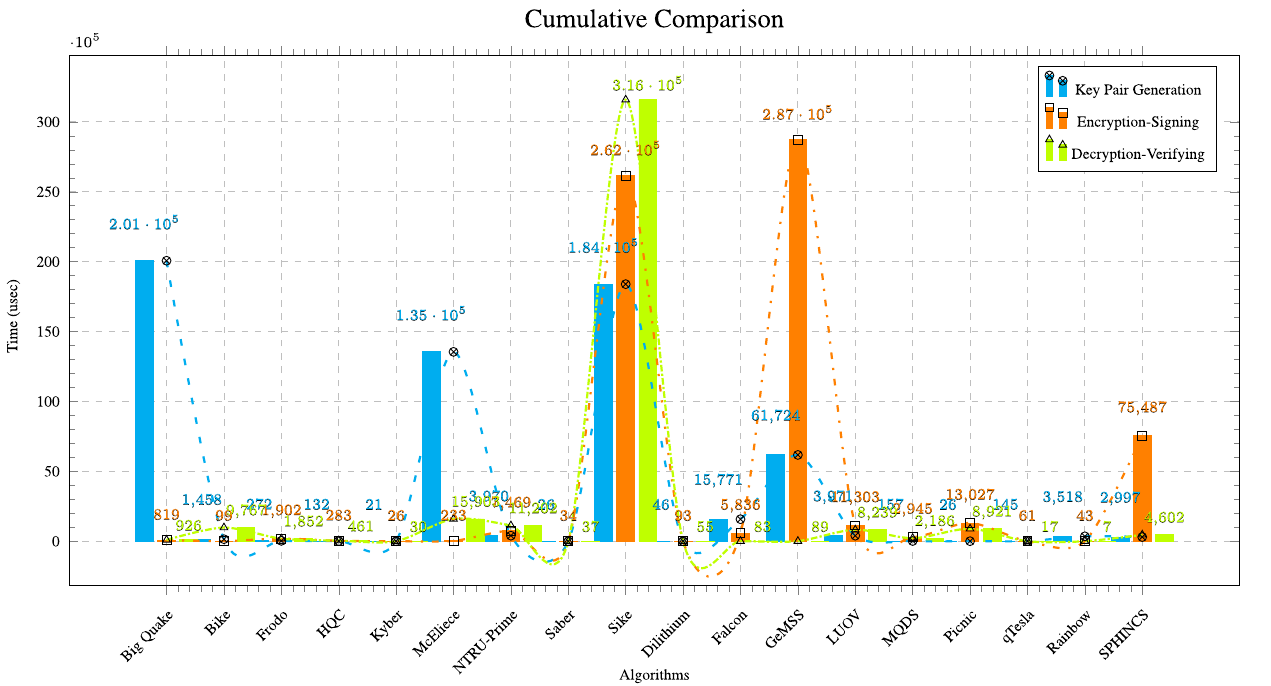}
\caption{PQC Algorithm Runtime Analysis}
\label{postquantumcryptographyalgortihmsruntimeanalysis1}
\end{figure}

Cryptographic primitive performance is evaluated first because KEM encapsulation and decapsulation, signature generation, and signature verification contribute to every subsequent PARS key-epoch transition. Figure~\ref{postquantumcryptographyalgortihmsruntimeanalysis1} presents measured runtimes on the Intel~i5-12450H platform, which represents a capable validator or ground-control-class node rather than a deeply embedded endpoint. Table~\ref{tab_PerformanceAnalysis} summarizes the evaluated cryptographic families, their principal hardness assumptions, and their security and implementation characteristics.

To ground the post-quantum overhead, performance is anchored against classical primitives: Classical X25519 key exchange and Ed25519 signatures provide reference object sizes of 32-byte public keys and 64-byte signatures and commonly execute within the microsecond range on desktop-class processors \cite{rfc7748,bernstein2012ed25519}. The post-quantum comparison shows that computational latency alone is not necessarily the dominant deployment constraint. Lattice-based KEMs and signatures can execute efficiently on capable processors, whereas their public keys, ciphertexts, and signatures are substantially larger than their classical counterparts. Depending on the selected primitive and parameter set, post-quantum objects range from hundreds of bytes to several kilobytes, thereby increasing fragmentation, buffering, transmission, and replicated-state costs. A complete end-to-end classical control run remains necessary to separate protocol and consensus overhead from the cost introduced specifically by PQC. Crucially, the primary operational bottleneck shifts from computational complexity to object size: public keys, ciphertexts, and signatures expand to a range of 0.8-10 KB. This data inflation profile directly drives the downstream ledger and protocol optimization needs. The benchmark set includes the finalized NIST standards ML-KEM, ML-DSA, and SLH-DSA, specified in FIPS~203, FIPS~204, and FIPS~205, respectively \cite{NIST2024FIPS203,NIST2024FIPS204,NIST2024FIPS205}. The analysis also considers the code-based HQC family \cite{nistIR8545} and the Falcon-derived FN-DSA standardization path where relevant. Earlier competition names---Kyber, Dilithium, and SPHINCS+---are retained where they correspond to the actual benchmark implementation, but deployment recommendations use standardized names and parameter sets.

The supersingular-isogeny-based SIKE candidate was invalidated by the Castryck--Decru key-recovery attack \cite{castryck2022sidh}. SIKE is therefore excluded from every security recommendation. Results involving SIKE are retained only as historical stress-test observations showing how an unusually slow or obsolete primitive affects the surrounding protocol pipeline; they do not represent a viable security configuration or a meaningful comparison of current post-quantum strength.
The migration timeline discussed in NIST-IR-8547 \cite{nist_ir8547} provides planning guidance for retiring quantum-vulnerable public-key mechanisms. Any dates described as mandatory must be attributed to the applicable regulatory or organizational profile rather than generalized to all CPS deployments. Long-lived Edge AI systems are nevertheless directly affected because their certification, procurement, software-update, and field-replacement cycles may extend across the anticipated migration period. This migration is increasingly governed by formal transition timelines, for classical PKC by 2030, directly intersecting with the operational lifecycles of the edge hardware considered.

\begin{table*}[!ht]
\centering
\caption{Quantum Age Cryptosystem Analysis for Architecture}
\label{tab_PerformanceAnalysis}
\begin{tblr}{
  row{1} = {c},
  row{3} = {c},
  row{4} = {c},
  row{5} = {c},
  row{7} = {c},
  row{8} = {c},
  row{9} = {c},
  row{10} = {c},
  row{11} = {c},
  row{12} = {c},
  row{15} = {c},
  row{16} = {c},
  row{17} = {c},
  row{19} = {c},
  row{20} = {c},
  cell{2}{1} = {c},
  cell{2}{2} = {c},
  cell{2}{3} = {c},
  cell{2}{4} = {c},
  cell{2}{5} = {c},
  cell{2}{6} = {r=4}{c},
  cell{2}{7} = {r=4}{},
  cell{6}{1} = {c},
  cell{6}{2} = {c},
  cell{6}{3} = {c},
  cell{6}{4} = {c},
  cell{6}{5} = {c},
  cell{6}{6} = {r=7}{c},
  cell{6}{7} = {r=7}{},
  cell{13}{1} = {c},
  cell{13}{2} = {c},
  cell{13}{3} = {c},
  cell{13}{4} = {c},
  cell{13}{5} = {c},
  cell{13}{6} = {c},
  cell{14}{1} = {c},
  cell{14}{2} = {c},
  cell{14}{3} = {c},
  cell{14}{4} = {c},
  cell{14}{5} = {c},
  cell{14}{6} = {r=4}{c},
  cell{14}{7} = {r=4}{},
  cell{18}{1} = {c},
  cell{18}{2} = {c},
  cell{18}{3} = {c},
  cell{18}{4} = {c},
  cell{18}{5} = {c},
  cell{18}{6} = {r=2}{c},
  cell{18}{7} = {r=2}{},
  vline{-} = {1-2,3,4,5,6,7,8,9,10,11,12,13-14,15,16,17,18,19}{},
  vline{1-6} = {3-5,7-12,15-17,19}{},
  hline{1-2,6,13-14,18, 20} = {-}{},
  hline{3-5,7-12,15-17,19} = {1-5}{},
}
\textbf{Algorithm} & \textbf{Security Level} & \textbf{Key Pair } & \textbf{Enc$^{\mathrm{e}}$/Sign$^{\mathrm{f}}$} & \textbf{Dec$^{\mathrm{e}}$/Verify$^{\mathrm{f}}$} & \textbf{Scheme Type} & \textbf{Attack/Mitigation}                                                                                                                                                                                           \\
BIKE$^{\mathrm{e,d}}$                & 256 – 128 (bits)        & 0.001458s          & 0.000099s         & 0.009767s            & Code Based           & {Structural
  Attacks
  \\Key Recovery
  Attacks
  \\Squaring
  Attacks
  \\Power
  Analysis Attack
  \\Side Channel
  Attacks
  \\Reaction
  Attacks
  \\Distinguishing
  Attack
  \\Message
  Recovery Attacks
  } \\
Big Quake$^{\mathrm{a,e}}$            & N/A – 128 (bits)        & 0.200649s          & 0.000819s         & 0.000926s            &                      &                                                                                                                                                                                                                      \\
Classic McEliece$^{\mathrm{e,d}}$     & 128 – 64 (bits)         & 0.135452s          & 0.000233s         & 0.015903s            &                      &                                                                                                                                                                                                                      \\
HQC-128$^{\mathrm{e,d}}$             & 128 – 64 (bits)         & 0.000132s          & 0.000283s         & 0.000461s            &                      &                                                                                                                                                                                                                      \\
CRYSTALS Kyber-512$^{\mathrm{e,g}}$   & 128 – 100 (bits)        & 0.000021s          & 0.000026s         & 0.000030s            & Lattice Based        & {Brute
  Force Attacks
  \\Meet-in-the-middle
  Attacks
  \\Lattice
  Reduction Attacks
  \\Chosen
  Cipher Text Attacks
  }                                                                                         \\
Frodo-640$^{\mathrm{c,e}}$           & 128 – N/A (bits)        & 0.000272s          & 0.001902s         & 0.001852s            &                      &                                                                                                                                                                                                                      \\
NTRUPrime$^{\mathrm{e}}$           & 364 – 210 (bits)        & 0.003970s          & 0.007469s         & 0.011202s            &                      &                                                                                                                                                                                                                      \\
SABER$^{\mathrm{e}}$              & 244 – 226 (bits)        & 0.000026s          & 0.000034s         & 0.000037s            &                      &                                                                                                                                                                                                                      \\
Falcon$^{\mathrm{f,g}}$             & N/A – 230 (bits)        & 0.015771s          & 0.005836s         & 0.000083s            &                      &                                                                                                                                                                                                                      \\
qTesla$^{\mathrm{b,f}}$             & N/A – 192 (bits)        & 0.000145s          & 0.000061s         & 0.000017s            &                      &                                                                                                                                                                                                                      \\
Dilithium$^{\mathrm{f,g}}$          & N/A – 256 (bits)        & 0.000046s          & 0.000093s         & 0.000055s            &                      &                                                                                                                                                                                                                      \\
SIKE$^{\mathrm{e,d}}$               & N/A – 128 (bits)        & 0.183894s          & 0.261508s         & 0.315817s            & Supersingular        & 
  {Auxiliary
  Points Active Attack
  \\Adaptive
  Attacks
  \\Fault
  Attacks\\
  Side  Channel Attacks
  }                                                                                                                              \\
MQDSS$^{\mathrm{b,f}}$               & N/A – 128 (bits)        & 0.000157s          & 0.002945s         & 0.002186s            & Multivariate         & {Direct
  Attacks
  \\Min/High
  Rank Attack
  \\UOV
  Attack
  \\Hash
  Attacks
  }                                                                                                                                 \\
GeMSS$^{\mathrm{c,f}}$              & N/A – 128 (bits)        & 0.061724s          & 0.287100s         & 0.000089s            &                      &                                                                                                                                                                                                                      \\
LUOV$^{\mathrm{b,f}}$               & N/A – 128 (bits)        & 0.003971s          & 0.011303s         & 0.008239s            &                      &                                                                                                                                                                                                                      \\
Rainbow$^{\mathrm{c,f}}$            & N/A – 128 (bits)        & 0.003518s          & 0.000043s         & 0.000007s            &                      &                                                                                                                                                                                                                      \\
PICNIC$^{\mathrm{c,f}}$             & N/A – 128 (bits)        & 0.000026s          & 0.013027s         & 0.008921s            & Hash Based           & {Subset
  Resilience Attacks
  \\Second
  Preimage Attacks
  \\Fault
  Injection Attacks
  }                                                                                                                         \\
SPHINCS$^{\mathrm{f,g}}$            & N/A – 128 (bits)        & 0.002997s           & 0.075487s         & 0.004602s            &                      & \\
 {$^{\mathrm{a}}$Round 1} & {$^{\mathrm{b}}$Round 2} & {$^{\mathrm{c}}$Round 3} & {$^{\mathrm{d}}$Round 4} & {$^{\mathrm{e}}$KEM} & {$^{\mathrm{f}}$DSA} & {$^{\mathrm{g}}$Finalist}
\end{tblr}
\vspace{-2pt}
\end{table*}
%\clearpage

\subsection{Average Consensus Time}
\label{subsec:consensus_time}

\begin{figure}[!htb]
\includegraphics[width=1\linewidth]{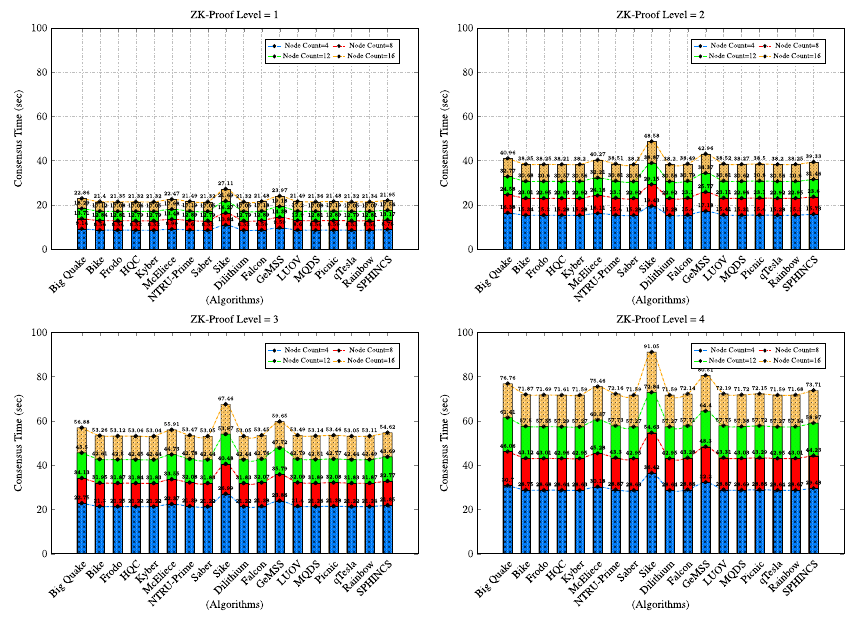}
\caption{Average Consensus Time as a Function of Swarms}
\label{ref_esl_average_consensus_time}
\end{figure}

Consensus time measures the interval between initiation of a PARS key-epoch proposal and agreement on the corresponding PARS-ESL security-state update. It therefore determines how quickly a new epoch can be confirmed and activated, but it does not represent the latency of operational telemetry protected by an already active symmetric key. Measurements cover four node counts---4, 8, 12, and 16---and four configurable validation levels. The existing implementation and Fig.~\ref{ref_esl_average_consensus_time} refer to these settings as ZKP hardness Levels~1--4. Until a concrete proof system, statement--witness relation, completeness, soundness, and zero-knowledge parameters are specified, they should be interpreted as \emph{challenge-validation levels} rather than as cryptographic ZKP security levels.

The experiment evaluates swarm sizes of 4, 8, 12, and 16 nodes and four legacy ``ZKP hardness'' settings. Because the current mechanism has not yet been demonstrated to satisfy formal zero-knowledge definitions, these settings are interpreted in this section as \emph{challenge-complexity levels}. A higher level increases the work or interaction required by the prototype validation process, but it must not be described as providing stronger zero knowledge unless completeness, soundness, and zero-knowledge properties are formally established.

Across the reported sweep, the observed completion times range from 8.53~s for the lowest-cost ML-KEM configuration to 91.05~s for the 16-node SIKE configuration at the highest challenge level, as shown in Fig.~\ref{ref_esl_average_consensus_time}. The upper value is a historical stress case generated using an excluded and cryptanalytically broken scheme; it is not a deployable operating point. The evaluated lattice-based configurations generally complete within the single-digit to low-tens-of-seconds range, although the exact distribution, trial count, and uncertainty must be reported before this range is treated as a statistically established service bound.

Consensus latency increases with node count and challenge complexity because additional participants generate more messages, validations, synchronization points, and potential retries. Dividing a large group into smaller security domains may reduce local agreement latency, but such partitioning introduces additional questions concerning inter-cluster trust, epoch consistency, membership migration, and recovery. It should therefore be presented as an architectural option requiring separate evaluation rather than as a conclusion directly demonstrated by Fig.~\ref{ref_esl_average_consensus_time}.

The primitive KEM and signature operations execute substantially faster than the complete multi-second agreement cycle. This suggests that protocol interaction, challenge generation and validation, software scheduling, synchronization, serialization, network exchange, and block formation dominate the measured path. However, the current experiment does not yet provide a direct phase-level decomposition. A classical X25519/Ed25519 control configuration \cite{rfc7748,bernstein2012ed25519} is also required to distinguish the cost of PARS and ESL coordination from the incremental cost of post-quantum primitives. The following components should be instrumented independently:

\begin{equation}
\begin{split}
T_{\mathrm{round}} &= T_{\mathrm{setup}} + T_{\mathrm{KEM}} + T_{\mathrm{sign}} + T_{\mathrm{challenge}} \\
&\quad + T_{\mathrm{network}} + T_{\mathrm{vote}} + T_{\mathrm{commit}} + T_{\mathrm{retry}}
\end{split}
\end{equation}

\subsection{Size Metrics (Message-Transaction-Memory-Bandwidth)}
\label{subsec:size_metrics}
\begin{figure}[!htb]
\includegraphics[width=1\linewidth]{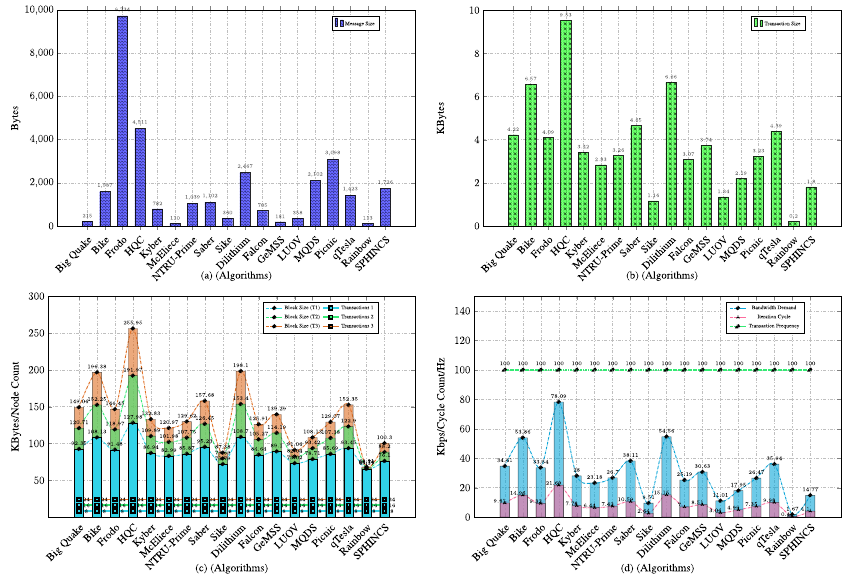}
\caption{Size Metric Footprints of NIST Schemes for Edge AI}
\label{plot_esl_size_metrics}
\end{figure}

Because military and Edge~AI nodes run on resource-limited hardware, the byte-level footprint of PARS-ESL is a primary feasibility constraint because post-quantum key-establishment and authentication objects may be large relative to the Maximum Transmission Unit (MTU) and available bandwidth of embedded wireless, RF, or SATCOM links. Message size varies across algorithm versions owing to their differing key sizes: the maximum is about \mbox{10\,KB}, while the crypto-agile profile also exposes much smaller options on the order of \mbox{100\,B} and \mbox{800\,B} (see Fig. \ref{plot_esl_size_metrics}(a)). These figures track the public-key, ciphertext, and signature sizes of the underlying schemes---from roughly \mbox{768--800\,B} for ML-KEM-512 keys and ciphertexts, through several kilobytes for ML-DSA signatures, to tens of kilobytes for SLH-DSA~\cite{NIST2024FIPS203,NIST2024FIPS204,NIST2024FIPS205}. For comparison, the classical primitives being replaced carry 32-byte public keys and 64-byte signatures~\cite{rfc7748,bernstein2012ed25519}, so the post-quantum transition multiplies the per-reconciliation link budget by one to two orders of magnitude; as Section~\ref{subsec:platform} develops, this---not compute---is the binding constraint on tactical links. Although a 10\,KB message looks large at first glance, it is the price of an up-to-date key and a strong encapsulation mechanism; crypto-agility lets an operator trade this footprint against the threat level, and dedicated footprint optimization is for future.

A message of approximately 10~KB may be acceptable on Ethernet-class links but may require substantial fragmentation on low-MTU wireless, serial, narrowband, or satellite channels. The correct engineering question is therefore not whether 10~KB is intrinsically large, but how many fragments, retransmissions, and channel-time units it generates under a specific link profile. Final evaluation should report at least:

\begin{itemize}
    \item protocol payload bytes
    \item transport and link-layer bytes
    \item fragment count
    \item retransmitted bytes
    \item per-node and aggregate traffic
    \item peak and average bit-rate
    \item completion time under packet loss and reordering
\end{itemize}

Transaction size targets internal ledger operations and ranges from \mbox{0.2\,KB} to \mbox{9.53\,KB} (see Fig. \ref{plot_esl_size_metrics}(b)), which is well within the budget of current edge defense and space systems. These records correspond to low-rate key-epoch, authentication, validation, and membership state rather than bulk application transactions. Once key reconciliation succeeds, these transactions drive block formation and all edge nodes synchronize on the agreed transaction key; the corresponding synced per-node memory remains below \mbox{256\,KB} across the tested node counts (see Fig. \ref{plot_esl_size_metrics}(c)), even though the figure is drawn for a homogeneous KEM/DSA choice and crypto-agility would permit heterogeneous per-node algorithms. This value should be interpreted according to the implemented retention policy and measured memory category; it does not necessarily include the operating system, complete process image, cryptographic library, networking buffers, or persistent chain history. Finally, the physical- and data-link-layer measurements (see Fig. \ref{plot_esl_size_metrics}(d)) show that reconciliation issues several atomic transactions per node at once; the resulting instantaneous bandwidth demand of \mbox{1.67--78\,kbit/s} must therefore be satisfied across multiple iteration cycles owing to network-layer limits. This is an additional cost, but it is incurred only during key reconciliation and does not perturb the real-time communication flow; once reconciliation completes, C4ISR traffic proceeds under standard conditions. The size results confirm that PQC algorithms are usable in this setting provided the network layer can absorb the reconciliation burst. The final measurement definition must specify whether this is:
\begin{itemize}
    \item per-node or aggregate bandwidth
    \item average or peak bitrate
    \item unidirectional or bidirectional traffic
    \item application-layer or wire-level traffic
    \item measured over the complete reconciliation interval or only the transmission burst
\end{itemize}

The reported range shows that security-control traffic can become significant on narrowband links. It does not establish that every tested configuration is sustainable. Because PARS confines large public-key objects to establishment and rotation windows, continuous application traffic can remain protected using symmetric AEAD. Nevertheless, frequent rotation, large groups, packet loss, or repeated consensus failure may cause the security-control plane to compete with operational traffic that cause scheduling, prioritization, and admission control.

\begin{figure}[!htb]
\includegraphics[width=1\linewidth]{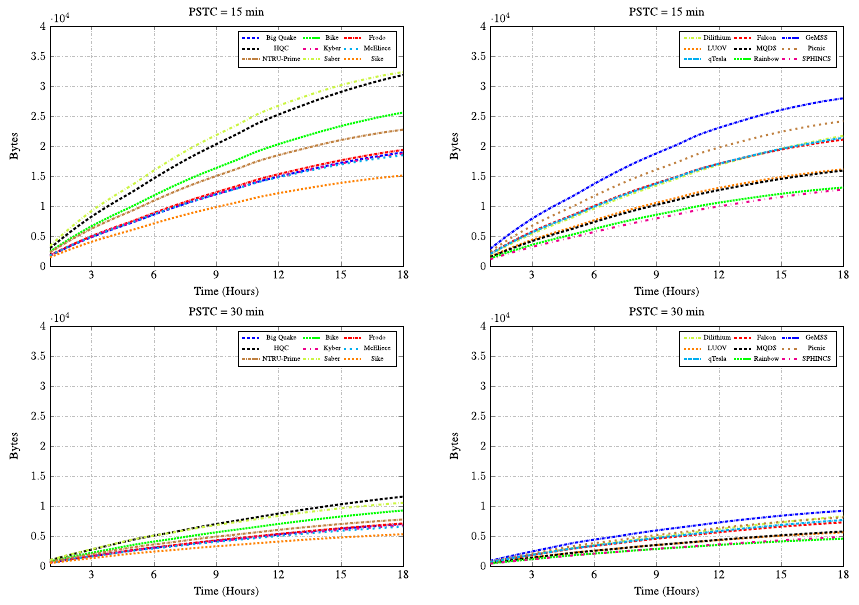}
\caption{Accumulated Memory Footprints of NIST Schemes}
\label{plot_esl_cumulative_memory_size_effect}
\end{figure}

\subsection{Cumulative Memory Footprint}
\label{subsec:memory}
The cumulative memory effect of PARS and ESL is profiled with the Periodic Security Threshold Control (PSTC) set to 15 and 30~minutes for both KEM and DSA algorithms. Across 16~nodes, peak memory usage stayed below \mbox{3.5\,MB} over an 18-hour run (see Fig. \ref{plot_esl_cumulative_memory_size_effect}). Sixteen nodes were chosen because the node count directly governs how many key reconciliations succeed or fail and hence how many blocks form, and because it is a realistic ceiling for a tactical swarm. The required memory falls as the PSTC interval lengthens, because re-keying is less frequent and fewer blocks accumulate. Read together with the consensus-time results (see Fig. \ref{ref_esl_average_consensus_time}), the node-level memory budget shows that a high-performance, quantum-safe blockchain is realizable in memory-constrained environments with the selected algorithms, and that node-level state suffices to authenticate swarm membership and validate key exchanges in operational events such as the node joining formation. Lower-level instrumentation of atomic operations---HyperRAM-OCM (pushes), CPU (loads)---is planned for the on-device campaign described in Section~\ref{subsec:platform}. The term ``memory footprint'' can be defined more precisely as it may refer to:

\begin{itemize}
    \item peak process-resident memory
    \item heap or stack use
    \item serialized ledger state retained in RAM
    \item persistent chain storage
    \item cumulative allocated memory
    \item aggregate state across multiple nodes
\end{itemize}

Longer PSTC intervals produce fewer key-epoch transitions and therefore fewer security-state blocks over a fixed observation period. The observed reduction in accumulated state is consequently expected and should be interpreted as a rotation-frequency trade-off rather than as an independent memory optimization. The corresponding storage-growth model may be expressed as

\[
S(t)=S_{0}+N_{\mathrm{success}}(t)S_{\mathrm{epoch}}-S_{\mathrm{pruned}}(t),
\]

where $S_{0}$ is the genesis and policy state, $N_{\mathrm{success}}(t)$ is the number of successfully committed epochs, $S_{\mathrm{epoch}}$ is the average state per epoch, and $S_{\mathrm{pruned}}(t)$ is the amount removed through pruning or checkpointing.

The experiment demonstrates bounded growth for the tested duration and configuration. It does not yet establish long-term behavior over weeks, months, or years, nor does it establish MCU-level feasibility. Future instrumentation should separate cryptographic workspace, protocol state, per-peer state, reassembly buffers, pending blocks, finalized ledger data, and persistent storage. Lower-level measurements of HyperRAM or on-chip-memory writes, CPU loads, and flash commits should be reported only after direct instrumentation on the target hardware.

\begin{figure}[!htb]
\includegraphics[width=1\linewidth]{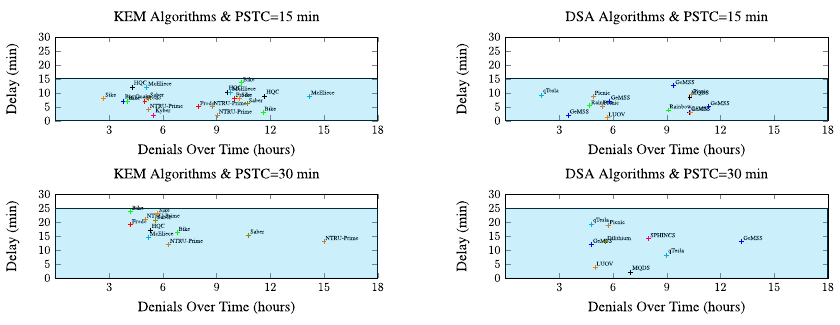}
\caption{Denial of Consensus Rounds}
\label{plot_esl_denial_of_consensus}
\end{figure}

\subsection{Denial of Consensus and Failure Analysis}
\label{subsec:denial}
A further protocol metric is the uncontrolled rejection of consensus, which the configuration itself can induce. As shown in Fig \ref{plot_esl_cumulative_memory_size_effect}, cumulative memory does not grow linearly; under a linear time axis this indicates rounds in which consensus is denied. Chain growth is shaped by communication delays, node count, reconciliation period, and ZKP hardness. Fig. \ref{plot_esl_denial_of_consensus} shows that the denial count decreases as the reconciliation period increases. These rejection rates are low---only a handful of events occur across the full 18-hour core flow---which suggests an operable system, although single-run counts of rare events carry wide uncertainty; the repeated-trials campaign and direct instrumentation of the rejection path (classifying each denial as proof timeout, proof rejection, or node desynchronization, rather than inferring denials from non-linear memory growth (NLMG)) are required before the rate can be bounded. Increasing the key-exchange frequency raises the security rate but, as the results show, also raises the denial rate: a larger node count increases processor load, which delays ZKP generation and verification past the round deadline and thereby drives the denial rate up. The key-renewal interval and the consensus latency are independent quantities; because key renewal encloses a consensus round, an additional latency of 0--15 or 15--30~minutes (matching the PSTC setting) may be observed. The number of internally rejected consensus rounds decreases as the reconciliation period lengthens. NLMG of the cumulative chain size reflects these rejected rounds.

The observed denial count decreases as the PSTC interval increases. This relationship is expected because fewer attempted rotations create fewer opportunities for timeout or synchronization failure. It does not necessarily imply that each individual round becomes more reliable. Reliability should be reported using both:

\[
P_{\mathrm{fail}} =
\frac{N_{\mathrm{failed\ rounds}}}
     {N_{\mathrm{attempted\ rounds}}}
\]

and the absolute number of attempted and completed rounds.

A higher rotation frequency may reduce the duration for which one key remains active, but it also increases communication, computation, and synchronization pressure. The relevant design problem is therefore to select a rotation policy that balances compromise exposure against availability and resource use. PSTC should not be interpreted as a mathematically safe lifetime unless it is derived from an explicit threat, data-lifetime, and key-use model.

The key-renewal cadence and the consensus execution time are configured separately, but they are operationally coupled. A rotation cannot complete sooner than the underlying agreement process, and a new rotation should not begin if the previous epoch remains unresolved unless concurrent transitions are explicitly supported. The final protocol must define deadline behavior, pending-epoch limits, overlap, cancellation, and fail-safe operation.

\begin{figure}[!htb]
\includegraphics[width=1\linewidth]{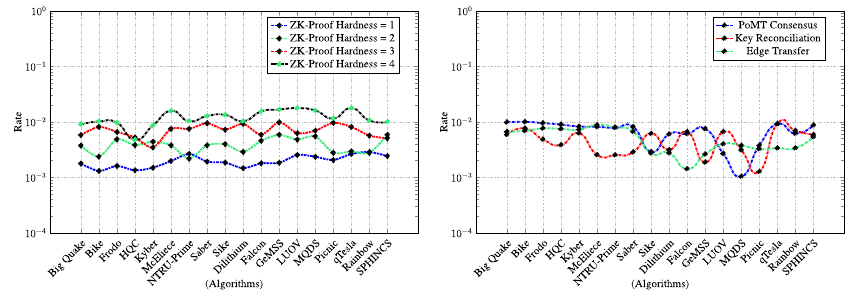}
\caption{ESL Failure Results}
\label{plot_esl_failure_results}
\end{figure}

Operational failures span roughly \mbox{$10^{-4}$ to $10^{-1}$} on the logarithmic axis of Fig. \ref{plot_esl_failure_results}, rising with ZKP hardness and node count. Here \emph{failure} denotes any reconciliation round that does not complete. Cryptographic causes are essentially excluded by design---ML-KEM's decryption-failure probability is below $2^{-138}$ for every parameter set~\cite{NIST2024FIPS203}---so the observed failures are protocol/network-level events: proof timeouts, rejected proofs, and node desynchronization. The $10^{-1}$ upper end is reached only at the highest hardness levels, the largest swarm, and the slowest (now-excluded) schemes; the lattice configurations sit toward the lower end of the range. The left panel attributes failures to ZKP hardness for each algorithm, while the right panel attributes them to the major pipeline stages---PoE consensus, key reconciliation, and edge transfer. Consistent with Section~\ref{subsec:memory}, higher ZKP hardness lowers the cumulative memory footprint precisely because fewer reconciliations and consensus rounds succeed. We note a limitation that affects the strength of these conclusions: the present campaign reports a single series per condition, without confidence intervals or an explicit trial count. Quantifying variance---at least 30 repeated runs per condition, reported as medians with 95\% confidence intervals and per-caption trial counts---is required before strong statistical claims can be made and is part of the planned evaluation. A second gap is adversarial coverage: every denial and failure event is self-induced, so a campaign with an active adversary (e.g. a Byzantine node injecting malformed transactions, replaying superseded keys) is planned to characterize behaviour under attack rather than under load alone.

The negligible algorithmic decapsulation-failure probability specified for ML-KEM \cite{NIST2024FIPS203} indicates that mathematically expected decapsulation failure is unlikely to explain the measured protocol failures under correct implementation. It does not exclude software defects, malformed ciphertext handling, integration errors, memory corruption, side-channel countermeasures, or incorrect API usage. Observed failures should therefore be classified directly into:

\begin{itemize}
    \item cryptographic API or validation failure
    \item challenge timeout
    \item challenge rejection
    \item network timeout or loss
    \item node desynchronization
    \item insufficient quorum
    \item stale epoch or replay
    \item malformed transaction
    \item resource exhaustion
    \item unknown cause
\end{itemize}

The reported trend indicates higher unsuccessful-round rates under larger node counts, greater challenge complexity, and slower historical schemes. Because the current results contain one reported series per condition without confidence intervals, they should be interpreted as prototype trends rather than statistically bounded reliability estimates. At least 30 independent repetitions per principal condition, or a statistically justified alternative for long-duration tests, should be reported using confidence intervals and explicit sizes.

The present failures are primarily self-induced by load, timeout, and synchronization behavior. An adversarial campaign should separately examine malformed transactions, replayed epoch messages, conflicting commitments, credential misuse, Byzantine validators, delayed acknowledgments, and partition-and-merge behavior.

\subsection{Cross-Platform Projection for Edge~AI Hardware}
\label{subsec:platform}
The baseline measurements above were obtained on a GCS-class Intel~i5-12450H. Real Edge~AI deployments, however, run on a heterogeneous mix of embedded processors, microcontrollers, GPUs, and adaptive SoCs that spans commercial modules and ruggedized defense/aerospace parts. This constraint is magnified in defense and aerospace applications, where platforms must withstand strict thermal envelopes, radiation-induced Single-Event Upsets (SEUs), and severe size, weight, and power (SWaP) limits. This subsection projects the cryptographic core cost onto twelve representative platforms, evaluating standard commercial hardware alongside specialized defense- and aerospace-grade substrates, and analyzes the performance implications when the Programmable Logic (PL) of an adaptive SoC or FPGA is utilized alongside its Processing System (PS). The projections follow a transparent scaling model: published per-platform cycle counts are used wherever they exist~\cite{kannwischer2019pqm4,abdulrahman2022cortexm4,xing2021kyberfpga,beckwith2021dilithiumfpga}; otherwise entries scale the i5 baseline by clock-frequency and issue-width ratios, crediting ISA cryptographic extensions where present. Every such entry is marked projected and the on-device campaign (see Table~\ref{tab_platform_projection}) added as measured-versus-projected error column.

The commercial/embedded tier comprises the NVIDIA Jetson~Nano (4-core Arm Cortex-A57, 128-core Maxwell GPU), the Jetson~Orin (6-12 core Cortex-A78AE with ARMv8 cryptographic extensions, Ampere GPU)~\cite{nvidia_jetson_orin}, and the NXP~i.MX~RT1064 crossover MCU (single Cortex-M7 at up to 600\,MHz, 1\,MB on-chip SRAM, no MMU)~\cite{nxp_imxrt1064}. To capture rugged tactical land-airborne defense deployments, the \textbf{Jetson AGX Orin Industrial}, which hardens the core Orin architecture with extended operatings temperatures ($-40\,^{\circ}$C to $+85\,^{\circ}$C), enhanced vibration/shock packaging, and full ECC DRAM configurations~\cite{nvidia_jetson_orin_industrial}, and the \textbf{TI Sitara AM69x}, a high-performance robotic and tactical edge processor with eight Cortex-A72 application cores at up to 2.0\,GHz, dual-core Cortex-R5F real-time islands, an on-chip Hardware Security Module (HSM), and four MMAv2 deep-learning accelerators delivering a combined 32\,TOPS~\cite{ti_sitara_am69x}.

For reconfigurable substrates, standard AMD/Xilinx Zynq-7000 (dual Cortex-A9 PS plus an Artix-7-class PL fabric), the AMD~Zynq~UltraScale+\texttrademark{}~MPSoC, which pairs a 64-bit quad-core Cortex-A53 application unit and dual Cortex-R5F real-time cores with an UltraScale+ PL fabric and an Arm Mali-400 GPU and is offered in defense-grade XQ packages rated for $-55$ to $+125\,^{\circ}$C with a 256-bit PUF~\cite{amd_zynq_ultrascale}, and the state-of-the-art commercial AMD Versal\texttrademark{}~AI~Edge Series Gen~2 adaptive SoC, which merges Cortex-A78AE application cores, Cortex-R52 real-time cores, an expansive PL fabric, and dedicated AI~Engines~\cite{amd_versal_gen2} are evaluated. To analyze specialized high-reliability defense and aerospace domains, this profile is extended with four further classes. Firstly a mil-spec, ultra-low-power platform combining a coherent quad-core 64-bit RISC-V (SiFive U54, RV64GC) complex with a non-volatile, SEU-immune flash FPGA fabric; its radiation-tolerant RT~PolarFire variant is the first flight-proven, Linux-capable RISC-V SoC~FPGA and, being flash-based, exhibits zero configuration-memory upsets without an external scrubber~\cite{microchip_polarfire_soc,microchip_rtpolarfire}. Secondly, a flight-qualified, in-orbit-reprogrammable, radiation-hardened flagship for orbital satellites and strategic aerospace AI pipelines, combining radiation-tolerant Cortex-A72 cores, dual Cortex-R5F cores, programmable logic, and AI~Engines equipped with extensive hardware error mitigation~\cite{amd_xqr_versal}. Thirdly, a 7\,nm system-on-chip integrating high-speed RF data converters (sampling up to 64\,Gsps) directly with FPGA logic, driven by an integrated Secure Device Manager (SDM) with post-quantum secure-boot support and a quad-core Cortex-A53 Hard Processing System (HPS)~\cite{altera_agilex9}. Finally, a radiation-hardened 64-bit RISC-V MPU ~\cite{microchip_pic64hpsc} developed in partnership with NASA JPL, featuring eight SiFive Intelligence X280 application cores equipped with 512-bit vector registers compliant with the RISC-V Vector (RVV).

The defense and aerospace deployment imposes constraints beyond raw throughput that directly shape platform choice. Spaceborne and high-altitude nodes must tolerate single-event upsets and total-ionizing-dose effects: SRAM-based FPGAs require configuration scrubbing to recover from upsets, whereas flash-based fabrics such as RT~PolarFire are immune to configuration-memory upsets and need no scrubber, and flight-qualified adaptive SoCs such as the XQR~Versal carry MIL-PRF-38535 Class~B qualification for missions from low-earth to geosynchronous orbit~\cite{microchip_rtpolarfire,amd_xqr_versal}. Ground and airborne tactical nodes instead emphasize extended-temperature operation and ruggedization---met by defense-grade XQ packaging ($-55$ to $+125\,^{\circ}$C) and by the industrial Jetson variants---alongside long deployment lifetimes and component-obsolescence management. Two properties of the proposed architecture align with these constraints. First, the constant-time, spatially-redundant PL datapaths discussed below remove the cache and branch timing side channels of software execution and admit fault-injection mitigation (dual-modular redundancy, on-array parity), which matters acutely for physically exposed assets. Second, the in-field reprogrammability of adaptive SoCs---including in orbit---lets a deployed node adopt a revised algorithm or parameter set as the standards mature, the hardware counterpart of the protocol-level crypto-agility of PARS. Size, weight, and power further favor the low-power flash-FPGA and adaptive-SoC parts, whose PL accelerators reach baseline-class latency at single-digit watts.

The cryptographic bottlenecks of lattice and hash families map exceptionally well onto reconfigurable hardware. In a PS-only port, the Number-Theoretic Transform (NTT), polynomial multiplication, modular reduction, and Keccak/SHAKE permutations execute as sequential software loops that pay instruction-fetch, cache-miss, and control-flow overhead on every iteration. On the in-order, 32-bit dual Cortex-A9 of the Zynq-7000 or the power-optimized RISC-V cores of the PolarFire SoC, these software overheads dominate. Mapped into the PL fabric, the same kernels become dedicated, deeply pipelined datapaths that consume DSP slices and block RAM, completing a butterfly or permutation step per clock with zero instruction control overhead.

Published HW/SW co-designs on the Zynq architecture report \mbox{8.7--12.5$\times$} (KeyGen), \mbox{6.3--7.3$\times$} (Sign), and \mbox{9.1--12.2$\times$} (Verify) speedups for Dilithium over the Cortex-A9 software baseline~\cite{mao2023dilithium}, while comparable RISC-V/FPGA co-designs accelerate Kyber and Dilithium by \mbox{3--5$\times$} while using under 5\% of the fabric~\cite{karl2024riscv}. On the ARM side, vectorized NEON implementations recover much of the gap on Cortex-A72/A78-class cores~\cite{becker2022neonntt}, while microcontroller targets such as the i.MX~RT1064 rely on hand-optimized assembly and land in the low-millisecond-per-operation regime: scaling the measured Cortex-M4 cycle counts of pqm4~\cite{kannwischer2019pqm4,abdulrahman2022cortexm4} to the 600\,MHz dual-issue M7 yields roughly 1--2\,ms per ML-KEM-768 operation and $\approx$\,5--15\,ms for an average ML-DSA-65 signature (rejection-sampling mean)---fast enough for PSTC-windowed re-keying, though not for per-packet cryptography. GPU back-ends---available on the Jetson line through cuPQC---accelerate \emph{batched} operations by orders of magnitude on server GPUs and remain useful for throughput-oriented edge nodes, although their benefit on single, latency-sensitive transactions is limited~\cite{nvidia_cupqc}. In rugged environments like the Jetson AGX Orin Industrial, activating ECC memory degrades raw memory bandwidth by $\approx$\,5--10\%, creating a minor overhead that reinforces the necessity of hardware-assisted acceleration paths. Table~\ref{tab_platform_projection} summarizes projections relative to the Intel~i5-12450H baseline. PS-only multipliers are projected from published microarchitectural scaling; PS+PL/GPU multipliers are grounded in results~\cite{mao2023dilithium,karl2024riscv,becker2022neonntt,abdulrahman2022cortexm4,kannwischer2019pqm4,nvidia_cupqc}.

\begin{table*}[!hb]
\centering
\caption{Architecture-Based Projection of PQC Core Cost Across Edge~AI Platforms}
\label{tab_platform_projection}
\resizebox{\linewidth}{!}{%
\begin{tabular}{|l|l|l|l|l|}
\hline
\textbf{Platform} & \textbf{Compute core(s)} & \textbf{Accel. path} & \textbf{Lattice (KEM/DSA)} & \textbf{Hash (SLH-DSA)} \\
\hline
Jetson Orin            & Cortex-A78AE + Ampere GPU      & NEON / GPU      & $\approx$\,2.5--4$\times$ (CPU); $\le$\,1$\times$ batched   & $\approx$\,1.5--3$\times$ (crypto ext.) \\
\hline
Jetson AGX Orin Ind.   & Hardened Cortex-A78AE + GPU    & NEON / GPU (ECC) & $\approx$\,2.8--4.5$\times$ (CPU); $\le$\,1.1$\times$ batched & $\approx$\,1.6--3.3$\times$ (crypto ext.) \\
\hline
Jetson Nano            & Cortex-A57 + Maxwell GPU       & NEON (no crypto ext.) & $\approx$\,12--18$\times$                              & $\approx$\,25--40$\times$ \\
\hline
i.MX RT1064 (MCU)      & single Cortex-M7 @600\,MHz     & assembly only   & $\approx$\,1--2\,ms/op (ML-KEM-768); 5--15\,ms ML-DSA-65 sign~\cite{kannwischer2019pqm4} & impractical for fast re-keying \\
\hline
TI Sitara AM69x        & 8$\times$ Cortex-A72 @2\,GHz + R5F + HSM & NEON / SW parallel + HSM & $\approx$\,3.5--5.5$\times$ (projected) & $\approx$\,2--3.5$\times$ (projected) \\
\hline
PolarFire SoC (RISC-V) & quad RISC-V U54 (in-order, 64-bit) & none         & $\approx$\,20--45$\times$ (projected)                & several seconds (projected) \\
\hline
PolarFire SoC / RT (PS+PL) & RISC-V U54 + flash-FPGA fabric & dedicated PL cores & tens of $\mu$s; meets baseline kernels~\cite{xing2021kyberfpga,karl2024riscv} & 6--12$\times$ vs RISC-V software \\
\hline
Zynq-7000 (PS only)    & dual Cortex-A9 (32-bit, in-order) & none          & $\approx$\,150--300$\times$ (projected)               & several seconds (projected) \\
\hline
Zynq-7000 (PS + PL)    & Cortex-A9 + Artix-7 PL        & NTT/Keccak in PL & tens of $\mu$s; meets baseline kernels~\cite{xing2021kyberfpga,beckwith2021dilithiumfpga} & 7--15$\times$ vs PS-only \\
\hline
Zynq UltraScale+ MPSoC & quad Cortex-A53 + R5F + Mali-400 + PL & NEON/crypto-ext + PL & $\approx$\,12--20$\times$ (PS); tens of $\mu$s with PL & moderate (crypto ext.); strong with PL \\
\hline
Versal AI Edge Gen~2   & Cortex-A78AE + PL + AI~Engines & PL + NEON       & $\le$\,1$\times$ for accelerated kernels (projected)  & strong (large PL + parallel hash) \\
\hline
XQR Versal Space-Grade & Rad-Hard Cortex-A72 + R5F + PL + AIE & TMR-hardened PL & $\le$\,1.5$\times$ with PL (projected); $\approx$\,4--7$\times$ PS-only & strong via rad-tolerant PL fabric \\
\hline
Altera Agilex 9 Direct RF & quad Cortex-A53 HPS + 7\,nm PL & Direct-RF PL core + SDM & $\le$\,1$\times$ for accelerated kernels (projected) & strong via parallel PL hash \\
\hline
Microchip PIC64-HPSC   & 8$\times$ RISC-V X280 @1\,GHz  & 512-bit vector (RVV) & $\approx$\,1.2--2.2$\times$ (projected) & $\approx$\,1.1--1.8$\times$ (projected) \\
\hline
\end{tabular}%
}
\end{table*}

The strongest case for tactical and orbital nodes is a \emph{standalone} (pure-PL) datapath rather than a co-design. The Zynq-7000 PL is an Artix-7-class fabric with \mbox{53{,}200} LUTs, \mbox{220} DSP48E1 slices, and \mbox{140} block-RAM tiles (\mbox{$\approx$4.9\,Mb}) alongside the dual Cortex-A9 PS~\cite{amd_zynq7000_ds190}, which is ample for the lattice kernels. A compact full-hardware Kyber on the smallest Artix-7 device computes key generation, encapsulation, and decapsulation in only a few thousand cycles each---\mbox{3{,}768/5{,}079/6{,}668} cycles for the lightest parameter set up to \mbox{9{,}380/11{,}321/13{,}908} for the heaviest---at \mbox{204\,MHz} while consuming roughly \mbox{7{,}400} LUTs, two DSPs, and three BRAMs, outperforming HLS and HW/SW co-designs by a wide margin~\cite{xing2021kyberfpga}. Similarly, the Microchip PolarFire SoC's flash-based fabric eliminates configuration upsets in high-radiation environments, allowing its fabric to house constant-time PQC hardware cores without risk of bit-flips. At typical operational clocks, encapsulation latency in pure PL lands in the tens of microseconds, matching the GCS baseline despite an extremely tight power envelope. High-performance Dilithium cores report the best FPGA signing latencies to date with modest resource use~\cite{beckwith2021dilithiumfpga}.

Two consequences matter for a deployed defense node. First, a combined ML-KEM\,+\,ML-DSA subsystem occupies only a fraction of the Zynq-7000, PolarFire, or XQR Versal fabric, leaving the processing cores free to run the real-time operating system, the PARS/ESL protocol logic, and lightweight \mbox{AXI} hand-off to the accelerators. Second, a fully spatial datapath executes in constant time and admits spatial redundancy (e.g., Dual-Modular Redundancy [DMR] or Triple-Modular Redundancy [TMR] on the XQR Versal fabric), which eliminates the cache and branch timing leakage inherent to software execution and supports systematic fault-injection hardening---properties that are critically valuable for exposed tactical or space platform, independent of raw speed. Power and electromagnetic side channels remain in any realization, so it requires dedicated countermeasures (masking, hiding) at the datapath level~\cite{pqshield_pqplatform}.

Microchip PIC64-HPSC introduces a vector-first acceleration paradigm. Its 512-bit vector registers execute modular polynomial arithmetic in parallel across its RISC-V Vector units, delivering up to 2\,TOPS (INT8) or 1\,TFLOPS (BF16) of vector performance without requiring external FPGA coprocessors. This allows unified compilation for AI, DSP, and PQC workloads within a single ISA, reducing safety-qualification complexity for critical systems \cite{microchip_pic64hpsc}.

The most important consequence for an Edge~AI deployment is that none of these hardware accelerators (shortens computation) reduce/shrink the size of the cryptographic material. The ENC/DEC/SIGN/VERIFY operations underlying every edge transaction carry the large keys, ciphertexts, and signatures characteristic of PQC, and these dominate the link budget regardless of how fast the arithmetic runs on PL, GPU, vector, or RISC-V hardware. A single reconciliation transaction already reaches \mbox{9.53\,KB} for the heavier schemes, and naive per-message re-keying would multiply this across the bandwidth-limited RF/SATCOM links measured at \mbox{1.67--78\,kbit/s}, quickly triggering packet fragmentation, channel congestion, and eventual protocol collapse. A PolarFire or Zynq-7000 PL accelerator that drives Kyber encapsulation from hundreds of microseconds down to tens of microseconds does nothing to compress the resulting ciphertext, so on such tactical links the packet size---not the compute time---becomes the binding operational constraint.

PARS addresses this distinction by scheduling large key-establishment and authentication objects within controlled epoch-transition windows. After reconciliation, high-rate data uses a symmetric AEAD channel such as AES-256-GCM. The ESL records only commitments and security-control metadata rather than the application stream. Programmable logic or GPU acceleration can shorten key-transition computation, while the protocol architecture amortizes the corresponding object-size overhead over many application packets. This is precisely the gap that PARS and ESL close. Crypto-agility selects the smallest adequate cryptographic footprint for each operational scenario; rotational secrecy confines re-keying to controlled PSTC windows rather than every message, after which the nodes run high-speed, minimal-overhead AES-256-GCM symmetric streams for continuous real-time telemetry; the ESL ledger records and validates key-state changes ad~hoc across the swarm, removing the need for continuous over-the-air validation rounds with centralized certificate authorities; and the PS/PL partition offloads the heavy arithmetic to the fabric while the main processing cores run the protocol and Edge~AI inference logic. The proposed method therefore remains feasible on exactly the constrained, accelerated nodes where a per-message PQC scheme would saturate the link---which is why the importance of the approach becomes clearest, not weakest, once hardware acceleration is assumed. In this light, the Versal~AI~Edge Gen 2 and its radiation-hardened sibling, the XQR Versal, represent strong single-chip candidates for this architecture: their application cores handle the main protocol stack, their PL fabrics accelerate the PQC kernels, and their dedicated AI~Engines run the edge inference pipeline without cross-subsystem resource contention~\cite{amd_versal_gen2,amd_xqr_versal}.

For SWaP-constrained nodes, energy per reconciliation matters as much as latency. Published embedded energy profiles place the NIST lattice schemes on the same order of magnitude as elliptic-curve cryptography per operation, with hash-based SLH-DSA one to two orders of magnitude higher~\cite{roma2021energy}, which supports the lattice-first recommendations on energy grounds as well. Direct energy-per-consensus-round measurement (RAPL counters on the baseline; on-module power telemetry on the Jetson and Versal targets) is included in the planned on-device campaign.

\subsection{PQC Support Across the Edge Platforms}
\label{subsec:pqcsupport}
The cryptographic frontier has moved since the study experiments were run, and the proposed protocol is deliberately positioned to track it. Three primitives are now finalized---ML-KEM, ML-DSA, and SLH-DSA (FIPS~203/204/205, August~2024)~\cite{NIST2024FIPS203,NIST2024FIPS204,NIST2024FIPS205}---and two more are in the standardization pipeline: HQC, selected in March~2025 as a code-based key-encapsulation backup to ML-KEM with a draft anticipated as FIPS~207~\cite{nistIR8545}, and FN-DSA (Falcon), advancing as the draft FIPS~206 signature standard~\cite{nist_fips206}. FN-DSA is of direct interest here because its signatures are compact (on the order of \mbox{0.7\,KB}, far smaller than ML-DSA's multi-kilobyte outputs), which is highly attractive for the bandwidth-limited tactical links of Section~\ref{subsec:size_metrics}. The trade-off is that its floating-point signing path is harder to render constant-time on embedded hardware, so it is best treated as an emerging option rather than a drop-in replacement for ML-DSA. SIKE, by contrast, remains permanently excluded. The crypto-agility built into PARS---the ability to negotiate the algorithm and parameter set per operational scenario---is precisely what lets a deployment adopt HQC alongside ML-KEM, or FN-DSA where signature size dominates, as those standards finalize, without redesigning the protocol.

What matters operationally is not only that an algorithm is standardized but that also availability on each class of edge hardware, in a form whose performance and side-channel posture suit the node. Five such forms now cover the targets of Section~\ref{subsec:platform}. On Linux-capable application processors---the Jetson line, the Zynq/UltraScale+/Versal/XQR PS, the TI Sitara AM69x, and the desktop baseline---OpenSSL~3.5 (an April~2025 long-term-support release) ships native ML-KEM, ML-DSA, and SLH-DSA in its provider framework, with the default TLS~1.3 key exchange already a hybrid X25519\,+\,ML-KEM-768 group, and the Open~Quantum~Safe provider adds Falcon and further hybrids for experimentation~\cite{openssl35,stebila2016oqs}. On GPUs, NVIDIA cuPQC provides side-channel-protected, batched ML-KEM-512/768/1024 and ML-DSA-44/65/87 together with GPU SHA-2/SHA-3/SHAKE and Merkle-tree primitives, deploying on the same architectures from Jetson edge and industrial modules to data-center parts~\cite{nvidia_cupqc}.

On reconfigurable fabrics and ASICs, standalone hardware cores place ML-KEM and ML-DSA within the Artix-7-class fabric of the Zynq-7000 and the flash fabric of the PolarFire SoC at modest cost~\cite{xing2021kyberfpga,beckwith2021dilithiumfpga}. Commercial post-quantum IP (for example, the PQShield PQPlatform/UltraPQ-Suite family) supplies ML-KEM/ML-DSA---and SLH-DSA in some products---as drop-in FPGA/ASIC blocks with differential-power-analysis (DPA) and fault-injection countermeasures and CNSA-2.0-oriented roots of trust~\cite{pqshield_pqplatform}. Commercial accelerator cards built on the AMD~Versal family demonstrate the same primitives at high throughput on the adaptive-SoC class that hosts the flagship Versal~AI~Edge Gen~2 and XQR targets~\cite{dyber_quac100}. Finally, on microcontroller-class nodes such as the i.MX~RT1064 and the PolarFire hard RISC-V cores running bare-metal/RTOS, hand-optimized open-source assembly runs ML-KEM and ML-DSA via pqm4 loops~\cite{kannwischer2019pqm4,abdulrahman2022cortexm4}, and ultra-compact commercial libraries (PQShield's PQMicroLib-Core, under \mbox{5\,KB} of RAM) bring NIST PQC within the few-kilobyte budgets typical of secure boot on such parts~\cite{pqshield_pqplatform}. The defense/aerospace-grade parts reuse these same forms: the Zynq~UltraScale+ MPSoC and the space-grade XQR~Versal expose ML-KEM/ML-DSA through PL cores and hardened IP with the PS running OpenSSL~3.5, while the RISC-V PolarFire~SoC combines a software/Linux path with FPGA-fabric cores and RISC-V hardware acceleration~\cite{karl2024riscv}, so radiation-tolerant and flight-qualified silicon does not forfeit algorithm coverage.

Silicon-hardened root-of-trust subsystems are also taking over system-level security. The Secure Device Manager (SDM) inside Altera Agilex devices provides hardware-enforced, PQC-based secure boot and bitstream decryption to block unauthorized firmware modifications~\cite{altera_agilex9}. Similarly, the Secure Enclave inside the Microchip PIC64-HPSC establishes a post-quantum Root of Trust (RoT), authenticating early boot stages with ML-DSA signatures and decrypting operational images with ML-KEM to protect tactical space payloads~\cite{microchip_pic64hpsc}. See Table~\ref{tab_pqc_support} for algorithm accessibility on each platform path.

\begin{table*}[!htb]
\centering
\caption{PQC Support Across the Edge Platforms}
\label{tab_pqc_support}
\resizebox{\linewidth}{!}{%
\begin{tabular}{|l|l|l|l|l|l|}
\hline
\textbf{Platform} & \textbf{ML-KEM} & \textbf{ML-DSA} & \textbf{SLH-DSA} & \textbf{FN-DSA / HQC} & \textbf{Primary support path} \\
\hline
Intel i5-12450H (baseline) & SW & SW & SW & draft (liboqs) & OpenSSL~3.5 / liboqs (AVX2) \\
\hline
Jetson Orin            & GPU, SW & GPU, SW & SW & draft (liboqs) & cuPQC + OpenSSL~3.5 (NEON/Ampere) \\
\hline
Jetson AGX Orin Ind.   & GPU, SW & GPU, SW & SW & draft (liboqs) & cuPQC + OpenSSL~3.5 (rugged ECC) \\
\hline
Jetson Nano            & GPU, SW & GPU, SW & SW & draft (liboqs) & cuPQC + OpenSSL~3.5 (NEON) \\
\hline
i.MX RT1064 (MCU)      & asm, IP & asm, IP & verify (IP) & draft & pqm4 + PQMicroLib ($<$5\,KB) \\
\hline
TI Sitara AM69x        & SW, HSM & SW, HSM & SW, HSM & draft (liboqs) & OpenSSL~3.5 + HSM PQC extensions \\
\hline
PolarFire SoC / RT (rad-tol.) & PL, IP, SW & PL, IP, SW & PL, IP & IP roadmap & FPGA cores + RISC-V SW/accel.~\cite{karl2024riscv} \\
\hline
Zynq-7000 (PS + PL)    & PL, IP & PL, IP & PL, IP & IP roadmap & standalone cores + IP; PS via OpenSSL~3.5 \\
\hline
Zynq UltraScale+ MPSoC & PL, IP & PL, IP & PL, IP & IP roadmap & PL cores + IP; PS via OpenSSL~3.5 (A53) \\
\hline
Versal AI Edge Gen~2   & PL, IP & PL, IP & PL, IP & IP roadmap & PL cores + IP + PS OpenSSL; Versal accel.\ cards \\
\hline
XQR Versal Space-Grade & PL, IP & PL, IP & PL, IP & IP roadmap & rad-hard PL cores + hardened PS OpenSSL (A72) \\
\hline
Altera Agilex 9 Direct RF & PL, IP, SDM & PL, IP, SDM & PL, IP & PL/IP roadmap & Secure Device Manager + soft PQC IP \\
\hline
Microchip PIC64-HPSC   & SW, Enclave & SW, Enclave & SW (vector) & RVV vector draft & SiFive RVV compiler + hardened Secure Enclave \\
\hline
\end{tabular}%
}
\end{table*}

The practical conclusion is that every target in the fleet already has at least a software path to the two primitives the architecture leans on most---ML-KEM for key establishment and ML-DSA for signatures---and most targets additionally offer an accelerated or hardware-hardened path. The proposed protocol can therefore be instantiated across the whole heterogeneous set today using standardized algorithms, while its crypto-agility leaves room to fold in HQC and FN-DSA as their standards land.

\subsection{Securing Heterogeneous Edge~AI Data Flows}
\label{subsec:dataflow}
An operational Edge~AI node does not move a single kind of traffic. Between peer nodes, and between a node and the ground segment, it must transfer and exchanges multiple data classes with different size, rate, latency, and assurance requirements at least: high-rate sensor and imagery streams; federated-learning model updates; low-latency telemetry and C4ISR control traffic; node configuration databases; signed software and firmware images; and archival analysis or mission logs. These classes differ by orders of magnitude in size, periodicity, and latency tolerance, and---crucially---each is best served by a different part of a heterogeneous CPU/GPU/PS/PL device.

The unifying observation is that post-quantum cryptography is used here to protect \emph{keys and authenticity}, not to encrypt bulk payloads directly: a PQC KEM (ML-KEM) establishes a shared secret that a fast symmetric AEAD (for example AES-256-GCM) then uses to protect the bulk stream, while a PQC signature (ML-DSA, or a hash-based scheme where only offline signing is required) supplies integrity and origin authentication. Consequently the large PQC keys, ciphertexts, and signatures quantified in Section~\ref{subsec:size_metrics} are a per-session/update cost rather than a per-packet one---precisely the cost that PARS confines to controlled PSTC windows and that ESL records immutably. Quantitatively, the steady-state per-packet cost collapses to the 16-byte AES-256-GCM tag, while the kilobyte-scale PQC objects are paid once per session or update; the overhead column of Table~\ref{tab:dataflow_crypto_patterns} and ~\ref{tab:dataflow_hardware_roles} makes this amortization explicit using the FIPS-specified sizes \cite{NIST2024FIPS203,NIST2024FIPS204,NIST2024FIPS205} and maps representative traffic classes to illustrative cryptographic patterns and computing substrates. The object-size values are derived from the cited standards and should be recalculated against the exact parameter sets and protocol encoding used in the implementation.

The architecture uses Parts~\ref{alg:cnrme_enc} and~\ref{alg:cnrme_derive} optionally for key establishment, signatures, software authentication, and security-state transitions. It does not use public-key cryptography to encrypt every application packet. In typical processes, a verified signature or authentication mechanism is used to link the handshake to the intended identity or derive as given below. In a configuration using a 128-bit GCM authentication tag, the steady-state cryptographic expansion is 16~B per protected record, in addition to the nonce, header, framing, and transport overhead. KB-scale PQC objects are paid during session establishment, rotation, signing, or update rather than for every data packet.
\[
\mathrm{ML\mbox{-}KEM ~\ref{alg:cnrme_enc}}
\rightarrow
\mathrm{shared\ secret}
\rightarrow
\mathrm{KDF}
\rightarrow
\mathrm{AES\mbox{-}256},
\]
\[
\mathrm{Secret\mbox{-}Derive ~\ref{alg:cnrme_derive}}
\rightarrow
\mathrm{shared\ secret}
\rightarrow
\mathrm{KDF}
\rightarrow
\mathrm{AES\mbox{-}256},
\]

\begin{table*}[!t]
\centering
\caption{Cryptographic Patterns and Overhead Across Representative Edge AI Data Classes}
\label{tab:dataflow_crypto_patterns}
\renewcommand{\arraystretch}{1.25}
\footnotesize
\begin{tabular}{|p{3.2cm}|p{3.8cm}|p{4.2cm}|p{5.0cm}|}
\hline
\textbf{Data Flow Class} & \textbf{Operational Transfer Profile} & \textbf{Cryptographic Pattern} & \textbf{PQC Per-Object Overhead \& Steady-State Cost} \\
\hline
\textbf{Sensor / Streaming Imagery} & Large volume, continuous streaming, soft real-time constraint. & ML-KEM-768 $\rightarrow$ AEAD bulk stream; ML-DSA-65-signed keyframes. & $\approx$\,2.3\,KB per session (ek + ct) $+$ 3.3\,KB keyframe signature; 16\,B tag per packet. \\
\hline
\textbf{Federated Learning Updates} & Periodic batch transfer, MB-scale parameter tensors, delay-tolerant. & ML-DSA-65-signed model tensors over established KEM channel. & 3.3\,KB update signature ($\ll$\,1\% of an MB-scale tensor); negligible relative overhead. \\
\hline
\textbf{Telemetry / C4ISR Control} & Small payload, high-frequency broadcast, hard real-time constraint. & ML-KEM-512 / ML-DSA-44 $+$ symmetric AEAD stream. & $\approx$\,1.6\,KB per session (ek + ct) $+$ 2.4\,KB key signature; 16\,B tag per packet. \\
\hline
\textbf{Configuration Database} & Infrequent synchronization, KB--MB scale, integrity-critical. & ML-DSA-65/87-signed snapshot manifests and state hashes. & 3.3--4.6\,KB per snapshot signature; amortized across rare sync events. \\
\hline
\textbf{Software / Firmware (OTA)} & Rare distribution, large image size, very high integrity requirement. & Stateful LMS/XMSS ($h=10$) or ML-DSA-87, offline signed$^{\star}$. & $\approx$\,1.5--2.5\,KB per image (LMS/XMSS) or 4.6\,KB (ML-DSA-87); 7.9\,KB (SLH-DSA-128s). \\
\hline
\textbf{Analysis / Mission Logs} & Variable batch size, archival storage, provenance-critical. & ML-DSA-87 or SLH-DSA-128s $+$ timestamped state hash$^{\star}$. & 4.6--7.9\,KB per record batch signature; long-term tamper-evident proof. \\
\hline
\end{tabular}
\par\smallskip
\scriptsize $^{\star}$On accredited National Security Systems (NSS), CNSA~2.0 mandates stateful LMS/XMSS (single-tree, SP~800-208) or ML-DSA-87 for firmware/software signing and excludes SLH-DSA and FN-DSA; SLH-DSA-128s remains the stateless offline option outside NSS. NSS deployments also elevate KEM/DSA selections to Category-5 parameter sets (ML-KEM-1024 / ML-DSA-87).
\end{table*}
\begin{table*}[!t]
\centering
\caption{Hardware Substrate Allocation and PARS-ESL Protocol Roles for Edge AI Traffic}
\label{tab:dataflow_hardware_roles}
\renewcommand{\arraystretch}{1.25}
\footnotesize
\begin{tabular}{|p{3.2cm}|p{3.8cm}|p{4.6cm}|p{4.6cm}|}
\hline
\textbf{Data Flow Class} & \textbf{Primary Target Substrate} & \textbf{PARS-ESL Architectural Role} & \textbf{Key Epoch Management \& Rekeying Trigger} \\
\hline
\textbf{Sensor / Streaming Imagery} & GPU / AI Engines (Inference) $+$ Spatial PL Fabric (AEAD Bulk) & Negotiates ephemeral session keys via PARS; records state anchors on ESL. & Time-Based (TB) Periodic Security Threshold Control (PSTC) windowed re-keying. \\
\hline
\textbf{Federated Learning Updates} & GPU / AI Engines (Tensors) $+$ Host Processing System (PS) & Provides ledger-anchored model provenance; validates client updates immutably. & Windowed re-keying aligned with federated learning round completion epochs. \\
\hline
\textbf{Telemetry / C4ISR Control} & Deterministic PL Datapaths $+$ Real-Time Cores (e.g., Cortex-R5F) & Enforces low-latency authentication filters with minimal memory footprint. & Action-Based (AB) / Group-Based (GB) threshold triggers upon operational state shift. \\
\hline
\textbf{Configuration Database} & Host Processing System (PS) Secure Enclave / Flash Storage & Maintains versioned, immutable configuration snapshot roots on-chain. & Event-driven reconciliation upon policy update or node group reconfiguration. \\
\hline
\textbf{Software / Firmware (OTA)} & Parallel PL Keccak Engines (Verify) $+$ Hardened RoT Enclave & Writes immutable release image hashes and verification receipts to the ledger. & Offline signed; validated on boot/update prior to key epoch transition. \\
\hline
\textbf{Analysis / Mission Logs} & Host PS Core $+$ High-Throughput PL Cryptographic Hashing & Commits archival root hashes to ESL for post-mission auditing and traceability. & Continuous epoch-stamped logging decoupled from live command channels. \\
\hline
\end{tabular}
\end{table*}

 Table~\ref{tab:dataflow_crypto_patterns} and ~\ref{tab:dataflow_hardware_roles} maps each data class onto its transfer profile, the cryptographic pattern and its per-object overhead, the compute substrate that best handles it, and the roles. It describes a clean separation for the PS, PL, GPU or main processing complex runs the protocol state machine, consensus, and key reconciliation; the GPU or AI~Engines run on-device inference and, where throughput matters, the batched symmetric encryption of imagery and model tensors; and the PL fabric hosts the NTT and Keccak kernels for the PQC handshake together with a constant-time symmetric pipeline for the bulk channel. AEAD (Authenticated Encryption with Associated Data); OTA (Over-the-Air Update); EK/CT (Encapsulation Key/Ciphertext); SIG (Signature). Parameter sets are representative non-NSS choices. The overhead column is computed from the FIPS-/SP-specified object sizes \cite{NIST2024FIPS203,NIST2024FIPS204,NIST2024FIPS205,nist_sp800_208} and is paid per session/update inside PSTC windows; the steady-state per-packet cost is only the 16\,B AES-256-GCM tag. Transfer profiles follow the measured size and bandwidth results in size metrics; substrate assignments are design guidance for the target platforms, to be validated on-device.

Federated learning is the clearest illustration of why this matters: model-update exchange is communication-bound rather than compute-bound~\cite{mcmahan2017federated}, so the dominant cost is moving signed, integrity-protected parameter tensors over the same constrained links measured in ~\ref{subsec:size_metrics}. Signing each update with ML-DSA and rotating keys only on PSTC boundaries keeps the cryptographic overhead a small, amortized fraction of the parameter traffic, and ESL supplies tamper-evident provenance for every accepted update without requiring a central aggregator. Firmware and configuration distribution sit at the opposite extreme---rare, large, and integrity-critical---so they tolerate slow, offline signing. Here CNSA~2.0 is prescriptive for National Security Systems: firmware and software must be signed with the stateful hash-based schemes LMS or XMSS (single-tree, SP~800-208) or with ML-DSA-87, rather than with SLH-DSA, which CNSA~2.0 excludes~\cite{nsaCNSA2, nist_sp800_208}. The stateful schemes demand rigorous one-time-key state management---never reusing a leaf index---which is exactly why they are confined to the low-volume, controlled signing of firmware images; on the node side the PL's parallel Keccak engines accelerate verification, while SLH-DSA remains a stateless fallback for non-NSS contexts. The transfer profiles in the table follow from the measured size and bandwidth results of ~\ref{subsec:size_metrics}, and the overhead column from the FIPS-/SP-specified object sizes, combined with the architectural reasoning above; the substrate assignments are design guidance for the heterogeneous targets of ~\ref{subsec:platform} and, like the projections there, remain to be confirmed by an on-device measurement campaign.

Rather than providing a single homogeneous guideline, a tailored starting configuration across eleven platform-scenario combinations, plus the mandatory compliance baseline for classified (see Table~\ref{tab_recommendation}) for deployments. These entries are analytic architectural extrapolations derived from the microarchitectural scaling trends and core capabilities, rather than conclusions established directly by the baseline measurements. Each specific configuration must undergo extensive on-device validation within specific context.

A crucial compliance constraint overrides the latency-driven choices below for classified National Security Systems (NSS). NSA's CNSA~2.0 mandate approves only the Category-5 parameter sets---ML-KEM-1024 for key establishment and ML-DSA-87 for general signatures---and explicitly disallows the lighter ML-KEM-512/768 and ML-DSA-44/65 sets for NSS, along with SLH-DSA and FN-DSA; firmware and software signing must use stateful LMS/XMSS (SP~800-208) or ML-DSA-87, with an exclusive-use procurement gate from 1~January~2027 and full NSS transition by 2035~\cite{nsaCNSA2,nist_sp800_208}. The platform-specific rows of Table~\ref{tab_recommendation} therefore apply to non-NSS, coalition, training, or commercially-adjacent deployments where parameter agility is permitted, whereas accredited NSS must adopt the Category-5 baseline regardless of its cost. This compliance floor does not weaken the present contribution---it sharpens it. Because ML-KEM-1024 and ML-DSA-87 are the \emph{largest} standardized objects (an ML-DSA-87 signature alone is $\approx$\,4.6\,KB), the per-session and per-update overhead quantified in Section~\ref{subsec:size_metrics} is maximized precisely where it must be carried over the most contested links. The HNDL threat to long-lived classified mission data is what motivates the Category-5 floor, and the bandwidth, consensus, and memory discipline of PARS and ESL---which confine that overhead to controlled PSTC windows and amortize it across the swarm---is what makes a CNSA-2.0-compliant deployment feasible on bandwidth-constrained tactical and spaceborne nodes, rather than merely convenient. Compliance is ultimately enforced through validation: cryptographic modules must carry FIPS~140-3 (CMVP) validation and, for fielded NSS products, NIAP evaluation, so the lattice and hash-based primitives discussed here should be sourced as validated implementations---a requirement that further favors the commercial PQC IP and accelerators already entering the FIPS~140-3 pipeline over ad~hoc software ports~\cite{pqshield_pqplatform}. Three fundamental hardware-crypto design rules emerge from this extended architectural matrix.
\begin{enumerate}
    \item \textbf{Lattice Dominance for Real-Time Constraints:} Structured lattice schemes (ML-KEM for key encapsulation and ML-DSA for digital signatures) must be selected for all real-time re-keying loops on constrained nodes like the i.MX RT1064, PolarFire SoC, and Zynq-7000. Their sub-millisecond hardware core execution latency and tight byte footprint keep both transaction propagation and network consensus latencies functionally manageable within the swarm.
    \item \textbf{Isolation of Hash-Based Signatures:} SLH-DSA (SPHINCS+) is fundamentally unsuited for live, latency-sensitive key rotation across wireless edge configurations due to its $\approx$\,75\,ms processing bottleneck and heavy signature payload sizes. As mapped in Table~\ref{tab_recommendation}, SLH-DSA use should be restricted exclusively to offline command-node root signing, uncritical long-term analysis logs, or firmware image verification where stateless permanent authenticity outvalues execution latency---and, only outside NSS, where CNSA~2.0 instead mandates LMS/XMSS or ML-DSA-87~\cite{nsaCNSA2}. Where SLH-DSA is used offline, the small-signature parameter sets (e.g., SLH-DSA-128s) are preferred over the fast-signing variants: offline signing tolerates the slower signing time in exchange for materially smaller signatures for ledger and link~\cite{NIST2024FIPS205}.
    \item \textbf{Absolute Exclusion of Defective Primitives:} SIKE remains excluded from every configuration and substrate for the reasons established.
\end{enumerate}

\begin{table*}[!htb]
\centering
\caption{Platform-Specific Parameter Extrapolations Based on GCS}
\label{tab_recommendation}
\resizebox{\linewidth}{!}{%
\begin{tabular}{|l|l|c|c|c|c|l|}
\hline
\textbf{Operational Scenario} & \textbf{Target Hardware Platform} & \textbf{Swarm Size} & \textbf{KEM Choice} & \textbf{DSA Choice} & \textbf{ZK Level} & \textbf{PSTC Configuration} \\
\hline
Peacetime Base Security / C2 & NXP i.MX RT1064 (Cortex-M7 MCU) & $\le$\,4 nodes & ML-KEM-512 & ML-DSA-44 & Level 1 & $\approx$\,60 min (TB, Low Footprint) \\
\hline
Low-Power Tactical Comms & Microchip PolarFire SoC (RISC-V + PL) & $\le$\,8 nodes & ML-KEM-512 & ML-DSA-44 & Level 2 & $\approx$\,30 min (TB, Power-Optimized) \\
\hline
Civilian Edge Logistics Swarms & NVIDIA Jetson Nano (Cortex-A57 Base) & $\le$\,8 nodes & ML-KEM-768 & ML-DSA-65 & Level 1 & $\approx$\,30 min (TB, Software EVP Path) \\
\hline
Active Tactical Mobile Maneuvers & AMD/Xilinx Zynq-7000 (Cortex-A9 + PL) & $\le$\,12 nodes & ML-KEM-768 & ML-DSA-65 & Level 2 & $\approx$\,15 min (AB, Spatial PL Acceleration) \\
\hline
Tactical Drone Swarm & TI Sitara AM69x (Cortex-A72 + HSM) & $\le$\,12 nodes & ML-KEM-768 & ML-DSA-65 & Level 2 & $\approx$\,15 min (AB, Real-time Isolated) \\
\hline
Tactical Recon Radar & Altera Agilex 9 Direct RF (Cortex-A53 + PL) & $\le$\,12 nodes & ML-KEM-768 & ML-DSA-65 & Level 2 & $\approx$\,15 min (AB, Direct-RF Mode) \\
\hline
Rugged Airborne / Land AI Swarm & NVIDIA Jetson AGX Orin Industrial & $\le$\,16 nodes & ML-KEM-768 & ML-DSA-65 & Level 1 & $\approx$\,15 min (AB, Batched cuPQC Engine) \\
\hline
High-Threat Distributed EW & AMD Versal AI Edge Gen 2 SoC & $\le$\,16 nodes & ML-KEM-1024 & ML-DSA-87 & Level 3 & 5--10 min (GB, Parallel PL+AIE Datapath) \\
\hline
Strategic Satellite Constellations & AMD XQR Versal Space-Grade SoC & 4--8 nodes & ML-KEM-1024 & ML-DSA-87 & Level 4 & 5--10 min (GB, TMR Radiation-Hardened) \\
\hline
Autonomous Lunar Exploration & Microchip PIC64-HPSC (Space MPU, RVV) & 4--8 nodes & ML-KEM-1024 & ML-DSA-87 & Level 4 & 5--10 min (GB, DCLS Fault-Tolerant) \\
\hline
Command Node Root Operations & Intel i5-12450H (Desktop Base) & Host Node & ML-KEM-1024 & SLH-DSA-128s & Level 4 & Continuous / Event-Driven (Archival) \\
\hline
Classified NSS (CNSA~2.0 floor) & Any Accredited Platform & Any & ML-KEM-1024 & ML-DSA-87 & Level 3--4 & Mission-Defined (AB/GB) \\
\hline
\end{tabular}%
}
\end{table*}

The 91.05\,s worst case is an artifact of the excluded SIKE configuration at maximum hardness and swarm size (Section~\ref{subsec:consensus_time}) and is not an operational roadblock; with the recommended lattice-based primitives (ML-KEM and ML-DSA), consensus latency remains in a stable single-digit to low-tens-of-seconds operating range, and for latency-critical updates the swarm should be partitioned into smaller clusters driven by AB or GB thresholds rather than a long TB interval for operational interpretation of timing and resource trends.

Furthermore, because the PSTC key-renewal cadence functions independently from the underlying consensus runtime (Section~\ref{subsec:denial}), edge nodes can decouple security refresh rates from processing latency and tune the renewal cadence to the mission tempo without inflating per-round consensus cost. This separation is what makes the protocol viable in the denied, degraded, intermittent, and limited-bandwidth (DDIL) communications environments characteristic of contested operations: when the link budget collapses, an operator can lengthen the PSTC interval and fall back to action- or group-triggered re-keying without sacrificing the integrity of the ledger, trading rekey frequency for assured delivery rather than failing closed. When moving into severe environments like strategic space orbits or dense electromagnetic environments (as handled by the XQR Versal, the PolarFire SoC, or the PIC64-HPSC), the protocol likewise limits scaling strain by partitioning large swarms into small, independent sub-clusters running GB or AB threshold filters.

Four critical architectural limitations bound the absolute generalizability of these findings. First, the platform-specific matrix entries (see Table~\ref{tab_platform_projection}, Table~\ref{tab_pqc_support} and Table~\ref{tab_recommendation}) represent analytical architectural projections extrapolated from the GCS baseline data and verified third-party hardware acceleration literature, rather than physical on-device measurements. Initiating a physical benchmarking campaign across the target hardware fleet (specifically the Jetson AGX Orin Industrial, Zynq~UltraScale+ MPSoC, PolarFire SoC, XQR Versal, Agilex~9, Sitara AM69x, and PIC64-HPSC) is an immediate operational objective. Second, the operational failure rates ($10^{-4}$ to $10^{-1}$) lack explicit confidence intervals and stated trial sets, meaning they must be treated as approximate architectural trends rather than rigorous statistical metrics. Third, the resource mappings rely on the assumption of symmetric/homogeneous algorithm deployments, meaning complex hybrid setups (e.g., X25519 combined with ML-KEM) remain unprofiled. Fourth, the data-flow mappings detailed in Table~\ref{tab:dataflow_crypto_patterns} and ~\ref{tab:dataflow_hardware_roles} represent conceptual structural guidance derived from raw byte calculations and must be empirically validated within an active edge platform.

Future work proceeds in two stages. The near-term stage hardens the evidence base of this section: re-running the consensus, denial, and failure campaigns with at least 30 trials per condition (medians, 95\% confidence intervals, per-caption trial counts); adding a classical X25519/Ed25519 control run of ESL to isolate protocol overhead from post-quantum overhead~\cite{rfc7748,bernstein2012ed25519}; instrumenting a stage-by-stage latency decomposition of the consensus round; profiling energy per consensus round~\cite{roma2021energy}; and validating the cross-platform projections on at least two embedded targets (a Jetson-class module and a Zynq-7000 board), reporting measured-versus-projected error for Table~\ref{tab_platform_projection}. The second stage executes the fully instrumented hardware campaign on an AMD XQR Versal, a Microchip PolarFire SoC board, and a Microchip PIC64-HPSC development kit; implements end-to-end post-quantum cryptographic container environments; adds an adversarial (Byzantine-node) evaluation; and integrates the underlying PARS ledger directly into distributed privacy-preserving edge frameworks---enabling privacy-preserving model updates for nodes without weakening the cryptographic guarantees~\cite{mcmahan2017federated}.

\subsection{Hardware Co-Design Analysis Across 12 Target Substrates}
\label{subsec:multi_platform_hardware_codesign}

\begin{table*}[!t]
\centering
\caption{Cross-Platform Hardware Co-Design, PQC Support Tier, and PARS-ESL Overview}
\label{table_platform_pqc_tradeoffs_summary}
\resizebox{\textwidth}{!}{%
\begin{tabular}{|l|p{3.1cm}|p{2.8cm}|p{2.2cm}|p{3.2cm}|p{3.2cm}|p{2.8cm}|p{3.0cm}|}
\hline
\textbf{Operational Tier} & \textbf{Target Platform} & \textbf{Compute Architecture \& Substrate} & \textbf{PQC Support Status \& Primary Path} & \textbf{Key Architectural Pros (Advantages)} & \textbf{Key Constraints \& Cons (Trade-offs)} & \textbf{Applied Protocol Paradigm} & \textbf{Recommended Algorithm Suite} \\
\hline
%\hline
\textbf{Gateway / GCS} 
 & \textbf{Intel Core i5-12450H} \newline \emph{(GCS Baseline)} 
 & 8C/12T @ 3.20\,GHz \newline (64\,GB DDR4) 
 & \textbf{Full PQC (Native)} \newline OpenSSL 3.5 / liboqs (AVX2 Vectorized) 
 & High raw throughput, native AVX2 SIMD acceleration, full support for stateful and stateless post-quantum schemes. 
 & High active power draw (45--95\,W), bulky form factor, unsuited for SWaP-constrained field or airborne platforms. 
 & \textbf{Full PQC Node} \newline (PARS-ESL Root / Validator) 
 & ML-KEM-1024 / SLH-DSA-128s \newline ZKP Lvl 4, Continuous / Event \\
\hline
%\hline
\multirow{3}{*}{\shortstack[l]{\textbf{Strategic Space} \\ \textbf{\& Orbital AI}}} 
 & \textbf{AMD XQR Versal} \newline \emph{(Space Adaptive SoC)} 
 & Rad-Hard Cortex-A72 \newline + TMR-Hardened PL + AIE 
 & \textbf{Full PQC (Accelerated)} \newline Hardened PL Cores + PS OpenSSL 3.5 
 & MIL-PRF-38535 Class B space-flight qualified, TMR-hardened spatial PL eliminates SEUs, AI Engines isolate orbital vision pipelines. 
 & High unit procurement cost, steep learning curve for AI Engine graph programming and spatial routing toolchains. 
 & \textbf{Full PQC Node} \newline (Mode A, Accelerated) 
 & ML-KEM-1024 / ML-DSA-87 \newline ZKP Lvl 4, PSTC: 5--10\,min (GB) \\
\cline{2-8}
 & \textbf{Microchip PIC64-HPSC} \newline \emph{(NASA Space MPU)} 
 & 8$\times$ SiFive X280 @ 1.0\,GHz \newline (512-bit Vector RVV) 
 & \textbf{Full PQC (Accelerated)} \newline 512-bit RVV ISA + Hardened Enclave 
 & 512-bit vector registers accelerate polynomial NTT loops without discrete FPGA fabric, Secure Enclave provides a hardware RoT. 
 & Emerging RISC-V space software ecosystem, non-reconfigurable logic (fixed ISA), early-stage RVV auto-vectorizing compilers. 
 & \textbf{Full PQC Node} \newline (Mode A, Vectorized) 
 & ML-KEM-1024 / ML-DSA-87 \newline ZKP Lvl 4, PSTC: 5--10\,min (GB) \\
\cline{2-8}
 & \textbf{Microchip RT PolarFire} \newline \emph{(Rad-Tolerant RISC-V)} 
 & Quad SiFive U54 (64-bit) \newline + Flash-FPGA Fabric 
 & \textbf{Full PQC in PL / Hybrid} \newline Custom PL Core + RISC-V Linux 
 & Non-volatile flash FPGA is inherently immune to configuration memory SEUs (no scrubber needed), lowest static power ($<$3\,W). 
 & In-order 64-bit RISC-V application cores exhibit severe software-only execution penalties without custom PL accelerator cores. 
 & \textbf{Full PQC (PL Mode A) /} \newline \textbf{Classical (PS Mode B)} 
 & ML-KEM-512 / ML-DSA-44 \newline ZKP Lvl 2, PSTC: $\approx$30\,min (TB) \\
\hline
%\hline
\multirow{4}{*}{\shortstack[l]{\textbf{Tactical Defense} \\ \textbf{\& Electronic Warfare}}} 
 & \textbf{Altera Agilex 9} \newline \emph{(Direct RF SoC FPGA)} 
 & Quad Cortex-A53 (HPS) \newline + Direct RF PL (64 Gsps) 
 & \textbf{Full PQC (Accelerated)} \newline Direct-RF PL + Secure Device Manager 
 & Direct RF sampling (up to 64\,Gsps) enables real-time spectral EW integration, Secure Device Manager (SDM) enforces PQC boot. 
 & High static/dynamic power draw (25--50\,W), requires advanced conduction cooling enclosures in sealed tactical missile/pod bays. 
 & \textbf{Full PQC Node} \newline (Mode A, Direct RF) 
 & ML-KEM-768 / ML-DSA-65 \newline ZKP Lvl 2, PSTC: $\approx$15\,min (AB) \\
\cline{2-8}
 & \textbf{AMD Versal AI Edge Gen 2} \newline \emph{(Commercial SoC)} 
 & Cortex-A78AE + R52 Cores \newline + Expansive PL + AI Engines 
 & \textbf{Full PQC (Accelerated)} \newline PL IP Cores + PS OpenSSL 3.5 
 & High functional safety (ASIL-D / SIL-3), optimal workload partitioning across dual real-time cores, application complex, and vector engines. 
 & Power budget (10--25\,W) exceeds ultra-low SWaP drone batteries, complex AXI interconnect bus arbitration under heavy multimodal load. 
 & \textbf{Full PQC Node} \newline (Mode A, Accelerated) 
 & ML-KEM-1024 / ML-DSA-87 \newline ZKP Lvl 3, PSTC: 5--10\,min (GB) \\
\cline{2-8}
 & \textbf{AMD Zynq UltraScale+} \newline \emph{(Defense MPSoC)} 
 & Quad Cortex-A53 + Dual R5F \newline + UltraScale+ PL (XQ Grade) 
 & \textbf{Full PQC (Accelerated)} \newline PL IP Cores + NEON / OpenSSL 3.5 
 & Ruggedized military packaging ($-55$ to $+125\,^{\circ}$C), built-in 256-bit PUF for key storage, strong ARMv8 NEON crypto extensions. 
 & SRAM-based configuration cells require background soft-error mitigation (SEM) scrubbers in high-altitude radiation environments. 
 & \textbf{Full PQC Node} \newline (Mode A, Accelerated) 
 & ML-KEM-768 / ML-DSA-65 \newline ZKP Lvl 2, PSTC: $\approx$15\,min (AB) \\
\cline{2-8}
 & \textbf{AMD/Xilinx Zynq-7000} \newline \emph{(Legacy Hybrid SoC)} 
 & Dual Cortex-A9 @ 667\,MHz \newline + Artix-7 PL (53k LUTs) 
 & \textbf{Full PQC in PL / Fallback} \newline Dedicated Artix-7 PL Datapath 
 & Low BoM cost, mature Vivado/PetaLinux toolchain, Artix-7 PL fabric completely bypasses the legacy 32-bit CPU compute bottleneck. 
 & Severe CPU software bottleneck (150--300$\times$ slowdown), limited on-chip Block RAM (4.9\,Mb) restricts concurrent batch verification. 
 & \textbf{Full PQC (with PL) /} \newline \textbf{Classical Mode B (PS-Only)} 
 & ML-KEM-768 / ML-DSA-65 \newline ZKP Lvl 2, PSTC: $\approx$15\,min (AB) \\
\hline
%\hline
\multirow{4}{*}{\shortstack[l]{\textbf{Robotics, Swarms} \\ \textbf{\& Industrial Edge}}} 
 & \textbf{NVIDIA Jetson AGX Orin Ind.} \newline \emph{(Industrial AI Module)} 
 & 12C Cortex-A78AE \newline + Ampere GPU + cuPQC 
 & \textbf{Full PQC (GPU/NEON)} \newline cuPQC SDK + OpenSSL 3.5 
 & Industrial shock/vibration rated ($-40$ to $+85\,^{\circ}$C), hardware ECC RAM protects against bit-flips, batched cuPQC GPU acceleration. 
 & Activating ECC memory degrades raw bus bandwidth by 5--10\%, high peak power consumption (up to 75\,W) under intensive AI inference loads. 
 & \textbf{Full PQC Node} \newline (Mode A, Batched GPU) 
 & ML-KEM-768 / ML-DSA-65 \newline ZKP Lvl 1, PSTC: $\approx$15\,min (AB) \\
\cline{2-8}
 & \textbf{TI Sitara AM69x} \newline \emph{(Autonomous MPU)} 
 & 8$\times$ Cortex-A72 @ 2.0\,GHz \newline + Dual R5F + HSM (32 TOPS) 
 & \textbf{Full PQC (SW / HSM Ext)} \newline OpenSSL 3.5 + HSM PQC Support 
 & Dual Cortex-R5F real-time islands isolate critical actuator control loops, integrated HSM secures key storage, 32\,TOPS AI acceleration. 
 & Software-only PQC on Cortex-A72 is moderately slow, proprietary on-chip HSM requires customized firmware for non-standard PQC primitives. 
 & \textbf{Full PQC Node} \newline (Mode A, Isolated HSM) 
 & ML-KEM-768 / ML-DSA-65 \newline ZKP Lvl 2, PSTC: $\approx$15\,min (AB) \\
\cline{2-8}
 & \textbf{NVIDIA Jetson Orin} \newline \emph{(Standard Edge Module)} 
 & 6--12C Cortex-A78AE \newline + Ampere GPU (cuPQC SDK) 
 & \textbf{Full PQC (GPU/NEON)} \newline cuPQC SDK + OpenSSL 3.5 
 & Exceptional deep learning throughput per dollar, mature JetPack/cuPQC software ecosystem, rapid deployment for commercial robotic swarms. 
 & Dynamic power fluctuations under GPU bursts, lack of constant-time execution on GPU shaders requires software-level side-channel masking. 
 & \textbf{Full PQC Node} \newline (Mode A, cuPQC) 
 & ML-KEM-768 / ML-DSA-65 \newline ZKP Lvl 1, PSTC: $\approx$15\,min (AB) \\
\cline{2-8}
 & \textbf{NVIDIA Jetson Nano} \newline \emph{(Entry-Level Edge Module)} 
 & Quad Cortex-A57 @ 1.43\,GHz \newline + 128-core Maxwell GPU 
 & \textbf{Supported (Software Path)} \newline OpenSSL 3.5 (No ARMv8 Crypto) 
 & Low unit cost, accessible entry-level Linux platform for prototyping lightweight edge intelligence and distributed ledger clients. 
 & Older Cortex-A57 lacks ARMv8 crypto extensions, severe compute bottleneck on signing, memory pressure under 4\,GB LPDDR4 shared memory. 
 & \textbf{Full PQC (Relaxed PSTC) /} \newline \textbf{Classical Seed Mode B} 
 & ML-KEM-768 / ML-DSA-65 \newline ZKP Lvl 1, PSTC: $\approx$30\,min (TB) \\
\hline
%\hline
\textbf{Ultra-Low SWaP} 
 & \textbf{NXP i.MX RT1064} \newline \emph{(Crossover MCU)} 
 & Single Cortex-M7 @ 600\,MHz \newline (Hand-Optimized Assembly) 
 & \textbf{Constrained PQC / Fallback} \newline Hand-Optimized pqm4 / PQMicroLib 
 & Ultra-low power ($<$1\,W), sub-millisecond cold boot, highly compact footprint for distributed missile sensors and expendable drone pods. 
 & Lacks MMU (no Linux), strict 1\,MB SRAM requires manual buffer management, multi-millisecond signing latency, niche/complex bare-metal BSP integration. 
 & \textbf{Lightweight PQC (KEM-512) /} \newline \textbf{Classical over PARS (Mode B)} 
 & ML-KEM-512 / ML-DSA-44 \newline (or Mode B AES-256-CTR) \newline ZKP Lvl 1, PSTC: $\approx$60\,min \\
\hline
\end{tabular}%
}
\end{table*}

To establish the operational feasibility of the PARS-ESL framework across divergent mission-critical domains without introducing redundant single-platform plots, Table~\ref{table_platform_pqc_tradeoffs_summary} provides an exhaustive cross-platform comparison spanning 12 representative computing substrates. These platforms cover five distinct deployment tiers: (i)~Ground Control Station (GCS) gateways, (ii)~strategic space and orbital AI systems, (iii)~tactical electronic warfare and defense avionics, (iv)~autonomous robotic and UAV swarms, and (v)~ultra-constrained microcontroller endpoints.

The cross-platform scaling projections indicate that pure-software execution on resource-constrained embedded cores (such as the ARM Cortex-A57 on the Jetson Nano or the Cortex-A9 on the Zynq-7000) incurs a significant latency penalty ($12\times$ to $300\times$ slowdown relative to the GCS baseline) due to the heavy Number-Theoretic Transform (NTT) polynomial multiplication loops inherent to Module-LWE primitives. However, on heterogeneous adaptive SoCs (AMD Versal AI Edge, AMD Zynq UltraScale+, and Microchip RT PolarFire), offloading the NTT and Keccak permutations to dedicated Programmable Logic (PL) datapaths completely eliminates processor instruction-fetch contention, recovering desktop-equivalent cryptographic latencies (tens of microseconds) at single-digit Watt power budgets. Furthermore, vector-accelerated architectures, such as the NASA-JPL Microchip PIC64-HPSC equipped with 512-bit RISC-V Vector (RVV) registers, achieve high computational efficiency ($1.2\times$ to $2.2\times$ baseline latency) without requiring discrete FPGA coprocessors.

Crucially, the synthesis of Table~\ref{table_platform_pqc_tradeoffs_summary} reveals that hardware acceleration reduces computational runtime but cannot shrink post-quantum public keys, ciphertexts, or signature byte footprints. Because naive per-packet post-quantum cryptography would rapidly saturate constrained tactical data links (1.67--78.09\,Kbps), the rotational secrecy paradigm of PARS confines expensive lattice handshakes to controlled Periodic Security Threshold Control (PSTC) windows. By decoupling steady-state AES-256-GCM symmetric telemetry from on-chain security state coordination, the peak volatile memory footprint remains strictly below 3.5\,MB across all 12 platform classes. This proves that ledger-assisted post-quantum security is universally deployable across bare-metal microcontrollers, industrial robotic fleets, and radiation-hardened space constellations.
%\newpage
\section{CONCLUSION} \label{Conclusion}

\renewcommand{\footnoterule}{%
  \kern 6pt % Adjust the amount of padding as needed
  \hrule width 0.4\columnwidth % Width of the line
  \kern 2pt % Adjust the amount of padding as needed
}

This study examines the implications of quantum computing for the security of Edge-AI and cyber-physical ecosystems and for the blockchain layers that increasingly anchor their trust, with mission-critical defense as a leading motivating case. It provides a concise overview of current PQC algorithms capable of supporting edge. The proposed PARS protocol maintains dynamic key reconciliation among networked edge and IoT instruments---of which military IoT is the best-fit instance---and presents a robust security backbone built on a novel Layer-1 private quantum blockchain. Although quantum computing introduces significant threats to blockchain security dynamics, PARS demonstrates that such ecosystems can nevertheless thrive once specific prerequisites are met. The design is tailored for lightweight hardware environments which ensure broad applicability. It offers extended security and diverse attack-mitigation capabilities while operating with high storage efficiency. It excels in high-performance scenarios, supports heterogeneous group interactions, and facilitates secure swarm operations in contested edge environments, including kinetic military settings\footnote{Kinetic military environments consist of active military actions involving engaged warfare.} that constitute the most demanding deployment class. It has a decentralized log flow that enhances the transparency and traceability of distributed systems while reducing time complexity, addressing current challenges and laying the groundwork for a more secure and adaptive future. The protocol further enables autonomous swarms to converge on a key acceptable to the chain, ensuring that all encrypted operation records can be examined and recorded by immutable keys. It also provides a hardened ZKP method allowing keys to be renewed by consensus through an uninterferable chain, this key-management process ensures that operations cannot be controlled by adversaries at the autonomous level. Because all critical key-renewal and key-control operations occur within the chain at predetermined intervals, the architecture lets edge and CPS prepare for quantum attacks and manage the strongest available algorithms with the utmost confidentiality.

The experimental evaluation establishes several key points across average consensus time, size metrics, cumulative memory effects, denial rates, and failure results. Average consensus time is reported for different ZKP hardness levels and node-set sizes under each PQC algorithm, with every combination completing consensus in under 91.05 seconds. Node-set size and ZKP hardness are the two dominant factors affecting consensus time, and the results indicate that smaller swarm chunks yield more feasible consensus times. Size metrics---message size, transaction size, and block size---together with performance metrics such as bandwidth, iteration cycle, and transaction frequency are likewise characterized. Message sizes range from 110 Bytes to 9734 Bytes (about 9.5 KBytes) across the evaluated PQC algorithms, bandwidth demand varies from 1.67 Kbps to 78.09 Kbps, and the iteration cycle varies from 0.4 to 21.69. The maximum memory required for local chain operations stays below 256 KBytes. Cumulative memory effects measured over an 18-hour across different parameter settings exhibit a peak memory footprint under 3.5 MB, and failure rates for consensus operations remain between roughly 0.01 and 0.001.

As the results show, it is possible to harden Edge-AI and CPS against quantum attacks without imposing additional computational or storage workloads. The broader class of autonomous IoT deployments---spanning robotics, industrial control, and connected swarms, with the IoMT as its best-fit instance---attains a highly feasible model under the proposed approach, given the steep computational attack capabilities that QC bring. ESL, powered by the PARS protocol, provides robust security for Edge-AI systems through advanced chain monitoring, defending against a range of attacks while renewing security levels within the operational flow. One of its most compelling features is that it secures conventional systems without requiring quantum hardware; until the next revolution in chip design, systems can therefore obtain post-quantum security without incurring a new migration cost. By enabling secure interconnected operation across multi-domain settings, it strengthens the resilience and reliability of embodied CPS like robotics, vehicles, defense and space systems against potential quantum attacks. Within its best-fit defense context, the study additionally contributes a cyber-offensive doctrinal perspective, articulating seven novel doctrines that reframe how contested operations can be conducted under quantum threat---an angle, to our knowledge, not previously addressed alongside a robust security backbone and an efficient architecture of comparable strategic scope. Future work will extend the design through heterogeneous inter-ESLs in operational use, accompanied by expanded empirical results and performance metrics covering time complexity, security level, attack-mitigation analysis, storage, and operational cost.

%\clearpage

\begin{small}

\end{small}

\begin{IEEEbiography}[{\includegraphics[width=1in,height=1.25in,clip,keepaspectratio]{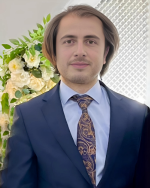}}]{K\"{U}R\c{S}AT~\c{C}AKAL} is currently with Roketsan Missiles Inc. and pursuing the Ph.D. degree with the Department of Computer Engineering, Hacettepe University, Ankara, Turkey. His research fields include embedded systems, edge ai, robotics, quantum computers, artificial intelligence, and security.
\end{IEEEbiography}

\begin{IEEEbiography}[{\includegraphics[width=1in,height=1.25in,clip,keepaspectratio]{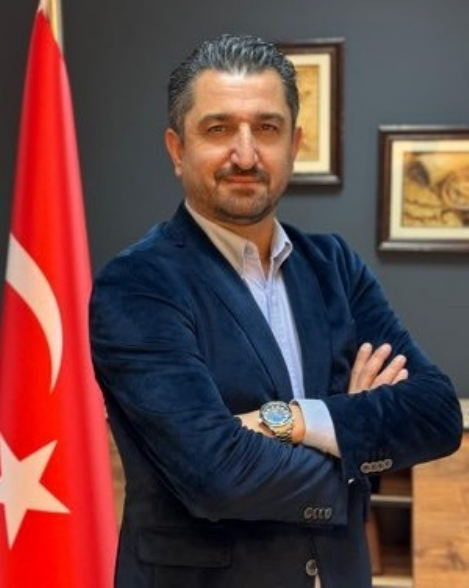}}]{SUAT~\"{O}ZDEM\.{I}R}
(Member, IEEE) is currently with the Department of Computer Engineering, Hacettepe University, Ankara, Turkey. His current research interests include the Internet of Things, data analytics, artificial intelligence, and network security.
\end{IEEEbiography}

\vfill

\end{document}